\documentclass[]{aa}  

\usepackage{graphicx}
\usepackage{txfonts}
\usepackage{lipsum}
\usepackage{subcaption}         
\usepackage{lscape}             
\usepackage{placeins}           
\usepackage{threeparttable}                                
\usepackage{hyperref}

\begin{document}

   \title{Asteroseismology of AI Velorum: a prototype of high-amplitude $\delta$ Scuti stars}


%
%
%

    \author{Przemys{\l}aw Walczak\corrauth{E-mail: przemyslaw.walczak@uwr.edu.pl}
    \and Jadwiga Daszyńska-Daszkiewicz
    \and Wojciech Niewiadomski
    \and Wojciech Szewczuk
    }  

   \institute{University of Wroc{\l}aw, Faculty of Physics and Astronomy, pl. Uniwersytecki 1, 50-137 Wroc{\l}aw, Poland}

   \date{Received September 30, 20XX}

 
\abstract
{AI Velorum is exceptional among high-amplitude $\delta$ Scuti (HADS) stars because of its multimode pulsational behaviour. In addition to the two radial modes, several non-radial modes have long been known. Although it is an excellent target for asteroseismic modelling, no detailed study of this prototype star has been performed to date.
}
{We determine the fundamental stellar parameters of AI Vel, including its age and evolutionary stage, and constrain the efficiency of envelope convection as well as convective core overshooting 
during the main-sequence phase. In addition, we test three widely used opacity data sets: OPAL, OP, and OPLIB.
}
{We analysed the \textit{TESS} light curve, ground-based Str\"omgren photometry, and high-resolution HARPS spectroscopy. Mode identification was performed using both photometric and spectroscopic methods. 
Empirical values of the bolometric flux amplitude (the parameter $f$) were derived for both radial modes.
Extensive seismic modelling was carried out within a Bayesian framework supported by Monte Carlo simulations.
}
{Our mode identification confirms that the two dominant frequencies correspond to radial modes. In addition, one non-radial mode is most consistent with an axisymmetric dipole or quadrupole identification. Within the explored parameter space, only seismic models computed with OPAL opacities reproduce the observed position of AI Vel in the HR diagram, whereas models based on OP and OPLIB opacities are systematically too cool and underluminous. All seismic models indicate a low efficiency of envelope convection, with the mixing-length parameter $\alpha_{\rm mlt}\lesssim0.5$. Including the non-radial mode substantially narrows the posterior distributions of the stellar parameters for a given mode identification. However, the remaining ambiguity between the dipole and quadrupole hypotheses introduces an additional uncertainty that is not reflected in the formal posterior errors. The seismic models indicate that AI Vel is an evolved, slowly rotating star in the hydrogen shell-burning phase.}
{
Our results demonstrate the diagnostic power of combining pulsation frequencies with empirical $f$-parameters in seismic studies of $\delta$ Scuti stars. The apparent preference for OPAL opacities found for AI Vel is consistent with results obtained for several similar HADS and SX Phoenicis stars studied so far, although this behaviour should not be interpreted as evidence for a general superiority of OPAL opacities.
}


   \maketitle
   \nolinenumbers

\section{Introduction}

Pulsating stars of the $\delta$ Scuti type have been known for decades and, in some cases, for more than a century. Notably, the variability of the prototype star, $\delta$ Scuti itself, was first reported in 1900 by \citet{1900ApJ....12..254C}.

In modern terminology, $\delta$ Scuti variables are defined as A0--F5 spectral-type stars \citep{1973A&A....23..221B,2000A&AS..144..469R} with luminosity classes ranging from V to III. They exhibit short-period brightness variations caused by both radial and non-radial pulsations in pressure ($p$), mixed, and gravity ($g$) modes. Their typical masses lie between 1.5 and 2.5\,M$_{\sun}$ \citep{2010aste.book.....A,2017ampm.book.....B}, and their pulsation periods range from approximately 0.5 to 6 hours \citep{2000ASPC..210....3B}. The pulsations are driven by the $\kappa$-mechanism operating primarily in the second helium ionization zone \citep{1962ZA.....54..114B,1965ApJ...142..868B,1963ARA&A...1..367Z,1971A&A....14...24C}.

Classical $\delta$ Scuti stars belong to Population I and are located predominantly on the main sequence, although they can also occur in more evolved evolutionary stages \citep{1998A&A...332..958B,2016MNRAS.460.1970B,2019MNRAS.485.2380M,2019MNRAS.490.4040A} or during the pre-main-sequence phase \citep{Zwintz2008,Zwintz2014,Zwintz2015}. A distinct subgroup, known as High-Amplitude $\delta$ Scuti (HADS) stars or dwarf Cepheids, is characterized by the photometric range in the Johnson $V$ band exceeding 0.3\,mag. These stars are usually more evolved, often during the hydrogen shell-burning phase, and typically exhibit slow rotation and pulsations dominated by one or two radial modes \citep{2000ASPC..210....3B}. Apart from their large amplitudes, however, HADS stars appear observationally similar to other $\delta$ Scuti variables and do not show clear differences in their fundamental stellar parameters \citep{1980ApJ...235..153B}.

$\delta$ Scuti stars follow a well-defined period--luminosity (P--L) relation \citep{1998A&A...331..989P,2010AcA....60....1P,2025RAA....25e5019L,2019MNRAS.486.4348Z}, which makes them useful distance indicators in stellar populations. For HADS stars the relation is even tighter, as their pulsation properties resemble those of classical Cepheids \citep{Poretti2008}.

In recent years, asteroseismology has become a powerful tool for probing the internal structure of $\delta$ Scuti stars. By analysing their pulsation spectra, it is possible to place strong constraints on stellar parameters such as mass, age, chemical composition, internal rotation, and the efficiency of convective energy transport \citep[e.g.][]{1995ASPC...76..596B,2000BaltA...9..149B,2000ASPC..210....3B,2018MNRAS.476.3169B,2019MNRAS.490.4040A,2019MNRAS.485.2380M,2022MNRAS.516.2080B,2024ApJ...960...94P,2024MNRAS.535.2927G,2024A&A...686A.174L,2025MNRAS.539.3381D}.

However, modelling $\delta$ Scuti stars remains challenging due to several physical uncertainties, including the treatment of convection in the outer envelope, rotational effects, and the choice of opacity data.

One of the most promising diagnostics for constraining some processes is the complex non-adiabatic $f$-parameter, defined as the ratio of the bolometric flux variation to the radial displacement at the stellar surface. The empirical determination of this parameter from multicolour photometry and radial velocity data provides a powerful test of stellar models, as the theoretical value of $f$ is highly sensitive to the efficiency of envelope convection and to the adopted opacity tables \citep{JDD2003}. Previous studies have demonstrated that comparisons between empirical and theoretical $f$ values can provide valuable constraints on convective energy transport in the stellar envelope as well as on the opacity data used in calculations.

Space-based photometry has further transformed the study of $\delta$ Scuti and HADS stars. In particular, \citet{2020Natur.581..147B} revealed remarkably regular high-frequency pulsation patterns in young $\delta$ Scuti stars, opening new possibilities for mode identification and seismic modelling. Space photometry has also shown that HADS stars, traditionally regarded as relatively simple radial pulsators, can exhibit rich spectra of low-amplitude non-radial modes \citep[e.g.,][]{Poretti2011}. Detailed seismic modelling of double-mode HADS stars has demonstrated that the combination of radial-mode frequencies with additional constraints, such as the non-adiabatic $f$-parameter and low-amplitude non-radial modes, can provide tight constraints on stellar parameters, convection, and opacity data \citep[e.g.,][]{2022MNRAS.512.3551D,2023MNRAS.526.1951D}. More recently, \citet{2025ApJ...995..188Y} used \textit{TESS} observations to investigate the pulsation properties and mode identification of a larger sample of HADS stars, further demonstrating the potential of high-precision space photometry for studying this class of pulsators.

In this context, AI Velorum is an especially valuable target. As the prototype of high-amplitude $\delta$ Scuti stars, it exhibits large-amplitude pulsations dominated by two radial modes, making mode identification relatively secure. At the same time, the presence of additional non-radial modes provides further seismic constraints. These properties make AI Velorum an ideal target for testing stellar structure and pulsation models.

In this work, we present a comprehensive asteroseismic analysis of AI Velorum using the most recent observational data available for this star. We combine \textit{TESS} photometry, ground-based multicolour observations, and high-resolution spectroscopy with modern stellar evolution and pulsation models. The modelling is carried out within a Bayesian framework supported by Monte Carlo simulations, enabling us to derive precise stellar parameters and investigate the effects of opacity data and convective efficiency on models of $\delta$ Scuti stars.

The paper is organized as follows. In Section\,\ref{Sect:AIvel}, we summarize the basic properties of AI Velorum and derive its fundamental parameters from spectroscopic analysis. Section~\ref{Sect:FA} describes the Fourier analysis of the available \textit{TESS} data for AI Velorum. In Section~\ref{Sect:MI}, we present the results of mode identification. The asteroseismic modelling is described in Section~\ref{Sect:Sim}. A summary and discussion are given in the final section.


\section{AI Velorum}
\label{Sect:AIvel}

\subsection{Previous studies and basic properties}

AI Velorum (AI Vel, HD\,69213) is the prototype of a class of HADS stars \citep{1992MNRAS.254...59W}.
Its visual magnitude in the Johnson $V$ passband is 6.98 mag \citep{2002yCat.2237....0D}. The spectral classification of AI Vel has changed over the years. It was classified as A4 at maximum brightness in the $V$ filter and F2 at minimum brightness \citep{1949ApJ...110..156H}. Later, it was classified as A1p at maximum luminosity and F1p at minimum luminosity \citep{1953ZA.....32...69G}.
In the General Catalogue of Variable Stars (GCVS) the range of spectral types is given as A2p-F2pIV/V \citep{2017ARep...61...80S}.

The first reports of AI Vel's pulsational variability were made by \citet{1952BAN....11..421W,1955BAN....12..223W}, who described the light variations by two periods: $P_0 = 0.111574$\,d and $P_1 = 0.0862073$\,d. \citet{1952BAN....11..421W} also suspected the presence of additional periodicities in the photometric data, specifically at $P_2 = 0.046460$\,d and $P_3 = 0.0317584$\,d, although he was unable to confirm their significance.  Forty years later, \citet{1992MNRAS.254...59W} confirmed the two dominant pulsation frequencies, 8.96260\,d$^{-1}$ ($P_0$) and 11.599779\,d$^{-1}$ ($P_1$) and reported additional modes based on data collected over a 38-year timespan. These included 22.521826\,d$^{-1}$, 15.972479\,d$^{-1}$, 9.13761\,d$^{-1}$ and 10.93000\,d$^{-1}$.

The parallax of AI Vel, $\pi = 9.970 \pm 0.019$\,mas, derived from {\it Gaia} DR3 data, corresponds to a distance of $100.40 \pm 0.19$\,pc. This value is in excellent agreement with the Hipparcos measurement of $\pi = 9.99 \pm 0.53$\,mas \citep{1997ESASP1200.....E}. The proper motion measured by {\it Gaia} is $26.222 \pm 0.019$\,mas\,yr$^{-1}$ in right ascension and $31.094 \pm 0.025$\,mas\,yr$^{-1}$ in declination. The brightness of the star in the {\it Gaia} $G$ band is 6.518\,mag \citep{2020yCat.1350....0G}.

The effective temperature derived from $uvby$ photometry \citep{1977PASP...89..339B} ranges between 7110 and 8520\,K.
Based on Gaia BP/RP spectra, the effective temperature is $T_{\rm eff} = 7188^{+14}_{-12}$\,K, the surface gravity is $\log g = 3.662^{+0.05}_{-0.03}$, and the metallicity is $[{\rm Fe/H}] = -1.330^{+0.003}_{-0.002}$. The extinction is estimated as $A_V = 0.059^{+0.007}_{-0.006}$\,mag in the Johnson $V$ band and $A_G = 0.053^{+0.006}_{-0.005}$\,mag in the {\it Gaia} $G$ band. The radial velocity is $V_{\rm rad} = 17 \pm 3$\,km\,s$^{-1}$. The Renormalized Unit Weight Error (RUWE) from {\it Gaia} DR3 is 0.940, indicating high-quality astrometric data \citep{2022yCat.1355....0G}. However, the Gaia metallicity estimate is likely biased due to pulsation-induced spectral variability and low-resolution BP/RP spectra. 

The reddening estimated from Hipparcos data is $E(B-V) = 0.10$, which, assuming a total-to-selective extinction ratio of $R_V = 3.15$, corresponds to $A_V = 0.32$\,mag in the Johnson $V$ band \citep{2018yCat.2354....0G}.

In the StarHorse catalogue \citep{2019A&A...628A..94A}, the distance to the star is estimated to be $100.92^{+0.38}_{-0.40}$\,pc. The extinction is reported as $A_V = 0.24$ mag in the Johnson $V$ band and $A_G = 0.22$ mag in the {\it Gaia} $G$ band. The effective temperature is given as $T_{\rm eff} = 7747.54^{+999}_{-1276}$\,K, and the surface gravity as $\log g = 3.862^{+0.081}_{-0.11}$. The metallicity is listed as $[{\rm Fe/H}] = -0.0486^{+0.35}_{-0.17}$, although the large uncertainty renders this estimate less reliable.

Several additional metallicity determinations for AI Vel can be found in the literature, although they often show considerable discrepancies. From Ca\,II triplet measurements, \citet{2024AA...685A..59M} derived $[{\rm Fe/H}] = +0.187 \pm 0.314$. Using Geneva photometric calibrations, \citet{2017MNRAS.469.3042N} obtained $[{\rm Fe/H}] = -0.15$. An independent spectroscopic analysis based on High Accuracy Radial velocity Planet Searcher (HARPS) observations \citep{2013A&A...550L..10G} was performed by \citet{2016AJ....152..207R}, who reported a metallicity of $[{\rm Fe/H}] = +0.11$. Similar inconsistencies are seen in estimates of interstellar extinction. For example, \citet{2006A&A...453..635M} reported $E(B-V) = 0.05$, which corresponds to $A_V = 0.15$\,mag.

\subsection{Spectroscopic analysis and fundamental parameters}
\label{Sect:Spectroscopic_analysis}

As discussed in the previous subsection, the effective temperature of AI Vel reported in the literature shows significant discrepancies. To obtain a more reliable estimate, we analysed a set of 104 high-resolution spectra  acquired with the HARPS spectrograph \citep{2000SPIE.4008..582P}. We adopted the HARPS-based spectroscopic parameters because the time-series spectra allow us to account explicitly for pulsation-induced variability rather than relying on phase-averaged measurements.

The observations were carried out over four consecutive nights, from 9 to 12 January 2011, with 26 spectra obtained each night \citep{2013A&A...550L..10G}. Each spectrum was recorded with an exposure time of 180\,s and a typical signal-to-noise ratio of approximately 160. The wavelength coverage spans 3780–6900\,\AA, with a spectral resolution of 0.01\,\AA. We normalised all spectra using the neural-network-based Suppnet software \citep{2022A&A...659A.199R}.

The radial velocities and effective temperatures for all spectra were determined using the iSpec software package \citep{2014A&A...569A.111B, 2019MNRAS.486.2075B}. For radial velocity measurements, we employed the HARPS\_SOPHIE.F0.360\_698nm atomic line list as a reference template.

The derived radial velocities range from 12.5 to 50.8\,km\,s$^{-1}$, with a mean value of 28.1(8)\,km\,s$^{-1}$. The typical uncertainties range from 0.06 to 0.2\,km\,s$^{-1}$. The large variations in radial velocity are attributed to the high-amplitude pulsations characteristic of AI Vel.
The discrepancy between our mean value and the radial velocity reported in the Gaia catalogue ($17\pm3$\,km\,s$^{-1}$) likely results from pulsation-induced variability combined with differences in temporal sampling between the Gaia and HARPS observations.

The effective temperature of AI Vel, determined from the wings of the hydrogen lines H$\alpha$, H$\beta$, and H$\gamma$, was found to vary between 6747 and 7391\,K, which gives $T_{\rm{eff}}=7069\pm322$\,K. The mean surface gravity was derived as $\log{g} = 4.1(1)$, and the microturbulent velocity as $\xi = 3.0(1)$\,km\,s$^{-1}$. These values of gravity and microturbulent velocity were obtained using the equivalent widths of selected \ion{Fe}{I} and \ion{Fe}{II} lines located in the wavelength range 4620–4635\,\AA\ and having well-determined oscillator strengths ($gf$) \citep{1998A&A...338.1041L}.

From well-isolated lines of Mg, Ca, Ti, and Fe, we derived a projected rotational velocity of $V_{\rm rot}\sin i = 12.5(1)$\,km\,s$^{-1}$ and an $\alpha$-element abundance ratio of $[\alpha/{\rm Fe}] = 0.33(1)$. The derived rotational velocity should be treated with caution, as line broadening may be significantly affected by stellar pulsations.

We determined the individual chemical abundances using spectral lines identified in the median-combined spectrum. Lines affected by telluric contamination were excluded from the analysis. 
The abundances of all analysed elements are listed in Table\,\ref{tab:AI_Vel_HARPS_elements_abundances} and shown in Fig.\,\ref{fig:AI_Vel_HARPS_abundances}, where they are plotted as a function of atomic number. 
\begin{table}[ht]
\begin{center}
\begin{small}
\caption{Chemical abundances of AI Vel derived from the median HARPS spectrum. The columns give the atomic number, $Z$, element, $[\mathrm{X/H}]$, its uncertainty $\sigma_{[\mathrm{X/H}]}$, and $[\mathrm{X/Fe}]$.}
\label{tab:AI_Vel_HARPS_elements_abundances}
\begin{tabular}{|r|l|r|r|r|}
\hline $Z$ & Element & [X/H] & $\sigma_{\rm [X/H]}$ & [X/Fe]  \\
\hline
6   &   C   &    -0.24   &   0.05   &   0.31  \\ 
8   &   O   &    0.13   &    0.16   &   0.68  \\ 
11   &  Na   &   -0.27   &   0.06   &   0.28  \\ 
12   &  Mg   &   -0.55   &   0.03   &   0.00   \\ 
14   &  Si   &   -0.22   &   0.04   &   0.33   \\ 
16   &  S   &    -0.31   &   0.09   &   0.24   \\ 
20   &  Ca   &   -0.37   &   0.02   &   0.18   \\ 
21   &  Sc   &   -0.16   &   0.03   &   0.39   \\ 
22   &  Ti   &   -0.22   &   0.01   &   0.33   \\ 
23   &  V   &    -0.30   &   0.10   &   0.25   \\ 
24   &  Cr   &   -0.36   &   0.02   &   0.19   \\ 
25   &  Mn   &   -0.58   &   0.03   &  -0.03   \\ 
26   &  Fe   &   -0.55   &   0.01   &   0.00   \\ 
28   &  Ni   &   -0.53   &   0.02   &   0.02   \\ 
30   &  Zn   &   -0.66   &   0.11   &  -0.11   \\ 
38   &  Sr   &   -0.11   &   0.11   &   0.44   \\ 
39   &  Y   &    -0.16   &   0.05   &   0.39   \\ 
40   &  Zr   &   0.06   &    0.08   &   0.63   \\ 
56   &  Ba   &   -0.40   &   0.06   &   0.15   \\ 
57   &  La   &   -0.01   &   0.14   &   0.54   \\
\hline
\end{tabular}
\end{small}
\end{center}
\end{table}
We found that the atmospheric abundances of most elements are slightly sub-solar. Notably low abundances are observed for iron-peak elements such as Mn, Cr, Fe, and Ni. In contrast, some elements—including O, Zr, and La—exhibit solar or even slightly super-solar abundances.
Despite the underabundance of iron-group elements, the global metallicity remains relatively high because the total metal mass fraction is dominated by light elements, particularly C and O. Taking the individual elemental abundances into account, we obtain a global metallicity of $Z = 0.011$.

For the derived effective temperature 
$\log(T_{\rm eff}/\mathrm{K}) = 3.850 \pm 0.020$, we adopt a bolometric correction of 
$BC = -0.096 \pm 0.033$ from Kurucz model atmospheres, using $\log{g}=4.1(1)$ and an assumed $\rm{[m/H]}=0.0(1)$. To account for uncertainties in interstellar reddening, we considered the range of extinction values reported in the literature, $A_V = 0.05$--0.33\,mag.
Using the distance 
$100.92^{+0.38}_{-0.40}$\,pc from the StarHorse catalogue, we derive a stellar luminosity of 
$\log(L/L_{\odot}) = 1.285\pm0.063$. 
This value is consistent with the period–luminosity relation for HADS stars \citep{Poretti2008}, given by
\[
M_V = -3.65 \log P - 1.83,
\]
which yields $M_V = 1.646$ for $P = 0.111574$\,d. 
Adopting $M_{\rm bol} = 1.55$, we obtain 
$\log(L/L_{\odot}) \approx 1.276$. The close agreement between both estimates supports the adopted distance and extinction.
A summary of the available determinations of the parameters of AI Vel is provided in Table\,\ref{tab:AI_Vel_parameters_summary}.

\begin{table*}
\centering
\caption{Selected determinations of the fundamental parameters of AI Vel.}
\label{tab:AI_Vel_parameters_summary}
\setlength{\tabcolsep}{5pt}
\begin{tabular}{lccccc}
\hline
Source &
$T_{\rm eff}$ [K] &
$\log g$ &
$[{\rm Fe/H}]$ &
$A_V$ [mag] &
$d$ [pc] \\
\hline
Gaia DR3 &
$7188^{+14}_{-12}$ &
$3.6623^{+0.051}_{-0.026}$ &
$-1.3295^{+0.0033}_{-0.0023}$ &
$0.0589^{+0.0066}_{-0.0056}$ &
$100.40\pm0.19$ \\

StarHorse &
$7748^{+999}_{-1276}$ &
$3.862^{+0.081}_{-0.110}$ &
$-0.049^{+0.35}_{-0.17}$ &
$0.24$ &
$100.92^{+0.38}_{-0.40}$ \\

Hipparcos & 
-- &
-- &
 -- &
0.33  &
-- \\

$uvby$ photometry \protect\citep{1977PASP...89..339B} &
7110--8520 &
-- &
-- &
-- &
-- \\

\protect\citet{2017MNRAS.469.3042N} &
-- &
-- &
$-0.15$ &
-- &
-- \\

\protect\citet{2016AJ....152..207R} &
-- &
-- &
$+0.11$ &
-- &
-- \\

\protect\citet{2024AA...685A..59M} &
-- &
-- &
$+0.187\pm0.314$ &
-- &
-- \\

This work &
$7069\pm322$ &
$4.1(1)$ &
$-0.55(1)$ &
$0.05$--$0.33$ &
$100.92^{+0.38}_{-0.40}$ \\
\hline
\end{tabular}
\end{table*}

\section{Frequencies from \textit{TESS} data}
\label{Sect:FA}

High-quality, nearly continuous space-based observations are available for AI Vel. The star was observed by the  \textit{TESS} satellite \citep{2015JATIS...1a4003R} in six Sectors: 34, 35, 61, 62, 88 and 89. 
The detailed information on the observations is given in Table\,\ref{tesstable}. 
\begin{table}[ht]
\setlength{\tabcolsep}{3pt}
\centering
    \caption{Summary of the \textit{TESS} observations of AI Vel. The columns list, from left to right, the sector number, observation dates, CCD detector number, cadence (in seconds), and the number of data points. All data were obtained with Camera\,3.}
        \label{tesstable}
    \begin{tabular}{crclccc} \hline
    Sector  & \multicolumn{3}{c}{ Dates (UTC)}            &  CCD & Cadence [s] & N\\ \hline
        34 & 14 Jan. &-& 8  Feb. 2021 &  1 & 120 & 17376 \\
        35 & 9  Feb. &-& 6  Mar. 2021 &  2 & 120 & 14146\\
        61 & 18 Jan. &-& 12 Feb. 2023 &  1 & 200 & 10636 \\
        62 & 12 Feb. &-& 10 Mar. 2023 &  2 & 200 & 10804\\ 
        88 & 14 Jan. &-& 11 Feb. 2025 &  1 & 200 & 11718 \\
        89 & 11 Feb. &-& 12 Mar. 2025 &  2 & 200 & 11411\\ 
        
         \hline
    \end{tabular}
\end{table}

\begin{figure}[ht]
     \centering
    \includegraphics[width=0.49\columnwidth]{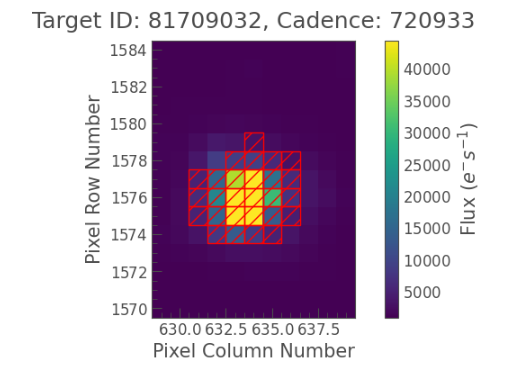}
    \includegraphics[width=0.49\columnwidth]{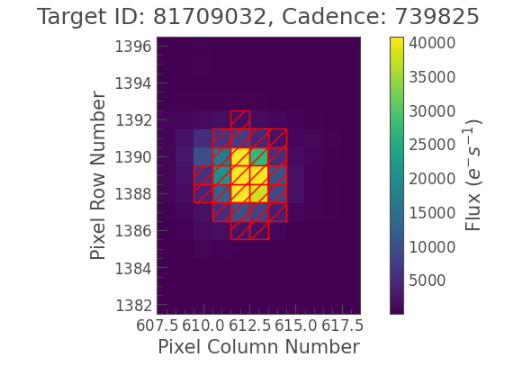}
    \includegraphics[width=0.49\columnwidth]{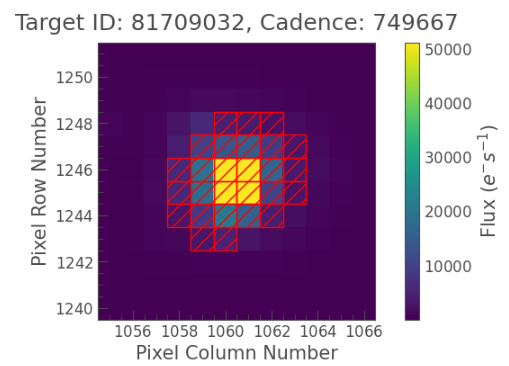}
    \includegraphics[width=0.49\columnwidth]{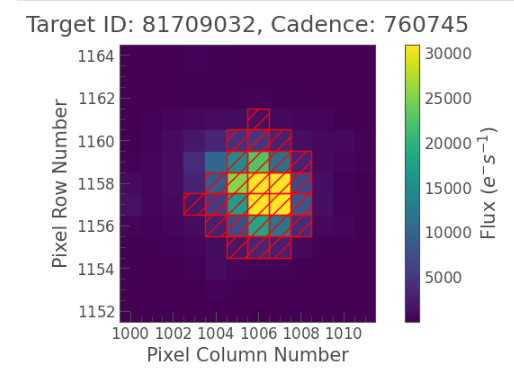}
    \caption{The star images captured by the \textit{TESS} satellite in Sector 34 (upper left), 35 (upper right), 61 (bottom left) and 62 (bottom right). The red squares mark the photometric aperture. The images were generated using the Python Lightkurve package. }
    \label{fig:tess_apertures}
\end{figure}

The \textit{TESS} target pixel files for Sectors 34, 35, 61, and 62 were downloaded from the Mikulski Archive for Space Telescopes\footnote{\url{https://archive.stsci.edu/}} (MAST). We used target pixel files processed by the SPOC pipeline and extracted the light curves with the \texttt{lightkurve} package \citep[see][\url{https://docs.lightkurve.org/}]{2018ascl.soft12013L}, using the corresponding SPOC aperture masks. The apertures adopted in our analysis are shown in Fig.\,\ref{fig:tess_apertures}. We also made use of the \texttt{astropy} \citep{astropy:2022} and \texttt{astroquery} \citep{2019AJ....157...98G} libraries. For Sectors 88 and 89, only Quick-Look Pipeline (QLP) light curves were available, and these were used in our analysis.

\begin{figure}[ht]
    \centering
    \includegraphics[width=\columnwidth]{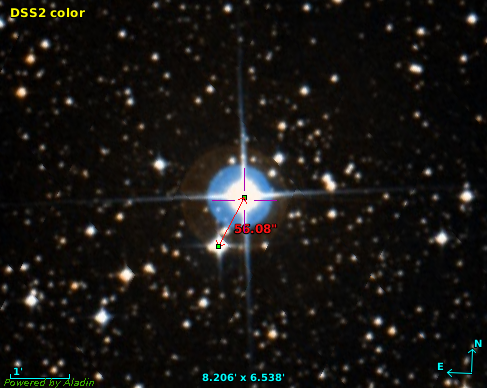}
    \caption{Image of the field surrounding AI Vel obtained from the Aladin Sky Atlas.}
    \label{fig:Ai_Vel_vicinity}
\end{figure}

Fig.\,\ref{fig:Ai_Vel_vicinity} presents an image of AI Vel and its surrounding field \citep[from Aladin sky atlas, v11.024,][]{2000A&AS..143...33B}. 
The brightest star in its vicinity is CD–44 4195, with a visual magnitude of $V = 10.47$\,mag. It is classified as an F0-type star \citep{2014yCat....1.2023S}, with an effective temperature of $T_{\rm eff} = 7224.3 \pm 8.9$\,K \citep{2022yCat.1355....0G}. Its surface gravity, $\log{g} = 3.9779^{+0.0042}_{-0.0037}$, is consistent with a main-sequence classification, and its subsolar metallicity is measured as $[{\rm Fe/H}] = -0.3218^{+0.0071}_{-0.0075}$. Based on Gaia DR3 data, the star has a radial velocity of $18.1\pm1.1$\,km\,s$^{-1}$ and a parallax of $\pi = 2.677\pm0.014$\,mas, corresponding to a distance of $377.7\pm1.8$\,pc.

CD–44 4195 is located at an angular separation of 56'' from AI Vel. Although its flux may fall partially within the \textit{TESS} photometric aperture, the level of contamination is expected to be minimal due to its significantly fainter brightness. Nonetheless, some of the low-amplitude frequencies detected in the \textit{TESS} light curve may originate from this nearby star.

The light curves from all sectors were subjected to detailed analysis. We computed Lomb-Scargle periodograms \citep{1976Ap&SS..39..447L, 1982ApJ...263..835S} using an algorithm based on non-equally spaced fast Fourier transforms \citep[e.g.,][]{2012A&A...545A..50L}. Subsequently, we applied a non-linear least-squares fitting to the following function:
\begin{equation}\label{eq:m_t}
    m(t)=\sum\limits_{i=1}^n A_i\sin{\left(2\pi(t-T_0) \nu_i+\varphi_i\right)}.
\end{equation}
Here, $m(t)$ represents the star's brightness, $t$ denotes time, and $\nu_i$, $A_i$ and  $\varphi_i$ are the adjustable parameters corresponding to frequency, amplitude, and phase, respectively. The reference time $T_0$ was set to the time of the first observation, determined separately for each analysed dataset.

Sectors 34 and 35 were observed consecutively, and thus were combined and analysed as a single dataset. The same approach was applied to sector pairs 61 and 62 as well as to 88 and 89. However, due to a gap of approximately two years between the observing periods, we chose not to merge all six sectors into one dataset, as the resulting spectral window would introduce aliases that complicate the frequency analysis. Additionally, possible variations in amplitudes and phases between the epochs could introduce additional uncertainties.

The combined data from Sectors 34+35 yielded about 31\,500 data points, while Sectors 61+62 and 88+89 provided roughly 21\,400 and 22\,300 points, respectively. The Rayleigh frequency resolution for all combined datasets is approximately 0.019\,d$^{-1}$. To assess the significance of detected peaks, we adopted a signal-to-noise ratio (S/N) threshold of 5.0, following the recommendation by \citet{2021AcA....71..113B}. The noise level was calculated as the mean amplitude within a 50\,d$^{-1}$ window centred on the frequency of interest.

\begin{figure}
   \begin{center}
      \includegraphics[width=0.9\columnwidth]{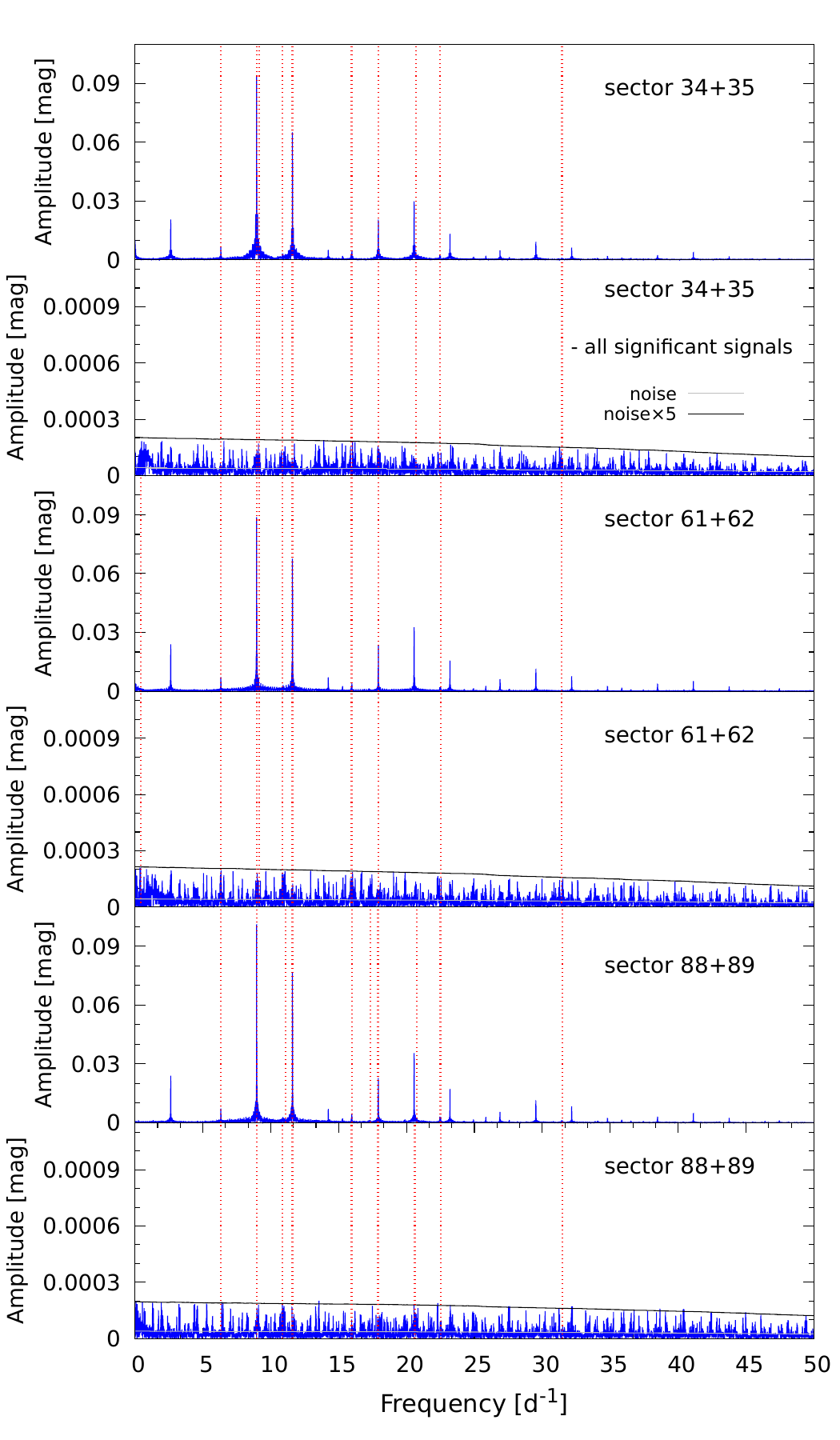}
     \caption{Fourier amplitude periodograms of AI Vel derived from the \textit{TESS} light curves in Sectors 34+35 (upper two panels), 61+62 (middle two panels), and 88+89 (bottom two panels). For each sector pair, periodograms computed from the original data and from the residuals after prewhitening all significant frequencies are shown.
     }
     \label{fig:fourier_compar}
  \end{center}
\end{figure}

The results of the Fourier analysis are presented in Table\,\ref{tab:TESS_freq_comb_sectors}, where only independent frequencies are listed. Periodograms of the original data and those obtained after prewhitening all significant signals are shown in Fig.\,\ref{fig:fourier_compar}. Red vertical lines indicate the independent frequencies. In the second, fourth, and sixth panels, the black solid lines mark the threshold corresponding to five times the signal-to-noise ratio.

In the combined data from Sectors 34+35, we detected four harmonics of $\nu_1$ and five harmonics of $\nu_2$. In Sectors 61+62, four harmonics were identified for both $\nu_1$ and $\nu_2$, while in Sectors 88+89 we found three harmonics of $\nu_1$ and five of $\nu_2$. In all datasets, numerous combination frequencies involving the independent modes were also present.

\begin{table}
\begin{threeparttable}
\centering
    \caption{Independent frequencies of AI Vel derived from \textit{TESS} data. The columns list the frequency identifier, frequency, amplitude, phase relative to the reference epoch $T_0$ adopted for each sector pair, and the signal-to-noise ratio.}
        \label{tab:TESS_freq_comb_sectors}
    \begin{tabular}{clclc} \hline
No. & Frequency  &   Amplitude&          $\varphi$   &     S/N          \\
   &   [d$^{-1}$]& [mag] & [rad] &  \\\hline
         \hline
\multicolumn{5}{c}{Sector 34+35 ($T_0=\text{BJD}-2457000.0=2228.75$) } \\
\hline 
1  & 8.9626240(20)$^\dagger$ & 0.093437(19) & 4.00332(21)     & 105.7      \\
2  & 11.599782(20)$^\dagger$ & 0.065320(16) & 4.34463(34)       & 113.9   \\
3  & 15.97823(11)$^\dagger$  & 0.004898(34) & 1.686(17)         & 36.7    \\
4  & 15.95443(12)  & 0.003637(57) & 0.914(15)         & 34.7    \\
5  & 22.45322(26)$^\dagger$  & 0.002273(68) & 2.447(31)         & 23.0  \\
6  & 9.13580(13)$^\dagger$   & 0.001594(14) & 3.288(22)         & 16.8    \\
7  & 17.9400(27)   & 0.001774(55) & 5.351(33)         & 19.5    \\
8  & 31.4524(46)   & 0.000510(24) & 1.131(68)         & 11.1    \\
9  & 10.8360(62)$^\dagger$   & 0.000437(14) & 3.04(10)          & 8.1    \\
10 & 6.30180(43)   & 0.000350(20) & 2.605(63)         & 6.7   \\
11 & 20.7136(68)   & 0.000214(16) & 0.34(11)          & 5.5   \\
\hline 
\multicolumn{5}{c}{Sector 61+62 ($T_0=\text{BJD}-2457000.0=2962.80$) }\\
\hline 
1  & 8.9626980(30)$^\dagger$ & 0.088688(18) & 4.09012(20)       & 106.3   \\
2  & 11.599775(40)$^\dagger$ & 0.067380(18) & 3.11499(26)       & 116.5  \\
3  & 15.97660(99)$^\dagger$  & 0.004005(20) & 1.5884(50)        & 33.6   \\
4  & 15.9468(17)   & 0.002878(20) & 1.6998(68)        & 25.7   \\
5  & 22.5162(85)$^\dagger$   & 0.001212(20) & 2.977(17)         & 18.9   \\
6  & 9.13600(17)$^\dagger$   & 0.001483(18) & 1.846(12)         & 14.5   \\
7  & 17.9424(69)   & 0.001081(21) & 2.841(20)         & 13.2   \\
8  & 10.8536(16)$^\dagger$   & 0.001401(18) & 1.757(13)         & 17.5  \\
9  & 0.45960(71)   & 0.000948(52) & 5.02(10)          & 10.7   \\
10 & 6.31107(34)   & 0.000781(22) & 2.870(57)         & 10.0   \\
11 & 31.41607(29)  & 0.000667(25) & 4.222(45)         & 10.9   \\
 \hline  
\multicolumn{5}{c}{Sector 88+89 ($T_0=\text{BJD}-2457000.0=3690.16$) }\\
\hline  
1  & 8.9625870(30)$^\dagger$ & 0.096808(37) & 4.13728(43) & 107.9      \\
2  & 11.599650(40)$^\dagger$ & 0.074370(27) & 4.84872(72) & 120.4   \\
3  & 15.973726(61)$^\dagger$ & 0.004688(27) & 4.841(12)   & 36.9     \\
4  & 17.90009(12)  & 0.002668(36) & 2.814(21)   & 28.1     \\
5  & 17.93526(19)  & 0.003400(82) & 2.216(30)   & 26.1     \\
6  & 22.52186(10)$^\dagger$  & 0.002657(26) & 4.668(20)   & 25.8    \\
7  & 10.85161(15)$^\dagger$  & 0.001842(38) & 5.760(24)   & 20.2   \\
8  & 15.95356(38)  & 0.001607(28) & 1.191(17)   & 14.5    \\
9  & 31.45314(30)  & 0.000873(26) & 1.465(59)   & 14.2    \\
10 & 6.30044(12)  & 0.000600(26) & 4.616(44)  &  10.3  \\
11 & 20.6118(77)   & 0.000333(25) & 0.589(75)   & 7.0    \\
\hline 
    \end{tabular}
    \begin{tablenotes}
\small
\item [$^\dagger$] Frequencies previously reported in the literature. 
\end{tablenotes}

\end{threeparttable}
\end{table}


Combining the datasets improved the frequency resolution, allowing us to resolve the close pair of frequencies at $\sim15.98$\,d$^{-1}$ and $\sim15.95$\,d$^{-1}$, which were not always clearly distinguishable in the individual sectors. The combined data also yielded a significantly higher signal-to-noise ratio, leading to reduced uncertainties in the derived frequencies, amplitudes and phases.

The two highest-amplitude frequencies, $\nu_1 = 8.963$\,d$^{-1}$ and $\nu_2 = 11.600$\,d$^{-1}$, were consistently detected across all analysed datasets. In the combined data from Sectors 34+35, the amplitude ratio is approximately $A_2/A_1 \approx 0.70$, while the third independent frequency shows a much lower amplitude, with $A_3/A_1 \approx 0.05$. All other detected frequencies have even smaller amplitudes. In the Sectors 61+62 dataset, the amplitude ratio increases slightly to $A_2/A_1 \approx 0.76$,  due to a decrease in the amplitude of the dominant mode accompanied by a simultaneous increase in the amplitude of the secondary mode. For Sectors 88+89 the amplitude ratio is on a similar level, $A_2/A_1 \approx 0.77$.

The dominant frequency, $\nu_1$, changed by approximately $7.4 \times 10^{-5}$\,d$^{-1}$ between Sectors 34+35 and 61+62. A slightly larger difference is observed between Sectors 61+62 and 88+89, amounting to $-1.11 \times 10^{-4}$\,d$^{-1}$. 
For the secondary frequency, $\nu_2$, the shift is $-7.0 \times 10^{-6}$\,d$^{-1}$ between Sectors 34+35 and 61+62, and $-1.25 \times 10^{-4}$\,d$^{-1}$ between Sectors 61+62 and 88+89. 
These variations are larger than the expected evolutionary frequency changes. For hydrogen shell-burning $\delta$ Scuti stars, the predicted secular variation of low-order radial-mode frequencies over a two-year interval (comparable to the gap between successive \textit{TESS} observations) is typically of the order of $2 \times 10^{-7}$ to $2 \times 10^{-5}$\,d$^{-1}$.
However, since AI Vel is a multiperiodic pulsator, such frequency shifts may instead arise from non-linear mode interactions or amplitude and phase modulation due to mode beating.

A portion of the \textit{TESS} light curve from Sector 61 is shown in Fig.\,\ref{fig:lc_model}, with the mean brightness value subtracted. The solid yellow line represents the model fitted to the combined 61+62 datasets using Eq.\,\ref{eq:m_t}. The model reproduces the observations well, capturing both the amplitude and phase of the variability.

Our analysis confirmed all previously reported frequencies and revealed several new ones. In particular, we detected signals at 6.3\,d$^{-1}$ and 17.94\,d$^{-1}$, both clearly present in the combined datasets from each pair of sectors. The $\sim31.4$\,d$^{-1}$ frequency also appears in all datasets. However, it is close to the combination frequency of the dominant mode at 8.962\,d$^{-1}$ and the 22.5\,d$^{-1}$ mode. The $\sim20.7$\,d$^{-1}$ frequency is detected  in the combined Sectors 34+35 and 88+89, while a low-frequency signal at 0.46\,d$^{-1}$, likely of instrumental origin, is present only in the 61+62 dataset.


\section{Mode identification}
\label{Sect:MI}

\subsection{Photometric method}

The frequency ratio of the two dominant modes, $\nu_1=8.963$\,d$^{-1}$ and $\nu_2=11.600$\,d$^{-1}$, suggests that they correspond to the radial fundamental (p$_1$) and first overtone  (p$_2$) modes, respectively. Nevertheless, to confirm this identification, we performed an independent mode analysis using available observations in the Str\"omgren $uvby$ photometric system \citep{1977PASP...89..339B}. 

We employed the method developed by \citet{JDD2003}, which allows for the simultaneous determination of the spherical degree $\ell$ and two pulsational parameters, $f$ and $\varepsilon$, using photometric amplitudes and phases in at least three passbands.

The parameter $f$ is defined as the ratio of the local bolometric flux perturbation to the radial displacement at the stellar surface and enters the flux variation as follows:
\begin{equation}
\frac{\delta \mathcal{F}_{\mathrm{bol}}}{\mathcal{F}_{\mathrm{bol}}} = \mathrm{Re} \left[ \varepsilon f Y_{\ell}^{m}(\theta, \phi) e^{-\mathrm{i}\omega t} \right],
\end{equation}
where $Y_{\ell}^{m}$ is the spherical harmonic, and $\omega$ is the angular pulsation frequency.
Theoretical values of the $f$-parameter are computed using linear non-adiabatic pulsation models, whereas empirical values are derived from multi-colour photometry.

The parameter $\varepsilon$ represents the intrinsic amplitude of the radial displacement caused by stellar pulsation and is defined as
\begin{equation}
\frac{\delta r}{r} = \mathrm{Re} \left[ \varepsilon Y_{\ell}^{m}(\theta, \phi) e^{-\mathrm{i}\omega t} \right].
\end{equation}

The complex amplitude of the relative flux variation in a given photometric passband $\lambda$ can be written as \citep[e.g.,][]{JDD2005}:
\begin{equation}\label{eq:campl}
\mathcal{A}_{\lambda} = \mathcal{D}_{\ell}^{\lambda}\,(\tilde{\varepsilon} f) + \mathcal{E}_{\ell}^{\lambda}\,\tilde{\varepsilon},
\end{equation}
where
$
\tilde{\varepsilon} = \varepsilon\, Y_{\ell}^{m}(i,0),
$
and $i$ denotes the inclination angle of the stellar rotation axis.

The coefficients $\mathcal{D}_{\ell}^{\lambda}$ and $\mathcal{E}_{\ell}^{\lambda}$ describe the sensitivity of the monochromatic flux to variations in effective temperature and surface gravity \citep{JDD2003,JDD2005}. They were computed using  Vienna model atmospheres \citep{2002A&A...392..619H}. We adopted the non-linear limb-darkening law of \citet{2000A&A...363.1081C}.

Next, using the four Str\"omgren passbands ($uvby$), we constructed a system of four complex equations which, for a given spherical degree 
$\ell$, can be solved for the complex parameters 
$\tilde{\varepsilon}$ and $f$. To identify the most probable mode degree, we evaluate the quality of the fit using the following goodness-of-fit statistic:
\begin{equation}
\chi^2 = \frac{1}{2N - N_{\mathrm{p}}} \sum\limits_{i=1}^N \frac{|\mathcal{A}_{\lambda_i}^{\mathrm{obs}} - \mathcal{A}_{\lambda_i}^{\mathrm{cal}}|^2}{|\sigma_{\lambda_i}|^2},
\end{equation}
where $\mathcal{A}_{\lambda_i}^{\mathrm{obs}}$ and $\mathcal{A}_{\lambda_i}^{\mathrm{cal}}$ are the observed and calculated complex amplitudes in the passband $\lambda_i$, $\sigma_{\lambda_i}$ is the associated observational uncertainty, $N$ is the number of passbands, and $N_{\mathrm{p}}$ is the number of free parameters. Since both $\tilde{\varepsilon}$ and $f$ are complex, $N_{\mathrm{p}} = 4$. Each passband provides two observational constraints (amplitude and phase), giving a total of $2N$ constraints.

If a clear minimum is found in the $\chi^2$ values, i.e. in the difference between the observed and calculated amplitudes and phases, the corresponding spherical degree $\ell$, along with the associated values of $\tilde{\varepsilon}$ and $f$, can be considered as the most probable. 

\begin{center}
   \begin{figure*}
     \centering
     \includegraphics[width=0.9\columnwidth]{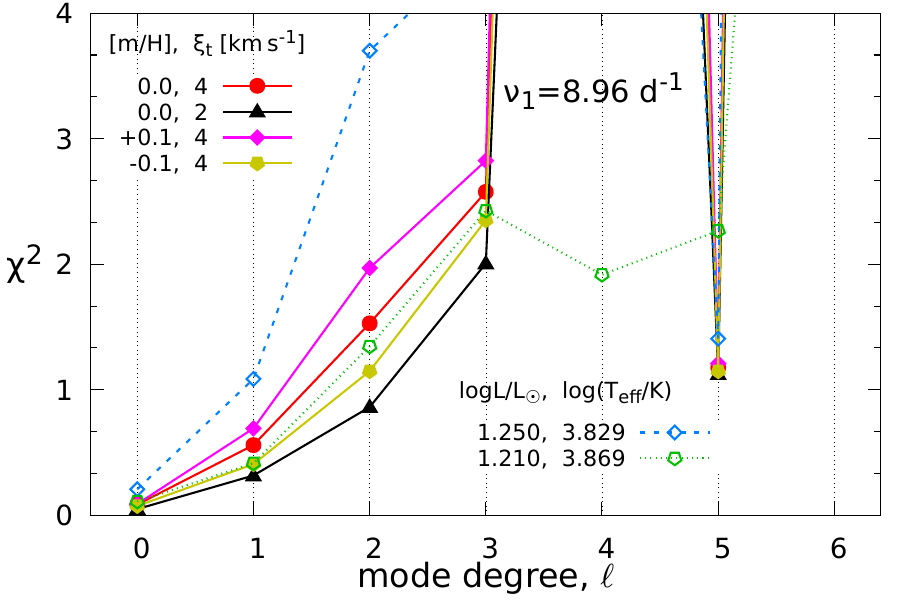}
     \includegraphics[width=0.9\columnwidth]{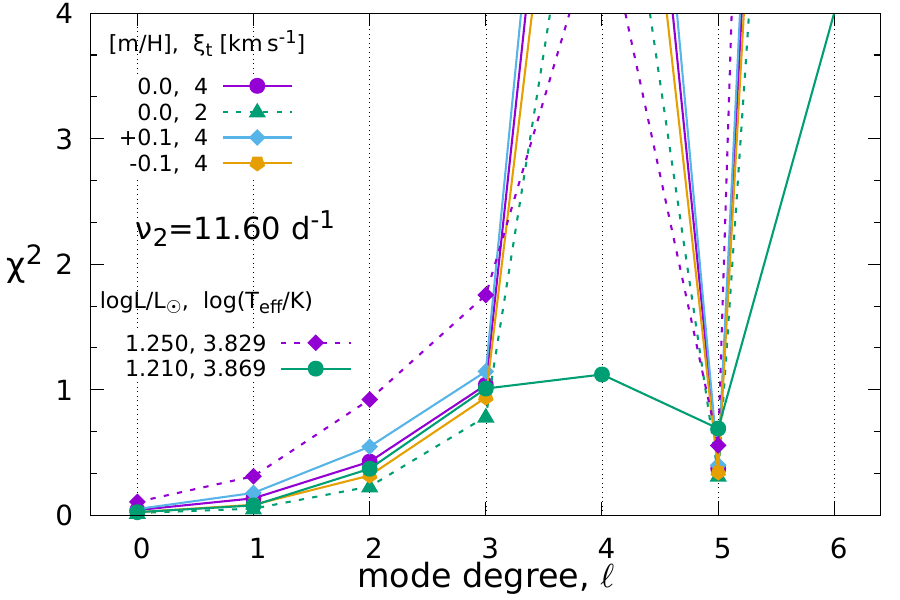}
     \caption{The $\chi^2$ values as a function of the mode degree, $\ell$. The left panel shows results for the dominant frequency, $\nu_1$, while the right panel corresponds to the secondary frequency, $\nu_2$. The effects of metallicity, [m/H], microturbulent velocity, $\xi_t$, and position in the HR diagram are illustrated. Unless otherwise indicated in the legend, the models assume [m/H]=0.0, $\xi_t=4$\,km\,s$^{-1}$, $\log(T_{\rm eff}/{\rm K}) = 3.849$, and $\log(L/L_{\odot}) = 1.230$.}
     \label{fig:mode_ident_chi2}
  \end{figure*}
 \end{center}

Reliable amplitudes and phases in the Str\"omgren $uvby$ passbands could be determined only for the two highest-amplitude modes, $\nu_1$ and $\nu_2$.
Fig.\,\ref{fig:mode_ident_chi2} presents the resulting $\chi^2$ values as a function of mode degree $\ell$. The left panel corresponds to the frequency $\nu_1$, while the right panel shows results for $\nu_2$. Photometric mode identification could not be reliably performed for the remaining frequencies due to their low amplitudes.

Since the discriminant $\chi^2$ depends to some extent on the adopted model atmospheres, we investigated the effects of varying the microturbulent velocity, $\xi_t$, and metallicity, $[{\rm m/H}]$, 
while keeping the stellar parameters ($T_{\rm eff}$, $\log g$) fixed to values corresponding to an evolutionary model located at the centre 
of the observational error box.

In a separate test, we also explored the sensitivity of the results to the adopted stellar parameters by considering two locations within the observational error box. We compared the $\chi^2$ values for two different pairs of ($\log T_{\rm eff}$, $\log L/L_{\sun}$), assuming fixed atmospheric 
parameters of ${\rm [m/H]}=0.0$ and $\xi_t=4$\,km\,s$^{-1}$.

In all tested configurations of the parameters, the $\chi^2$ minimum consistently occurs at $\ell = 0$ for both dominant frequencies, indicating that they are most likely radial modes. According to the frequency ratio, $\nu_1/\nu_2~\sim0.77$, the most probable radial  orders are $n=1$ (p$_1$) and $n=2$ (p$_2$) for $\nu_1$ and $\nu_2$, respectively.

An additional advantage of the photometric method of mode identification is that it provides empirical estimates of the $f$-parameter for the two highest-amplitude modes of AI Vel. These empirical values can be compared with theoretical predictions, which are highly sensitive to several stellar and pulsational properties, including:
\begin{itemize}
\item pulsation frequency, $\omega$,
\item the shape of the eigenfunctions, $\ell(\ell + 1)/\omega^2$,
\item stellar parameters, such as mass and effective temperature,
\item chemical composition (e.g., $X$, $Z$, and element mixture),
\item opacity data, $\kappa$,
\item efficiency of subphotospheric convection, described by the mixing-length parameter $\alpha_{\mathrm{MLT}}$.
\end{itemize}

The $f$-parameter therefore serves as an additional seismic observable, complementary to the pulsation frequencies, and provides strong constraints on various stellar properties. In particular, for $\delta$ Scuti stars, $f$ is highly sensitive to the efficiency of convective energy transport in the outer envelope, commonly parameterized by the mixing-length parameter $\alpha_{\mathrm{MLT}}$ \citep{JDD2003}. This quantity is included as a constraint in our asteroseismic analysis, described in detail in Section\,\ref{Sect:Sim}. 

\subsection{Spectroscopic method}

As described in Sect.\,\ref{Sect:Spectroscopic_analysis}, we analysed 104 HARPS spectra of AI Vel. Several spectral lines show clear time-dependent variations in their profiles, which we attribute to stellar pulsations.

\begin{center}
   \begin{figure}
     \centering
      \includegraphics[width=0.9\columnwidth]{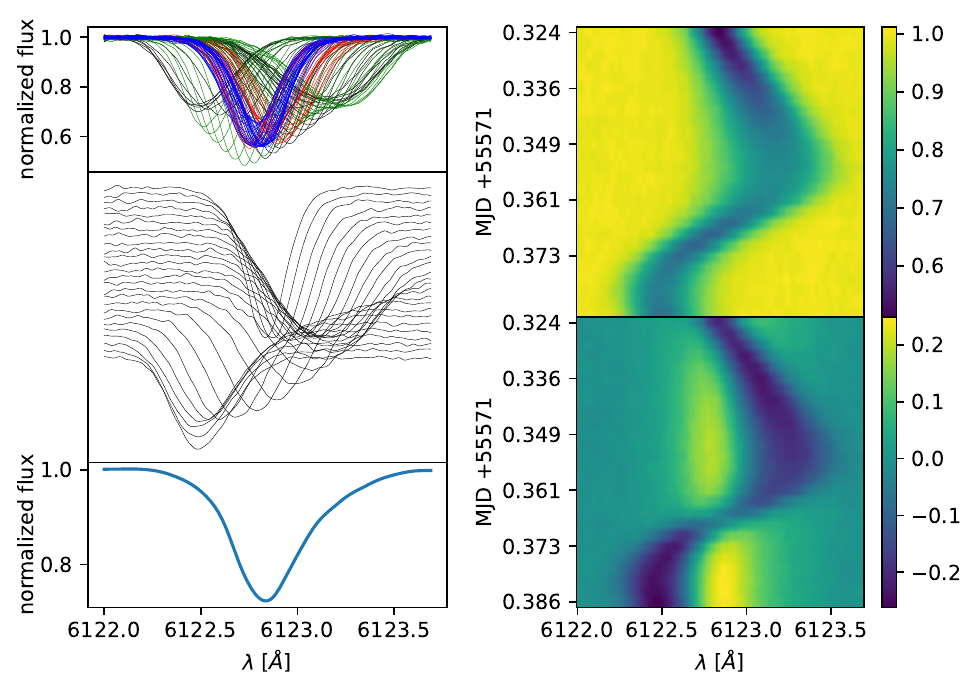}

     \caption{Variability of the \ion{Ca}{I} line in the HARPS spectra. Left panels: line profiles from all four nights (upper; successive nights shown in red, black, green, and blue), profiles from the second night with vertical offsets for clarity (middle), and the mean line profile (bottom). Right panels: phase-dependent variations in line intensity (upper) and residuals obtained after subtracting the mean profile (bottom).}
     \label{fig:AI_Vel_line_night_two}
  \end{figure}
 \end{center}

Figure~\ref{fig:AI_Vel_line_night_two} illustrates the variability of the \ion{Ca}{I} line. The upper-left panel shows line profiles from all spectra obtained over four nights. The middle-left panel presents profiles from the second night only, shifted vertically for clarity, while the bottom-left panel displays the mean line profile derived from all spectra.

The right panels show the phase-dependent behaviour of the line intensity. The upper-right panel presents a colour-coded map of intensity variations, whereas the bottom-right panel shows the residuals after subtracting the mean profile, emphasizing the variability pattern.

\begin{center}
   \begin{figure}
    \centering
     \includegraphics[width=0.85\columnwidth]{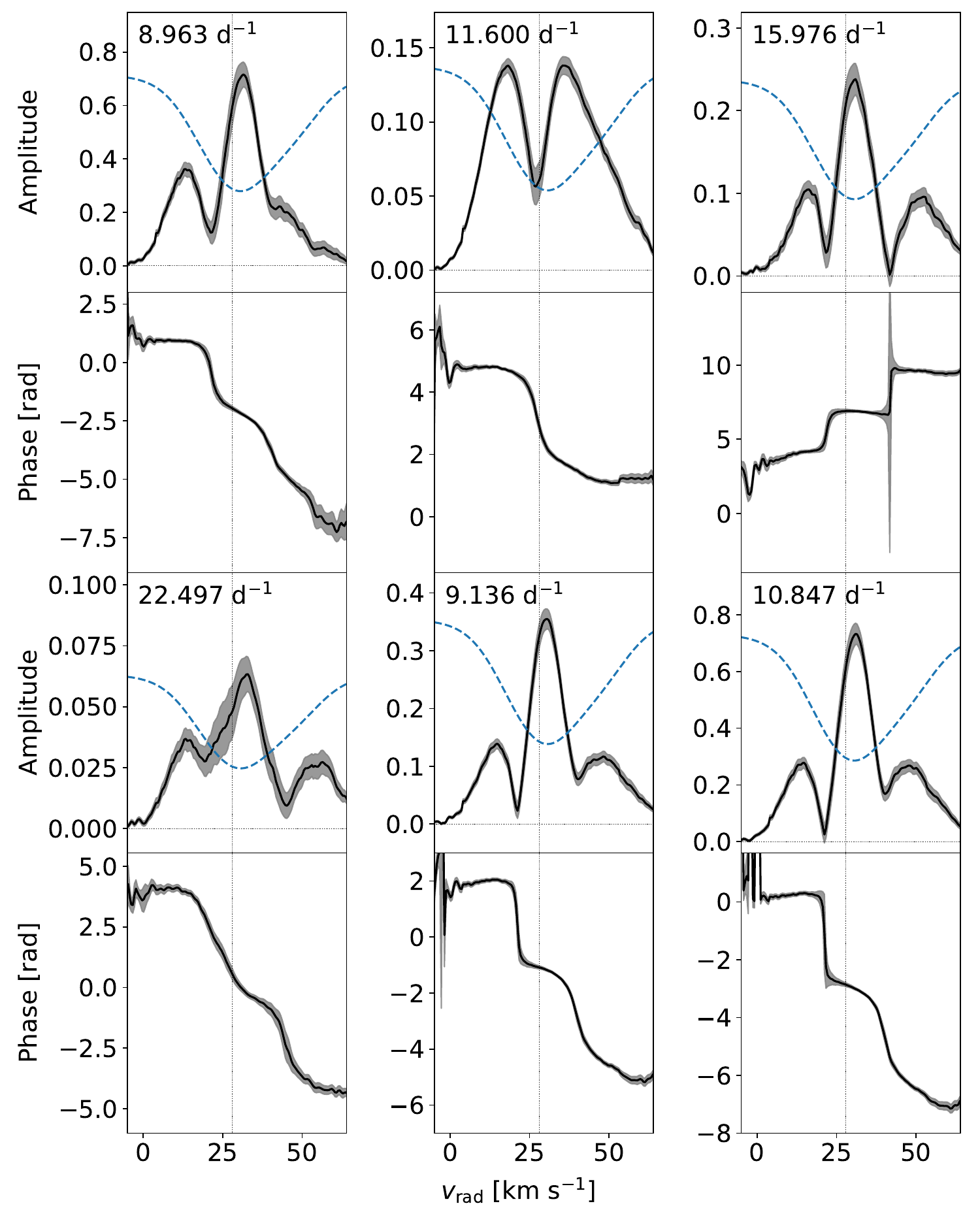}
     \caption{Amplitude and phase variations across the \ion{Ca}{I} 6122.217\,\AA\ line profile in the HARPS spectra of AI Vel for the selected pulsation modes. The first and third rows show the amplitude distributions across the line profile, while the second and fourth rows present the corresponding phase variations. The blue line indicates the mean spectral line profile of \ion{Ca}{I}.}
     \label{fig:AI_Vel_line_profiles}
  \end{figure}
 \end{center}

For each wavelength point across the profiles of five  lines of Ca and Fe, we fitted sinusoidal functions to determine the amplitudes and phases corresponding to the six independent frequencies with the highest photometric amplitudes: 8.963\,d$^{-1}$, 11.600\,d$^{-1}$, 15.976\,d$^{-1}$, 22.497\,d$^{-1}$, 9.136\,d$^{-1}$, and 10.847\,d$^{-1}$. As an example, Fig.\,\ref{fig:AI_Vel_line_profiles} presents the results for the \ion{Ca}{I} line at 6122.217\,\AA.

Photometric mode identification presented in the previous subsection showed that the frequencies 8.963\,d$^{-1}$ and 11.600\,d$^{-1}$ correspond to radial modes. As shown in Fig.\,\ref{fig:AI_Vel_line_profiles}, the pronounced asymmetry in the amplitude profile of the dominant frequency closely resembles the asymmetry observed in the amplitude profile of the 9.136\,d$^{-1}$ frequency. The latter is unlikely to be a radial mode, as its frequency lies between two securely identified radial modes and does not match the expected sequence of radial overtones; it must therefore correspond to a non-radial mode. This clear asymmetry can be naturally explained by the mutual interaction of the two closely spaced frequencies (9.136 - 8.963 = 0.173\,d$^{-1}$). Given the short time span of the spectroscopic observations, any contribution of hypothetical orbital motion to the measured radial velocities can be considered negligible.

The fitted amplitudes, phases, and zero points for each frequency were then used in our attempts to identify the corresponding pulsation modes. We employed the \texttt{FAMIAS} software \citep{2006A&A...455..227Z,2008CoAst.155...17Z}, which performs mode identification by determining the degree, $\ell$, and azimuthal order, $m$, of the observed frequencies.

We computed synthetic line profiles across the visible stellar surface, accounting for temperature variations and geometric deformation. The total synthetic spectra were obtained by integrating the contributions from all surface elements at the same observational phases as the real data. For a given combination of ($\ell$, $m$), theoretical amplitude and phase profiles were then calculated and compared with the observed ones.

Due to the close spacing and mutual interference between the 8.963\,d$^{-1}$ and 9.136\,d$^{-1}$ frequencies, we were unable to reproduce the amplitude and phase profiles of the dominant mode. However, we confirmed the photometric mode identification for the 11.600\,d$^{-1}$ frequency. 

In addition, we successfully modelled the amplitude, $A(\lambda)$, and
phase, $\phi(\lambda)$, profiles of the 15.976\,d$^{-1}$ frequency.
The spectroscopic analysis favours an axisymmetric mode, $m=0$, but
does not provide a unique constraint on the spherical degree, $\ell$.
The observed amplitude and phase profiles are compatible with several
values of $\ell$. Because geometrical cancellation becomes increasingly
strong for higher-degree modes, low-degree solutions are more likely.
We therefore consider dipole ($\ell=1$) and quadrupole ($\ell=2$)
identifications as the most plausible possibilities. For the remaining
frequencies, the mode identification remained inconclusive.

Thus, the spectroscopic analysis provides a partial mode identification
for the 15.976\,d$^{-1}$ frequency, constraining it to an axisymmetric,
low-degree non-radial mode. In the following section, we investigate
both the dipole and quadrupole hypotheses and show how the inclusion of
this additional frequency affects the seismic constraints.

\section{Asteroseismic modelling}
\label{Sect:Sim}

AI Vel is an excellent candidate for asteroseismic analysis, offering a rare combination of high-quality observational constraints. These include two securely identified radial pulsation modes, $\nu_1$ and $\nu_2$, the corresponding empirical values of the non-adiabatic $f$-parameter, and a well-constrained position in the HR diagram. In addition, the spectroscopic analysis indicates that the frequency at 15.976\,d$^{-1}$ corresponds to an axisymmetric, $m=0$, most likely low-degree non-radial mode. In the following analysis, we therefore consider both dipole, $\ell=1$, and quadrupole, $\ell=2$, hypotheses for this frequency.

To perform the seismic modelling, we applied a Bayesian analysis based on Monte Carlo simulations. The likelihood was assumed to be Gaussian and is given by \citep[e.g.][]{2005A&A...436..127J,2006A&A...458..609D,2017MNRAS.467.1433R,2021RAA....21..226J}:
\begin{equation}\label{eq:GL}
\mathcal{L}(\mathcal{O}|\boldsymbol{\theta}) =
\prod_i \frac{1}{\sqrt{2\pi\sigma_i^2}}
\exp\left[
-\frac{(\mathcal{O}_i-\mathcal{M}_i)^2}{2\sigma_i^2}
\right],
\end{equation}
where $\mathcal{O}_i$ denotes the observed quantities, $\mathcal{M}_i$ the corresponding model predictions, and $\sigma_i$ are the uncertainties adopted for the individual constraints. The vector $\boldsymbol{\theta}$ represents the set of adjustable model parameters. In our analysis, these comprise the stellar mass $M$, initial hydrogen abundance $X_0$, metallicity $Z$, initial rotational velocity $V_{\rm rot,0}$, convective core overshooting parameter $\alpha_{\rm ov}$, and mixing-length parameter $\alpha_{\rm mlt}$. For all adjustable parameters, we adopted uniform priors over the ranges explored by the model grid. This methodology has been successfully applied in our previous seismic analyses of BP Pegasi \citep{2022MNRAS.512.3551D}, AE Ursae Majoris and RV Arietis \citep{2023MNRAS.526.1951D}, V2367 Cygni \citep{2024MNRAS.532.1140D}, five SX Phoenicis stars in $\omega$ Centauri \citep{2025MNRAS.539.3381D}, and BL Camelopardalis \citep{2026arXiv260819076S}.

The evolutionary tracks were computed using the Warsaw--New Jersey stellar evolution code \citep{1969AcA....19....1P,1999AcA....49..119P}. Unless stated otherwise, the models adopt the solar chemical mixture of \citet{AGSS09} (AGSS09) and the OPAL2005 equation of state \citep{2002ApJ...576.1064R,1996ApJ...456..902R}. Convective core overshooting is included following the prescription of \citet{2008MNRAS.385.2061D}.


Our previous studies have shown that, for $\delta$ Scuti stars, the choice of opacity tables can significantly affect the results \citep[e.g.][]{2023ApJ...942L..38D}. Therefore, we performed simulations using three different opacity datasets: OPAL \citep{OPAL}, OP \citep{OP} and OPLIB \citep{OPLIB1, OPLIB2}. At lower temperatures, $\log T < 3.95$, all three datasets were supplemented with the low-temperature opacity tables of \citet{2005ApJ...623..585F}.

Stellar pulsations were computed using the linear non-adiabatic pulsation code developed by \citet{Dziembowski1977}. The code includes both first- and second-order rotational corrections to the pulsation frequencies within the framework of perturbation theory. 
The calculations were performed using the frozen-convection approximation, which assumes that the convective flux remains unchanged throughout the pulsation cycle. This approximation is expected to be appropriate when convective energy transport in the stellar envelope is relatively inefficient, as is likely the case for a moderately massive star such as AI Vel. Nevertheless, even weak convection may influence the theoretical values of the $f$-parameter. 

\subsection{Two radial modes}
\label{sect:seismic_models}

In the first approach to the seismic modelling of AI Vel, we considered only two radial modes. The model predictions were compared with eight observational constraints: the effective temperature $T_{\mathrm{eff}}$, luminosity $L/L_{\odot}$, the frequencies of the two radial modes $\nu_1$ and $\nu_2$, and the corresponding non-adiabatic $f$-parameters, namely Re$f(\nu_1)$, Im$f(\nu_1)$, Re$f(\nu_2)$, and Im$f(\nu_2)$.

We computed a large grid of models (approximately 200\,000 in total for each opacity source), from which posterior probability distributions were derived. The resulting histograms are presented in Fig.\,\ref{fig:simulation_his}, and the median values of model parameters are listed in Table\,\ref{tab:simulation_models}.
\setlength{\tabcolsep}{4pt}
    \begin{table*}
    \begin{center} 		
            \caption{Parameters  derived from simulations including two radial modes. The table columns contain: mass, $M$, metallicity, $Z$, initial hydrogen abundance, $X_0$, initial surface rotational velocity, $V_{\rm{rot,0}}$, current surface rotational velocity, $V_{\rm{rot}}$, convective mixing length parameter, $\alpha_{\rm{mlt}}$, convective core overshooting parameter, $\alpha_{\rm{ov}}$, effective temperature, $\log T_{\rm{eff}}$, luminosity, $\log L/L_{\odot}$,  age since the zero-age main sequence, Age, gravity, $\log{g}$, radius, $R$,  and amplitudes of relative radius variations due to pulsation in the radial fundamental, $\varepsilon(\nu_1)$, and first overtone, $\varepsilon(\nu_2)$, modes as well as the instability parameters for the radial fundamental, $\eta(\nu_1)$, and first overtone, $\eta(\nu_2)$, modes. The atmospheric microturbulent velocity, $\xi_{\rm{t}}$, was fixed at 4 km\,s$^{-1}$.}
	      \label{tab:simulation_models}           	
            \begin{tabular}{cccccccccc}
            \hline\hline
              Opacity & $M [M_{\sun}]$       & $Z$       & $X_0$       & $V_{\mathrm{rot,0}}$       & $V_{\mathrm{rot}}$ & $\alpha_{\mathrm{mlt}}$  & $\alpha_{\mathrm{ov}}$       & $\log{T_{\mathrm{eff}}}$  & $\log{L/L_{\sun}}$\\ \hline
 OPAL & $1.737^{+0.042}_{-0.032}$ &  $0.01392^{+0.00120}_{-0.00088}$ &  $0.7105^{+0.0079}_{-0.0073}$ &  $12.3^{+7.5}_{-8.7}$ &  $11.6^{+7.0}_{-8.1}$ &  $0.467^{+0.094}_{-0.088}$ &  $0.094^{+0.055}_{-0.056}$ &  $3.8591^{+0.0017}_{-0.0015}$ &  $1.2595^{+0.011}_{-0.0088}$ \\
 &&&&&&&&&
\\
OP & $1.566^{+0.024}_{-0.025}$ &  $0.01419^{+0.00064}_{-0.00084}$ &  $0.7207^{+0.0053}_{-0.011}$ &  $5.7^{+9.6}_{-5.7}$ &  $5.4^{+9.2}_{-5.4}$ &  $0.331^{+0.026}_{-0.028}$ &  $0.146^{+0.066}_{-0.038}$ &  $3.8179^{+0.0017}_{-0.0014}$ &  $1.0634^{+0.010}_{-0.0069}$  \\ 
 &&&&&&&&&
\\
OPLIB & $1.696^{+0.034}_{-0.040}$ &  $0.0178^{+0.0013}_{-0.0015}$ &  $0.7160^{+0.0084}_{-0.0094}$ &  $6.6^{+9.8}_{-6.6}$ &  $6.3^{+9.2}_{-6.3}$ &  $0.413^{+0.042}_{-0.033}$ &  $0.098^{+0.048}_{-0.029}$ &  $3.8276^{+0.0015}_{-0.0019}$ &  $1.1250^{+0.0085}_{-0.0084}$\\ 
\end{tabular}
     
\begin{tabular}{cccccccc}
\hline\hline
                 Opacity & Age [Gyr] & $\log{g}$    &            $R [R_{\sun}]$     & $\varepsilon(\nu_1)$ & $\varepsilon(\nu_2)$& $\eta(\nu_1)$ & $\eta(\nu_2)$ \\ \hline

OPAL &  $1.398^{+0.078}_{-0.068}$ &  $3.8055^{+0.0038}_{-0.0029}$ &  $2.728^{+0.021}_{-0.016}$ &    $0.01029^{+0.00017}_{-0.00013}$ &  $0.00480^{+0.00002}_{-0.00002}$ &  $0.051^{+0.006}_{-0.006}$ &  $0.073^{+0.004}_{-0.005}$ \\ 
\\
OP &  $2.099^{+0.13}_{-0.067}$ &  $3.7913^{+0.0024}_{-0.0022}$ &  $2.633^{+0.013}_{-0.015}$ &  $0.0221^{+0.0011}_{-0.0011}$ &  $0.0089^{+0.0005}_{-0.0005}$ &  $0.109^{+0.001}_{-0.001}$ &  $0.085^{+0.003}_{-0.003}$ \\
\\

OPLIB  &   $1.710^{+0.076}_{-0.057}$ &  $3.8029^{+0.0032}_{-0.0038}$ &  $2.704^{+0.017}_{-0.021}$ &  $0.0194^{+0.0013}_{-0.0012}$ &  $0.0077^{+0.0005}_{-0.0005}$ &  $0.099^{+0.001}_{-0.001}$ &  $0.092^{+0.001}_{-0.001}$  \\ 
\hline
            \end{tabular}
       \end{center}
   \end{table*}

We adopted the median rather than the mean as the representative value for each parameter, since many of the posterior distributions are significantly non-Gaussian (for example, initial hydrogen abundance or rotational velocity in Fig.\,\ref{fig:simulation_his}). In such cases, the median provides a more reliable and stable estimate of the central value.

Uncertainties were estimated as the difference between the median and the 16th and 84th percentiles of the posterior distributions. For a Gaussian distribution, the interval between the 16th and 84th
percentiles corresponds approximately to the conventional $1\sigma$
confidence interval.

All models listed in Table\,\ref{tab:simulation_models} reproduce the observed frequencies of the radial modes within the adopted uncertainties. Instead of using the formal uncertainties derived from the \textit{TESS} data analysis, we adopted a frequency uncertainty of 0.001\,d$^{-1}$. This value reflects both the numerical precision of the computed pulsation frequencies and the frequency variations observed between different \textit{TESS} sectors.

\begin{figure}
    \centering
    \includegraphics[width=\columnwidth]{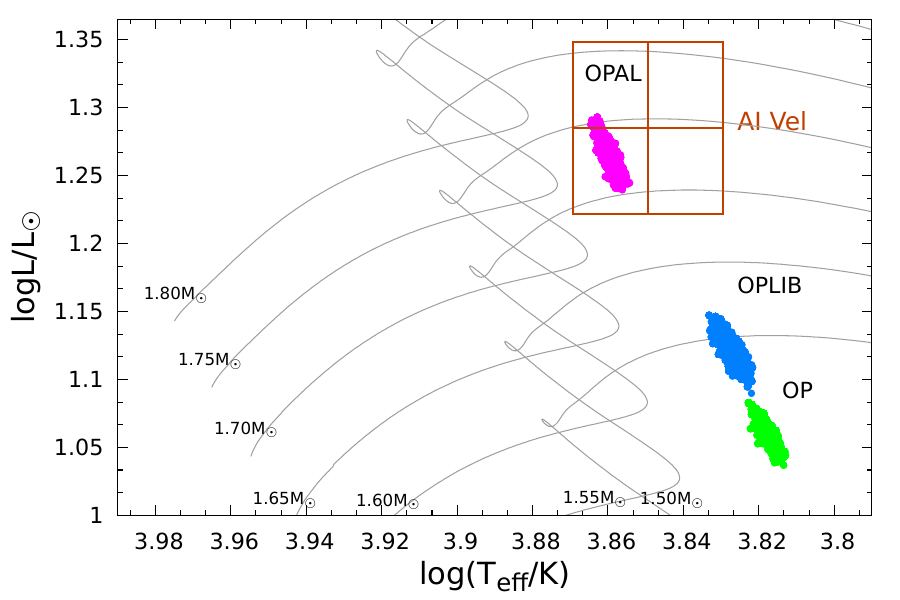}
    \caption{The HR diagram showing the position of AI Vel. The orange rectangle indicates the observational error box. Seismic models computed using OPAL, OP, and OPLIB opacity tables are also plotted for comparison (see text for details). The evolutionary tracks were calculated with $X_0=0.7$, $Z=0.010$, $V_{\rm{rot,0}}=10$\,km\,s$^{-1}$, $\alpha_{\rm{mlt}}=1.0$ and $\alpha_{\rm{ov}}=0.0$.}
    \label{fig:HR_models}
\end{figure}

In our calculations, we used three opacity data sources: OPAL, OP, and OPLIB. A significant result is that only the models computed with OPAL opacities simultaneously reproduce both the effective temperature and luminosity of AI Vel. In contrast, our seismic models based on OP and OPLIB opacities lie well below the observational error box.
Figure~\ref{fig:HR_models} presents the Hertzsprung–Russell (HR) diagram, where the position of AI Vel is marked with the seismic models obtained from our simulations. As demonstrated, only the OPAL-based models are consistent with the observed location of AI Vel. Similar behaviour has previously been reported for other A- and F-type pulsating stars \cite[e.g.,][]{2023ApJ...942L..38D,2025MNRAS.539.3381D}.

For the OPAL models, the derived stellar mass is $\sim1.737\,M_{\odot}$, with a metallicity close to the solar value ($Z\sim 0.0139$). The initial hydrogen content aligns with $X_0 \sim 0.71$, and the inferred rotational velocity is low, $V_{\rm{rot}} \in (4,19)$\,km\,s$^{-1}$, indicating that AI Vel is a slow rotator.

The efficiency of convective energy transport in the outer layers is found to be very low, with a mixing-length parameter $\alpha_{\rm{mlt}} \sim 0.5$. Similarly, the extent of convective core overshooting during the main-sequence phase is small, with $\alpha_{\rm{ov}} \sim0.1$, suggesting limited mixing beyond the core boundary. The estimated stellar age is 1.4\,Gyr and the models indicate that AI Vel is in the hydrogen shell-burning phase of post-main-sequence evolution. The central hydrogen abundance is fully depleted, with $X_c = 0$.

Both considered radial modes are predicted to be excited. To assess mode stability, we used the normalized instability parameter $\eta$ \citep{Stellingwerf1978}, which ranges from $-1$ (strongly damped) to $+1$ (strongly excited). For the dominant frequency $\nu_1$, we obtained $\eta \sim 0.05$, and for the secondary frequency $\nu_2$, we got $\eta \sim 0.07$, confirming that both modes are unstable and self-excited in the models.

The OP and OPLIB models differ significantly from the OPAL models. They are characterised by higher metallicities, $Z\in(0.014,0.019)$, and indicate larger initial hydrogen abundances, $X_0\in(0.716,0.72)$. 

Models computed with all opacity tables are slow rotators, with $V_{\rm rot} \lesssim 19$\,km\,s$^{-1}$. In all cases, the models require low efficiency of convective energy transport, as well as a small extent of convective core overshooting.

\begin{center}
   \begin{figure}[ht]
     \centering
     \includegraphics[width=0.9\columnwidth]{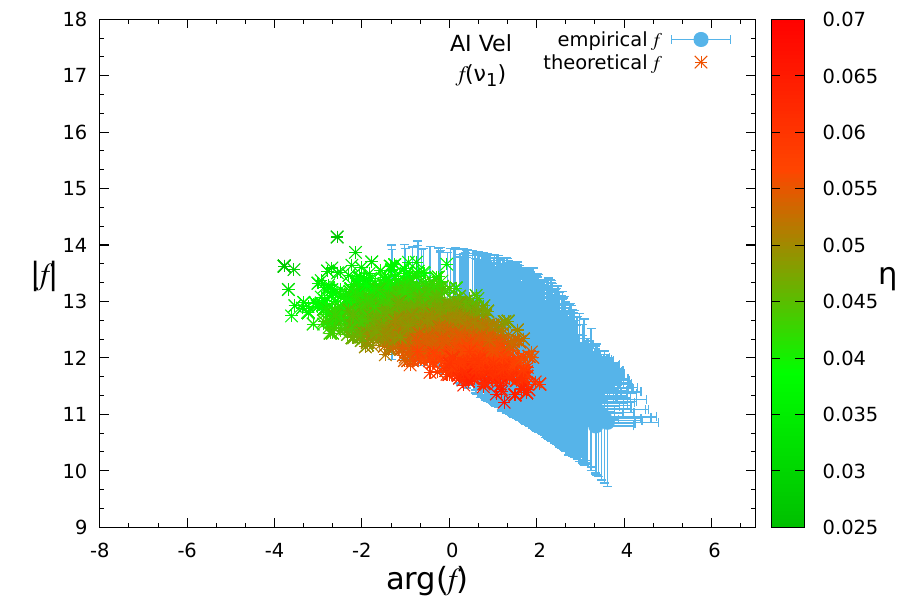}
     \includegraphics[width=0.9\columnwidth]{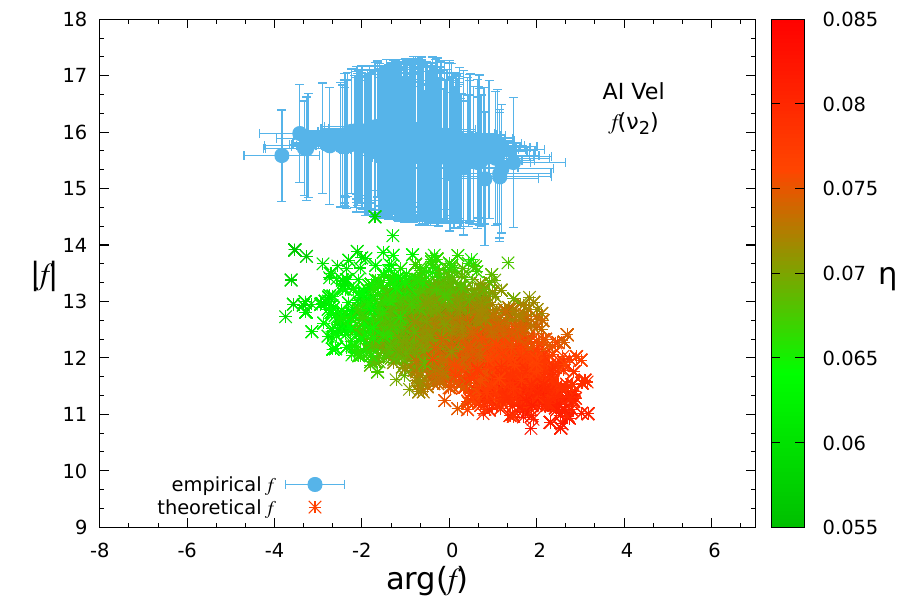}
     \caption{Empirical and theoretical values of the non-adiabatic $f$-parameter for the radial fundamental (upper  panel) and first overtone (bottom panel) modes. Models from the OPAL simulation are plotted. Different colours correspond to different values of the instability parameter, $\eta$.}
     \label{fig:empir_f}
  \end{figure}
 \end{center}

Figure\,\ref{fig:empir_f} presents the empirical and theoretical values of the $f$-parameter in the $|f|$--$\arg(f)$ plane. The quantity $|f|$ represents the amplitude of the bolometric flux variations, while the phase $\arg(f)$ corresponds to the phase lag between variations in stellar flux and radius.  The agreement between the observed and predicted values for $\nu_1$ (upper panel) is generally good, with most models falling within the $1\sigma$ uncertainties of $f$. In contrast, for the radial first overtone mode (bottom panel), a significant discrepancy is evident for $|f|$. All models used in this simulation were computed assuming a microturbulent velocity of $\xi_t = 4$\,km\,s$^{-1}$. The instability parameter $\eta$ is colour-coded. As noted earlier, all considered modes are unstable; however, a correlation between $\eta$ and the $f$-parameter is apparent. Interestingly, better agreement between the empirical and theoretical $f$ values tends to occur for modes with higher $\eta$.

Because of the small frequency difference, the theoretical $f$ values for the two radial modes are very similar. In contrast, the empirical 
amplitude, $|f|$, derived for $\nu_2$ is significantly larger. This discrepancy may be related to uncertainties in the determination of the 
empirical $f$-parameter for this mode. Alternatively, it may indicate the presence of an additional physical effect not accounted for in our 
analysis.

\subsection{Adding the non-radial mode}

Including a non-radial mode in the seismic modelling provides an additional constraint that can substantially narrow the posterior distributions of the stellar parameters, provided that the mode identification is correct. In this analysis, we incorporated the frequency $\nu_3=15.976$\,d$^{-1}$. The spectroscopic analysis indicates that this frequency corresponds to an axisymmetric mode ($m=0$), whereas its spherical degree $\ell$ could not be uniquely constrained. We therefore considered two plausible low-degree identifications: a dipole mode, $(\ell,m)=(1,0)$, and a quadrupole mode, $(\ell,m)=(2,0)$.

For each of these hypotheses, we performed an additional seismic simulation including $\nu_3$ as an observational constraint. Only OPAL opacities were considered at this stage, because, within the explored parameter space, our models computed with OP and OPLIB opacities lie well outside the observational error box in the HR diagram.

The inclusion of this mode leads to a substantial tightening of the posterior distributions for each assumed mode identification, as summarized in Table\,\ref{tab:simulation_models_l1_p2}. Most of the derived stellar parameters remain consistent with those obtained in the previous simulations, but their formal statistical uncertainties are considerably reduced. However, although the formal uncertainties decrease for a given mode identification, the true mode degree remains uncertain. Therefore, this apparent improvement in precision does not reflect the total uncertainty of the stellar parameters, as it does not account for the systematic differences between the alternative mode-identification hypotheses.

\begin{table*}[ht]
\setlength{\tabcolsep}{2pt}
   \begin{center} 	   
            \caption{Same as Table~\ref{tab:simulation_models}, but for simulations including the non-radial mode $\nu_3$ and restricted to OPAL opacities. 
            The microturbulent velocity, $\xi_t$, inferred from the  modelling is also listed. Different modelling assumptions are considered, including dipole- and quadrupole-mode identifications of $\nu_3$, the GN93 chemical mixture, and Kurucz model atmospheres.
            }
            \label{tab:simulation_models_l1_p2}	        
		      \begin{tabular}{cccccccccc}
              \hline\hline
            Opacity &$M [M_{\odot}]$       & $Z$       & $X_0$       & $V_{\mathrm{rot,0}}$     & $V_{\mathrm{rot}}$ & $\alpha_{\mathrm{mlt}}$  & $\alpha_{\mathrm{ov}}$       & $\log{T_{\mathrm{eff}}}$ & $\log{\frac{L}{L_{\odot}}}$   \\\hline
            \multicolumn{10}{c}{$(\ell=1,m=0)$}
            \\
            OPAL & $1.7258^{+0.0047}_{-0.020}$ &  $0.01335^{+0.00031}_{-0.00024}$ &  $0.7132^{+0.0035}_{-0.0077}$ &  $14.4^{+7.7}_{-5.2}$ &  $13.5^{+7.2}_{-4.9}$ &  $0.158^{+0.12}_{-0.095}$ &  $0.0690^{+0.0040}_{-0.0029}$ &  $3.8578^{+0.0013}_{-0.0013}$ &  $1.2515^{+0.0059}_{-0.0058}$ \\

             \multicolumn{10}{c}{$(\ell=2,m=0)$}
            \\
            OPAL & $1.7114^{+0.0052}_{-0.0079}$ &  $0.01375^{+0.00017}_{-0.00013}$ &  $0.7052^{+0.0026}_{-0.0055}$ &  $8.8^{+14.9}_{-5.9}$ &  $8.3^{+14.0}_{-5.5}$ &  $0.17^{+0.14}_{-0.11}$ &  $0.0785^{+0.0020}_{-0.0024}$ &  $3.8581^{+0.0012}_{-0.0013}$ &  $1.2509^{+0.0050}_{-0.0057}$ \\

            \multicolumn{10}{c}{GN93 $(\ell=1,m=0)$}
            \\ 
            OPAL & $1.7102^{+0.0088}_{-0.0023}$ &  $0.01391^{+0.00029}_{-0.00003}$ &  $0.6999^{+0.0007}_{-0.0084}$ &  $18.4^{+18.9}_{-7.1}$ &  $17.2^{+17.8}_{-7.6}$ &  $0.28^{+0.13}_{-0.15}$ &  $0.0813^{+0.0029}_{-0.0038}$ &  $3.869^{+0.005}_{-0.028}$ &  $1.297^{+0.019}_{-0.11}$  \\

            \multicolumn{10}{c}{Kurucz $(\ell=1,m=0)$}
            \\ 
            
            OPAL & $1.7367^{+0.0022}_{-0.0081}$ &  $0.01438^{+0.00010}_{-0.00031}$ &  $0.7135^{+0.0012}_{-0.0022}$ &  $10.7^{+2.2}_{-5.7}$ &  $10.0^{+2.1}_{-5.6}$ &  $0.082^{+0.054}_{-0.033}$ &  $0.087^{+0.005}_{-0.002}$ &  $3.854^{+0.001}_{-0.016}$ &  $1.240^{+0.005}_{-0.065}$   \\  
         \end{tabular}
       
		\begin{tabular}{cccccccccc}
         \hline\hline
                 Opacity &Age [Gyr] & $\log{g}$    &            $R [R_{\sun}]$     &      $\varepsilon(\nu_1)$ & $\varepsilon(\nu_2)$ &  $\eta(\nu_1)$ & $\eta(\nu_2)$ & $\eta(\nu_3)$ & $\xi_{\rm{t}}$ \\\hline                 
                 \multicolumn{10}{c}{$(\ell=1,m=0)$}
                 \\
                 OPAL & $1.395^{+0.014}_{-0.016}$ &  $3.8041^{+0.0006}_{-0.0012}$ &  $2.7221^{+0.0029}_{-0.011}$ &  $0.0103^{+0.0002}_{-0.0001}$ &  $0.0050^{+0.0001}_{-0.0001}$ &  $0.0365^{+0.0064}_{-0.0057}$ &  $0.0577^{+0.0047}_{-0.0033}$ &  $0.0637^{+0.0067}_{-0.0016}$ &  $3.19^{+0.28}_{-0.19}$ \\
                 \multicolumn{10}{c}{$(\ell=2,m=0)$}
                 \\
 
                 OPAL & $1.391^{+0.020}_{-0.017}$ &  $3.8031^{+0.0008}_{-0.0004}$ &  $2.7140^{+0.0032}_{-0.0041}$ &  $0.0102^{+0.0002}_{-0.0001}$ &  $0.0050^{+0.0001}_{-0.0001}$ &  $0.0382^{+0.0070}_{-0.0045}$ &  $0.0588^{+0.0061}_{-0.0027}$ &  $0.0645^{+0.0092}_{-0.0014}$ &  $3.18^{+0.34}_{-0.21}$ \\ 
                 
                 \multicolumn{10}{c}{GN93 $(\ell=1,m=0)$}
                 \\        

                 OPAL & $1.218^{+0.039}_{-0.038}$ &  $3.8024^{+0.0004}_{-0.0022}$ &  $2.7151^{+0.0038}_{-0.0052}$ &  $0.0103^{+0.0002}_{-0.0002}$ &  $0.0050^{+0.0001}_{-0.0001}$ &  $-0.003^{+0.007}_{-0.011}$ &  $0.0389^{+0.0054}_{-0.0074}$ &  $0.0714^{+0.0099}_{-0.0022}$ &  $2.69^{+0.27}_{-0.36}$ \\ 
                 
            \multicolumn{10}{c}{Kurucz $(\ell=1,m=0)$}  
            \\         

                 OPAL & $1.442^{+0.008}_{-0.012}$ &  $3.8059^{+0.0005}_{-0.0002}$ &  $2.7267^{+0.0015}_{-0.0062}$ &  $0.0113^{+0.0002}_{-0.0008}$ &  $0.0054^{+0.0001}_{-0.0004}$ &  $0.043^{+0.038}_{-0.003}$ &  $0.060^{+0.012}_{-0.002}$ &  $0.060^{+0.001}_{-0.018}$ &  $2.00$   \\ 
                          \hline
            \end{tabular}
       \end{center}
   \end{table*}

The posterior distributions of selected parameters are presented in Fig.\,\ref{fig:simulation_his_dipole_sel} (for the dipole mode hypothesis) and in Fig.\,\ref{fig:simulation_his_quadrupole} (for the quadrupole mode hypothesis). We emphasize that the quoted uncertainties reflect the statistical spread within the adopted model grid and do not account for systematic uncertainties associated with the input physics or the ambiguity in the identification of the non-radial mode.

Including this mode increases the total number of observational constraints to nine: $\log{T_{\rm eff}}$, $\log{L/L_{\sun}}$, $\nu_1$, $\nu_2$, Re$f(\nu_1)$, Im$f(\nu_1)$, Re$f(\nu_2)$, Im$f(\nu_2)$, and $\nu_3$.
This additional constraint allows us to treat the microturbulent velocity, $\xi_t$, as a free parameter in the derivation of the empirical $f$-parameter. Although $\xi_t$ is not a stellar-structure parameter, it influences the empirical determination of $f$ through the adopted model atmospheres. However, since the value derived from the HARPS spectroscopic analysis is $\xi_t = 3.0(1)$\,km\,s$^{-1}$, we restricted our analysis to models with $\xi_t \lesssim 3.5$\,km\,s$^{-1}$.

\begin{center}
   \begin{figure}[ht]
    \centering
      \includegraphics[width=0.9\columnwidth]{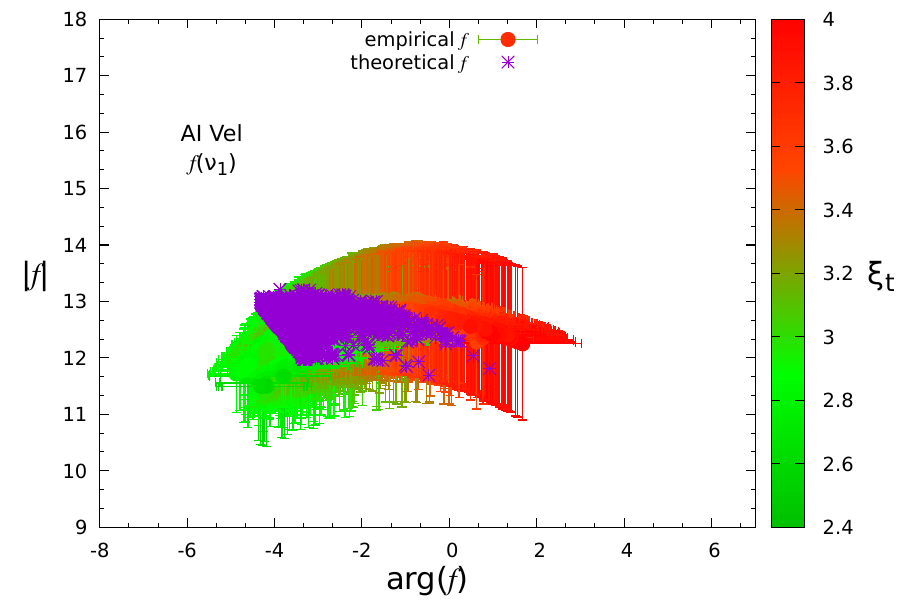}
     \includegraphics[width=0.9\columnwidth]{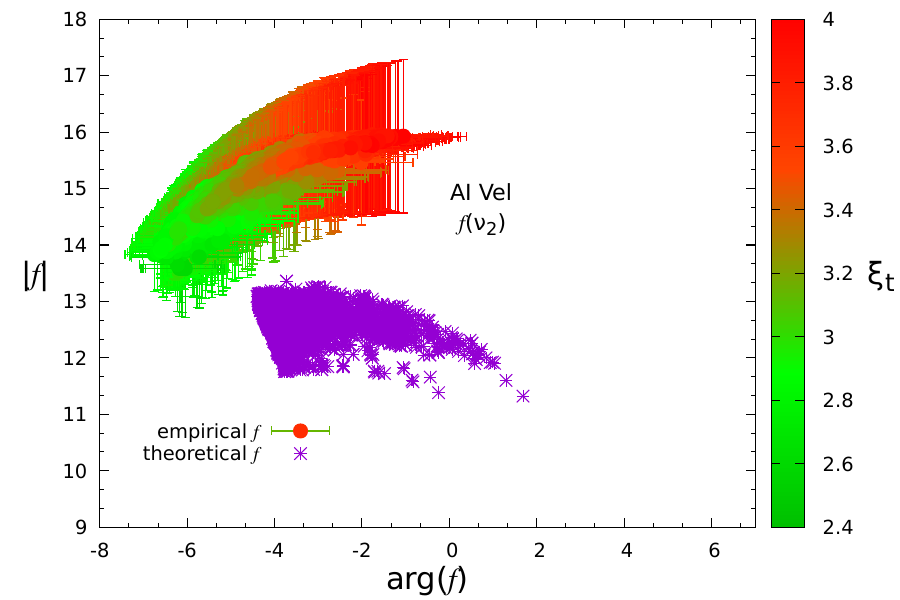}
     \caption{Empirical and theoretical values of the non-adiabatic $f$-parameter for the radial fundamental (upper panel) and first overtone
(bottom panel) modes for the seismic models obtained under the dipole-mode hypothesis for $\nu_3$. The models were computed with OPAL opacities. Different colours correspond to different values of the microturbulent
velocity, $\xi_t$.}
     \label{fig:empir_f_l1}
  \end{figure}
 \end{center}

Figure\,\ref{fig:empir_f_l1} compares the empirical and theoretical values of the $f$-parameter for the dipole-mode hypothesis. The colour coding indicates the value of $\xi_t$ adopted in deriving the empirical $f$ values.
The corresponding results for the quadrupole-mode  hypothesis are shown in
Fig.\,\ref{fig:empir_f_l2}.

For models fitting two radial modes and additionally the $\nu_3$ frequency as a dipole mode, the main difference, apart from the microturbulent velocity, is found in the mixing-length parameter, which is significantly reduced to $\alpha_{\rm mlt} \approx 0.16$. For the remaining parameters, we primarily observe a reduction in their spreads, leading to smaller uncertainties.

The radial order that best reproduces the observed dipole mode falls in the range $n = 9$--11 in our simulations. In hydrogen shell-burning models, these frequencies correspond to gravity modes, since most of their kinetic energy is concentrated in the deep interior of the star. The ratio of the kinetic energy contained within the gravity-mode propagation cavity to the total kinetic energy is 
$E_{\rm kin,grav}/E_{\rm kin} =
0.94^{+0.06}_{-0.19}$.
Consequently, the frequency at 15.976\,d$^{-1}$ can be identified with the g$_9$, g$_{10}$, or g$_{11}$ mode.

In Fig.\,\ref{fig:mode_In} we plotted the mode inertia, highlighting the different nature of the radial modes and the mixed dipole mode. For the radial modes, the inertia amplitude is distributed over a large region between $\log{(T/\rm{K})}\sim6.5$ and $\log{(T/\rm{K})}\sim4.5$. The dipole mode $g_{10}$, identified with the frequency $\nu_3$, is basically insensitive to that part of the star. The highest inertia of the dipole mode is concentrated in the stellar core, where the eigenfunction exhibits ten closely spaced radial nodes. As a result, the g$_{10}$ mode is particularly sensitive to the 
structure near the convective core boundary established during the main-sequence phase of stellar evolution. The model parameters correspond to those listed in Table\,\ref{tab:simulation_models_l1_p2} labelled as $(\ell=1,m=0)$.



\begin{center}
   \begin{figure}
     \centering
     \includegraphics[width=0.9\columnwidth]{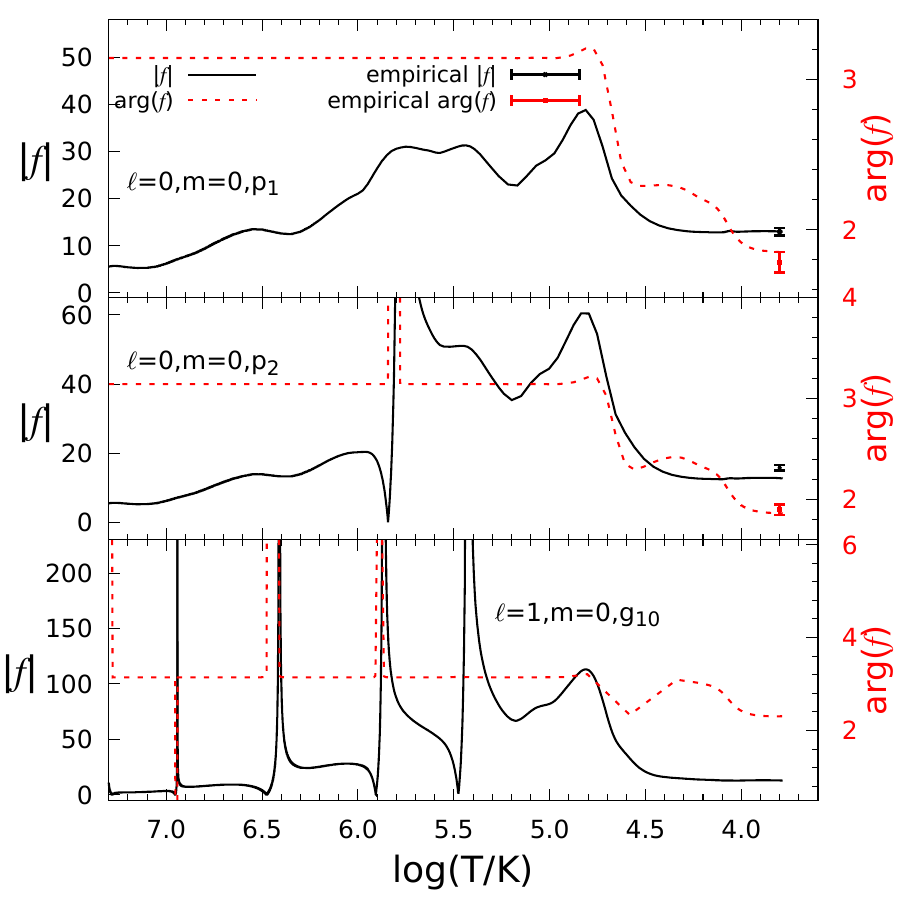}
     \caption{Profiles of the eigenfunction $f$  for the seismic model of AI Vel, whose parameters are listed in Table\,\ref{tab:simulation_models_l1_p2}. The left-hand axis shows the amplitude $|f|$, while the right-hand axis indicates the argument, $\arg(f)$. The upper panel corresponds to the radial fundamental mode, the middle panel to the first overtone radial mode, and the bottom panel to the dipole mode $g_{10}$. The empirical $f$ values for the two radial modes are also marked for comparison.}
     \label{fig:empir_f_l1_profile}
  \end{figure}
 \end{center}

The eigenfunction $f$ is shown in detail in 
Fig.\,\ref{fig:empir_f_l1_profile}, where both its amplitude, $|f|$, and phase, $\arg(f)$, are plotted as functions of temperature within a 
seismic model of AI Vel. The quantity referred to as the $f$-parameter corresponds to the value of this eigenfunction evaluated at the 
photosphere. 

From top to bottom, the panels in Fig.\,\ref{fig:empir_f_l1_profile} show the $f$-eigenfunction for the radial fundamental mode, the radial first overtone, and the dipole g$_{10}$ mode, respectively. For the two radial modes, the empirical values of the $f$-parameter, corresponding to the value of the eigenfunction at the photosphere, are also marked for comparison.
The empirical $f$-parameter for the radial fundamental mode is reproduced well by the theoretical model. In contrast, for the first overtone mode, the amplitude of the theoretical $f$-parameter is slightly underestimated.

\begin{center}
   \begin{figure}
   \centering
    \includegraphics[width=0.9\columnwidth]{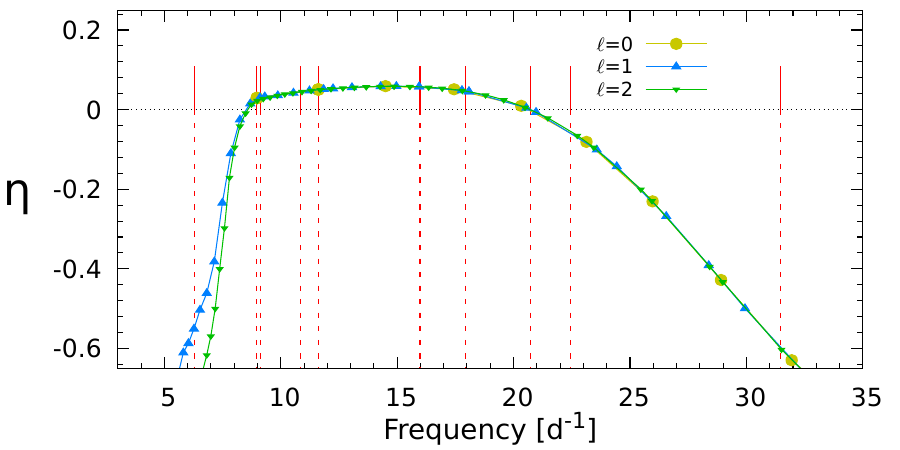}
     \caption{ Instability parameter, $\eta$, as a function of frequency for the seismic model of AI Vel, whose parameters are listed in Table\,\ref{tab:simulation_models_l1_p2}. Vertical lines indicate the independent frequencies detected in AI Vel.}
     \label{fig:nu_eta}
  \end{figure}
 \end{center}

The instability parameter as a function of frequency for mode degrees $\ell = 0$, 1, and 2 is shown in Fig.\,\ref{fig:nu_eta}. Vertical lines 
indicate the independent frequencies detected in AI Vel. Most of the observed frequencies fall within the unstable regime.
An intriguing feature is the presence of the low-frequency mode $\nu_{10} = 6.3$\,d$^{-1}$. If this frequency corresponds to an independent pulsation mode, it remains strongly damped in our models. A 
similar situation occurs for the highest detected frequency, $\sim31.45$\,d$^{-1}$, whose instability parameter is approximately $-0.6$. It is therefore possible 
that these two frequencies are not independent modes but instead represent unresolved combination frequencies. 
On the other hand, the 6.3\,d$^{-1}$ signal can be reproduced by an axisymmetric dipole g$_{28}$ mode, while the $\sim31.45$\,d$^{-1}$ frequency may correspond to an axisymmetric quadrupole p$_2$ mode. The difficulty in exciting these modes could therefore indicate that the opacity profile adopted in the models is not fully adequate and may require modifications \citep[e.g.,][]{2019MNRAS.485.3544W,2024MNRAS.529.4176W}. A detailed investigation of this issue, however, is beyond the scope of the present paper.

Details of the considered seismic model structure are presented in Appendix\,\ref{Sect:app:sm}, Fig.\,\ref{fig:kappa_L}. The upper panel shows the mean opacity coefficient, $\log{\kappa}$, along with its derivatives with respect to temperature, $\partial\log{\kappa}/\partial\log{T}$, and density, $\partial\log{\kappa}/\partial\log{\rho}$. In the bottom panel, the left-hand axis presents the energy generation rates from the pp chain ($\varepsilon_{\rm pp}$) and the CNO cycle ($\varepsilon_{\rm CNO}$), while the right-hand axis shows the hydrogen ($X$) and helium ($Y$) abundances. The model is clearly in the hydrogen shell-burning phase of evolution, with the CNO cycle as the dominant source of nuclear energy.

Finally, we consider the frequency $\nu_4=15.952$\,d$^{-1}$, which lies close to the dipole axisymmetric mode $\nu_3=15.976$\,d$^{-1}$. Still, the separation between the two frequencies is larger than the Rayleigh
frequency resolution. This frequency could represent a rotationally split component of a dipole mode, with $\nu_3$ and $\nu_4$ forming two members of the same triplet.

Finally, we consider the frequency $\nu_4=15.952$\,d$^{-1}$, which lies close to $\nu_3=15.976$\,d$^{-1}$. Still, the separation between the two frequencies is larger than the Rayleigh
frequency resolution. This frequency could represent a rotationally split component of a dipole mode, with $\nu_3$ and $\nu_4$ forming two members of the same triplet.

The observed separation between $\nu_3$ and $\nu_4$ is
approximately 0.024\,d$^{-1}$. For our seismic models, the Ledoux constant of the relevant dipole modes is
$C_{n=9\text{--}11,\ell=1}=0.49^{+0.07}_{-0.10}$, while the inferred rotational frequency is $\nu_{\rm rot}=0.098^{+0.052}_{-0.035}$\,d$^{-1}$.
The corresponding first-order rotational splitting is approximately 0.025--0.087\,d$^{-1}$. Thus, the observed separation lies close to the lower limit of the range predicted by our models. Although this does not
exclude the possibility that $\nu_3$ and $\nu_4$ are components of the same rotationally split multiplet, the identification remains tentative. Moreover, no convincing third component of the expected triplet is detected in the periodogram. We therefore did not include $\nu_4$ as an additional seismic constraint.

For the quadrupole-mode hypothesis, we obtained seismic parameters broadly similar to those derived under the dipole-mode hypothesis (Table\,\ref{tab:simulation_models_l1_p2}). However, for several parameters, the shift between the two solutions is comparable to, or larger than, the formal uncertainties inferred from the individual posterior distributions. This comparison demonstrates that uncertainties associated with mode identification introduce a systematic component that is not captured by the statistical posterior uncertainties.

The quadrupole mode that best reproduces the observed frequency has a radial order of $n=15$ and corresponds to the gravity mode $g_{15}$. The distribution of its mode inertia is similar to that of the dipole $g_{10}$ mode shown in Fig.\,\ref{fig:mode_In}. Most of its inertia is concentrated in the deep stellar interior, with only a small contribution from the outer layers. The $f$-eigenfunction of the quadrupole $g_{15}$ mode is also very similar to that of the dipole $g_{10}$ mode presented in the bottom panel of Fig.\,\ref{fig:empir_f_l1_profile}. Thus, despite their different spherical degrees and radial orders, the dipole and quadrupole modes reproducing the same observed frequency have broadly similar properties in the seismic models of AI Vel.

At present, we find no statistically significant preference for either the dipole or quadrupole identification of $\nu_3$. Both hypotheses reproduce the available observational constraints comparably well, and the spectroscopic mode identification does not uniquely constrain the spherical degree $\ell$. Therefore, the $(\ell,m)=(1,0)$ solution should not be regarded as uniquely preferred. We use the dipole solution below only as a representative seismic model for illustrating the internal structure and pulsational properties of AI Vel.

To investigate the effect of the adopted chemical mixture, we also computed models using the elemental abundance mixture GN93 \citep{GN93}. The resulting model parameters are listed in Table\,\ref{tab:simulation_models_l1_p2}. The overall stellar properties remain similar, although several parameters shift by amounts exceeding their formal posterior uncertainties. The most pronounced difference is found for the stellar age, which decreases from
about 1.40 to 1.22\,Gyr.

We also investigated the influence of the adopted model atmospheres. Instead of the Vienna model atmospheres, we used the Kurucz models \citep{2003IAUS..210P.A20C}. The results are given in Table\,\ref{tab:simulation_models_l1_p2}. This calculation differs from the others because of limitations of the available atmosphere grid. In particular, we were restricted to a single value of the microturbulent velocity, $\xi_t=2$\,km\,s$^{-1}$, since this was the only value available over the required metallicity range. The resulting seismic models have somewhat higher masses and slightly higher metallicities than those obtained with the Vienna atmospheres. More importantly, however, these models provide a poorer match to the central empirical $f$ values for both radial modes (see Fig.\,\ref{fig:empir_f_Kurucz}).

Nevertheless, our tests with different model atmospheres show that the adopted flux derivatives can significantly affect the results of the seismic modelling. The derived parameters should therefore be interpreted with some caution, particularly because our pulsation calculations rely on the frozen-convection approximation and neglect nonlinear pulsation effects. The inferred efficiency of envelope convection in AI Vel is low, suggesting that the limitations of the frozen-convection approximation may have a relatively small impact. On the other hand, AI Vel pulsates in two high-amplitude radial modes, and nonlinear effects may therefore be non-negligible. Such effects could potentially contribute to the discrepancy between the empirical and theoretical $f$-parameters. A detailed nonlinear pulsation analysis, however, is beyond the scope of the present study.

Another important aspect of the empirical determination of the $f$-parameter is the reliability of its uncertainties. In our standard approach, the uncertainties of $f$ are obtained as formal errors from the least-squares solution of Eq.\,\ref{eq:campl}. The observational uncertainties of the amplitudes and phases enter this procedure primarily as weights and therefore have only a limited effect on the resulting formal uncertainties of the empirical $f$-parameters.

To obtain a more realistic estimate of these uncertainties, we additionally performed a Bayesian analysis in which posterior distributions of the empirical $f$-parameter were determined. Because these calculations are computationally demanding and require a large number of iterations to achieve convergence, we restricted this test to a representative sample of 10\,000 models. The result is nevertheless clear: the formal uncertainties obtained from the least-squares method are systematically about two to three times smaller than those inferred from the Bayesian posterior distributions.

When these larger uncertainties are taken into account, the empirical and theoretical $f$-parameters for both radial modes become statistically consistent within the estimated uncertainties. However, the central values of the empirical $f$-parameters remain essentially unchanged. Consequently, a systematic offset between the empirical and theoretical values of $|f|$ for the first overtone is still present, although its statistical significance is reduced.

An important conclusion from the comparison of different modelling assumptions is that the resulting systematic variations in the inferred stellar parameters can exceed the formal uncertainties derived from the Bayesian posterior distributions. The latter should therefore be interpreted as statistical uncertainties within a given set of modelling assumptions rather than as estimates of the total uncertainty of the inferred parameters.

\section{Discussion and Conclusions}

The case of AI Vel demonstrates that even a limited set of pulsation constraints, consisting of two radial modes and one additional low-degree non-radial mode, can provide detailed insight into the internal structure of a star when 
combined with Bayesian asteroseismic modelling. Our analysis highlights the importance of pulsation amplitudes and phases, which offer unique 
constraints on the efficiency of convective energy transport through the empirical $f$-parameter.

By fitting both the frequencies of the radial modes and the corresponding non-adiabatic $f$-parameters, we were able to tightly constrain key stellar parameters, including mass, initial hydrogen abundance, metallicity, rotational velocity, age, the mixing-length parameter, and the convective core overshooting parameter. Additionally, the presence of a non-radial mode allowed us to further refine the derived stellar parameters and place constraints on the atmospheric microturbulent velocity. The inclusion of this additional frequency substantially reduced the formal posterior uncertainties. However, this improvement depends on the adopted identification of the non-radial mode. In particular, the dipole and quadrupole hypotheses yield slightly different stellar parameters, with differences that can exceed the formal uncertainties obtained for either hypothesis. The quoted posterior uncertainties should therefore not be interpreted as the total uncertainties of the inferred stellar parameters. Nevertheless, the overall seismic picture remains consistent across the two mode-identification hypotheses, supporting the conclusion that AI Vel is an evolved, slowly rotating star.

One unresolved issue concerns the systematic offset between the empirical and theoretical $f$-parameter for the first overtone mode. Although the two values become statistically consistent when the larger uncertainties inferred from our Bayesian analysis are taken into account, the central empirical value of $|f|$ remains systematically higher than the theoretical prediction. This residual discrepancy may point to limitations in the atmospheric models, the frozen-convection approximation, or neglected nonlinear pulsation effects, and deserves further investigation.

Another noteworthy result is that the metallicity inferred from seismic modelling, $Z \sim 0.014$, is close to the solar value and slightly higher than the atmospheric metallicity derived from spectroscopic analysis ($Z \sim 0.011$). Although the seismic parameter space explored in this work included low-metallicity models, no satisfactory solutions were found for metallicities below the solar value.
The significant underabundance of iron, as well as other elements such as Ni, Zn, and Mg, may indicate that diffusion processes are active in AI Vel. In particular, gravitational settling of heavy elements could lead to reduced metal abundances in the stellar atmosphere \citep[e.g.,][]{2016A&A...589A.140D}. The relatively slow rotation rate inferred from the seismic modelling is consistent with this interpretation, as weak rotational mixing may allow diffusion processes to operate more efficiently.

AI Vel is another $\delta$ Scuti star for which we find relatively low  efficiency of convective energy transport in the outer layers. For $\delta$ Scuti itself, we derived that the mixing-length parameter $\alpha_{\rm mlt}$ is generally smaller than 1.0 \citep{2021MNRAS.505...88D}. In our earlier studies, we obtained $\alpha_{\rm mlt} \approx 0.6$ for BP Peg, $0.4$ for AE UMa, and $0.5$ for RV Ari. For the Population II star SX Phe, we found $\alpha_{\rm mlt} \approx 1.0$ \citep{2023ApJ...942L..38D}, while for RY Lep the value is $\alpha_{\rm mlt} \approx 0.3$ \citep{2026ApJ..1002..169N} as opposed to the solar value $\sim 1.8$.

The modelling of AI Vel also highlights the critical role of stellar opacity data in asteroseismic studies. Within the explored parameter space and adopted modelling assumptions, only models computed with OPAL opacities reproduce the observed properties
of AI Vel consistently. Similar results were obtained in our previous studies of other $\delta$ Scuti pulsators, including BP Peg, AE UMa, and RV Ari.
This behaviour, however, is not restricted to $\delta$ Scuti stars. A number of Population II SX Phoenicis variables exhibit the same preference for OPAL opacities. In particular, SX Phe itself, as well as the $\omega$ Centauri stars V194, V220, and V225, can only be modelled successfully using OPAL opacity tables \citep{2025MNRAS.539.3381D}. In contrast, some stars, including $\delta$ Scuti itself and the $\omega$ Centauri variables NV326 and V237, can be reproduced equally well with OPAL, OP, or OPLIB opacity datasets.

\begin{center}
   \begin{figure}
     \centering
     \includegraphics[width=0.95\columnwidth]{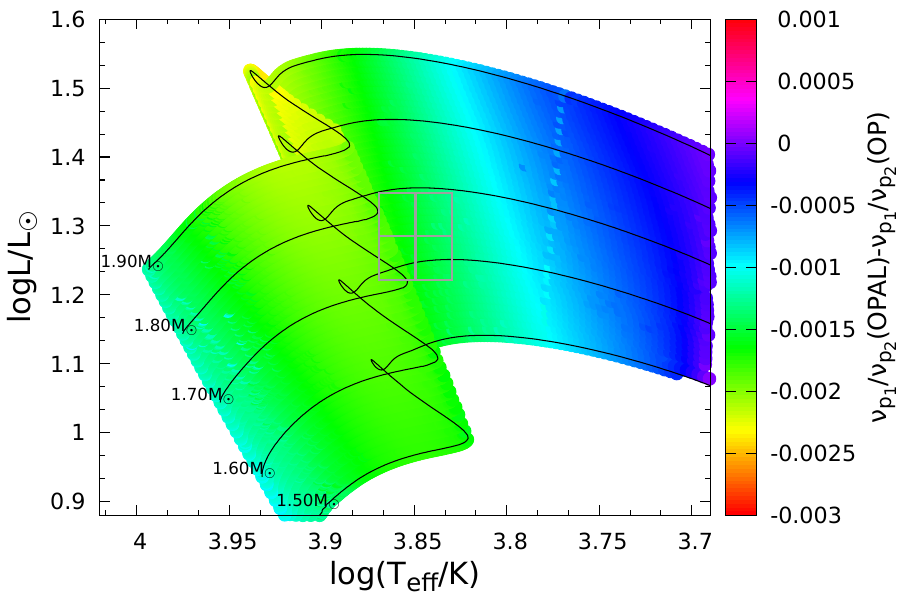}
     \includegraphics[width=0.95\columnwidth]{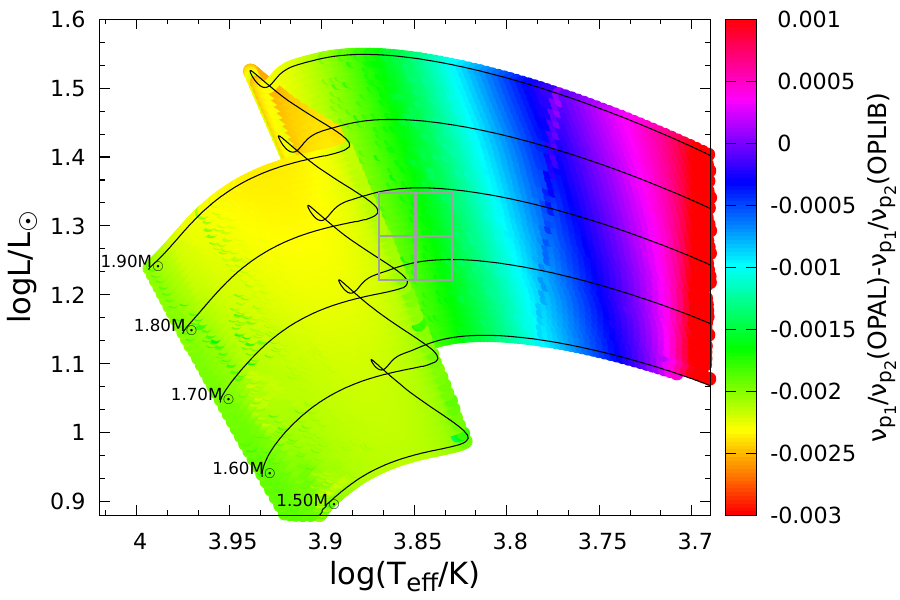}
     \includegraphics[width=0.95\columnwidth]{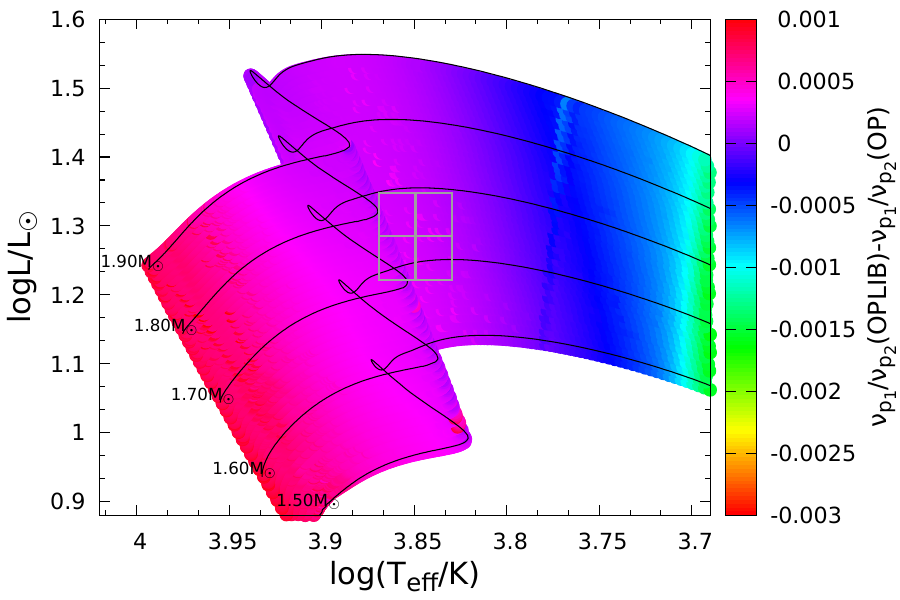}
     \caption{HR diagram showing the colour-coded differences in the frequency ratio $\nu_{p_1}/\nu_{p_2}$ between models computed with different opacity tables. Upper panel: OPAL minus OP, $\nu_{p_1}/\nu_{p_2}(\mathrm{OPAL}) - \nu_{p_1}/\nu_{p_2}(\mathrm{OP})$. Middle panel: OPAL minus OPLIB, $\nu_{p_1}/\nu_{p_2}(\mathrm{OPAL}) - \nu_{p_1}/\nu_{p_2}(\mathrm{OPLIB})$.  Bottom panel: OPLIB minus OP, $\nu_{p_1}/\nu_{p_2}(\mathrm{OPLIB}) - \nu_{p_1}/\nu_{p_2}(\mathrm{OP})$. Here, $\nu_{p_1}$ is the frequency of the radial fundamental mode and $\nu_{p_2}$ the frequency of the first overtone.
     }
     \label{fig:HR_opacity_nu_ratio}
  \end{figure}
 \end{center}

This diversity suggests that the opacity dependence of seismic models is a complex problem without a simple or universal explanation. Nevertheless, we identified a possible correlation between the opacity effect and a star’s position in the HR diagram. 
Figure\,\ref{fig:HR_opacity_nu_ratio} presents HR diagrams in which the colour scale represents the difference in the frequency ratio of the radial fundamental and first overtone modes between models computed with different opacity datasets: 
$\nu_{\rm p_1}/\nu_{\rm p_2}(\mathrm{OPAL}) -
\nu_{\rm p_1}/\nu_{\rm p_2}(\mathrm{OP})$ (upper panel), 
$\nu_{\rm p_1}/\nu_{\rm p_2}(\mathrm{OPAL}) -
\nu_{\rm p_1}/\nu_{\rm p_2}(\mathrm{OPLIB})$ (middle panel), and 
$\nu_{\rm p_1}/\nu_{\rm p_2}(\mathrm{OPLIB}) -
\nu_{\rm p_1}/\nu_{\rm p_2}(\mathrm{OP})$ (bottom panel). Here, $\nu_{\rm p_1}$ and $\nu_{\rm p_2}$ denote the frequencies of the radial fundamental and first overtone modes, respectively. The models shown in the figure were computed assuming $X_0 = 0.70$, $Z = 0.010$, $\alpha_{\rm mlt} = 0.5$, $\alpha_{\rm ov} = 0.1$, and an initial rotational velocity of $V_{\rm rot,0} = 10$\,km\,s$^{-1}$.

Near the zero-age main sequence, the OPAL frequency ratio is smaller than both its OP and OPLIB counterparts. The difference grows during evolution until the star reaches the end of the main sequence phase. From this point, the differences decrease, reaching zero. In more advanced evolutionary phases, the OPAL frequency ratio becomes larger than those from the OP and OPLIB opacity data. The smooth variation of these differences across the HR diagram clearly indicates a systematic offset between the opacity datasets and underscores that the opacity issue in stellar modelling remains unresolved.

\begin{acknowledgements}
The work was financially supported by the Polish National Science Centre grant 2023/50/A/ST9/00144.

Calculations were carried out using resources provided by Wrocław Centre for Networking and Supercomputing (http://wcss.pl),  grant no.  265.  

This work has made use of data from the European Space Agency (ESA) mission
{\it Gaia} (\url{https://www.cosmos.esa.int/gaia}), processed by the {\it Gaia}
Data Processing and Analysis Consortium (DPAC,
\url{https://www.cosmos.esa.int/web/gaia/dpac/consortium}). Funding for the DPAC
has been provided by national institutions, in particular the institutions
participating in the {\it Gaia} Multilateral Agreement.

This paper includes data collected by the {\em TESS} mission. 
We acknowledge the use of \textit{TESS} High Level Science Products (HLSP) produced by the Quick-Look Pipeline (QLP) at the \textit{TESS} Science Office at MIT, which are publicly available from the Mikulski Archive for Space Telescopes (MAST). Funding for the \textit{TESS} mission is provided by NASA's Science Mission Directorate.

This research made use of Lightkurve, a Python package for Kepler and \textit{TESS} data analysis.

Some of the data presented in this paper were obtained from the Mikulski Archive for Space Telescopes (MAST) at the Space Telescope Science Institute. The specific observations analysed can be accessed via \url{https://doi.org/10.17909/t9-nmc8-f686} and via
\url{https://doi.org/10.17909/0cp4-2j79}. STScI is operated by the Association of Universities for Research in Astronomy, Inc., under NASA
contract NAS5–26555. Support to MAST for these data is provided by the NASA Office of Space Science via grant NAG5–7584 and by other grants and contracts.
\end{acknowledgements}

%
   \bibliographystyle{bibtex/aa} 
   \bibliography{bibtex/PWbiblio} 
\begin{appendix}




\onecolumn
\section{Spectral analysis}

\begin{figure*}[ht]
    \centering
    \includegraphics[width=\textwidth]{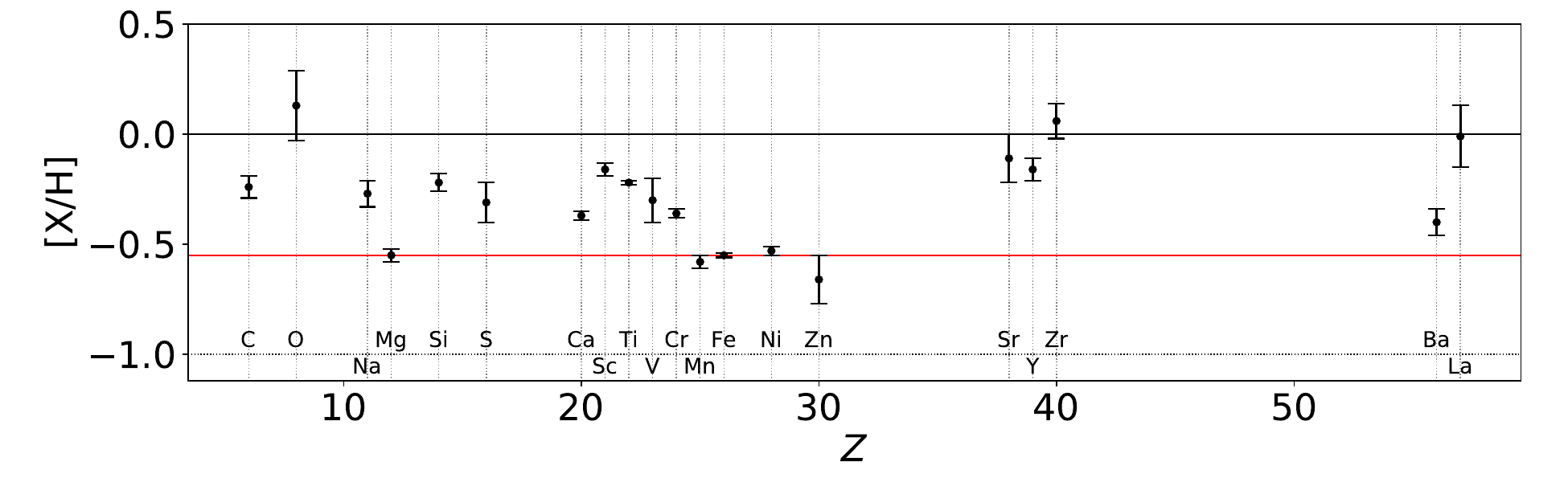}
    \caption{Chemical abundances of AI Vel derived from the median HARPS spectrum, plotted as a function of atomic number. The horizontal red line indicates the iron abundance, [Fe/H].}%
    \label{fig:AI_Vel_HARPS_abundances}%
\end{figure*}

In Fig.\,\ref{fig:AI_Vel_HARPS_abundances}, we present the relative abundances of 20 chemical elements derived from the HARPS spectra. Compared to solar values, several iron-group elements—most notably Mn, Fe, Ni, and Zn—show a clear underabundance. Most of the remaining elements are also slightly sub-solar. In contrast, a few elements, including O, Zr, and La, exhibit solar or mildly enhanced abundances.

\section{\textit{TESS} data}

\begin{figure*}[ht]
   \begin{center}
     \centering
     \includegraphics[width=1\columnwidth]{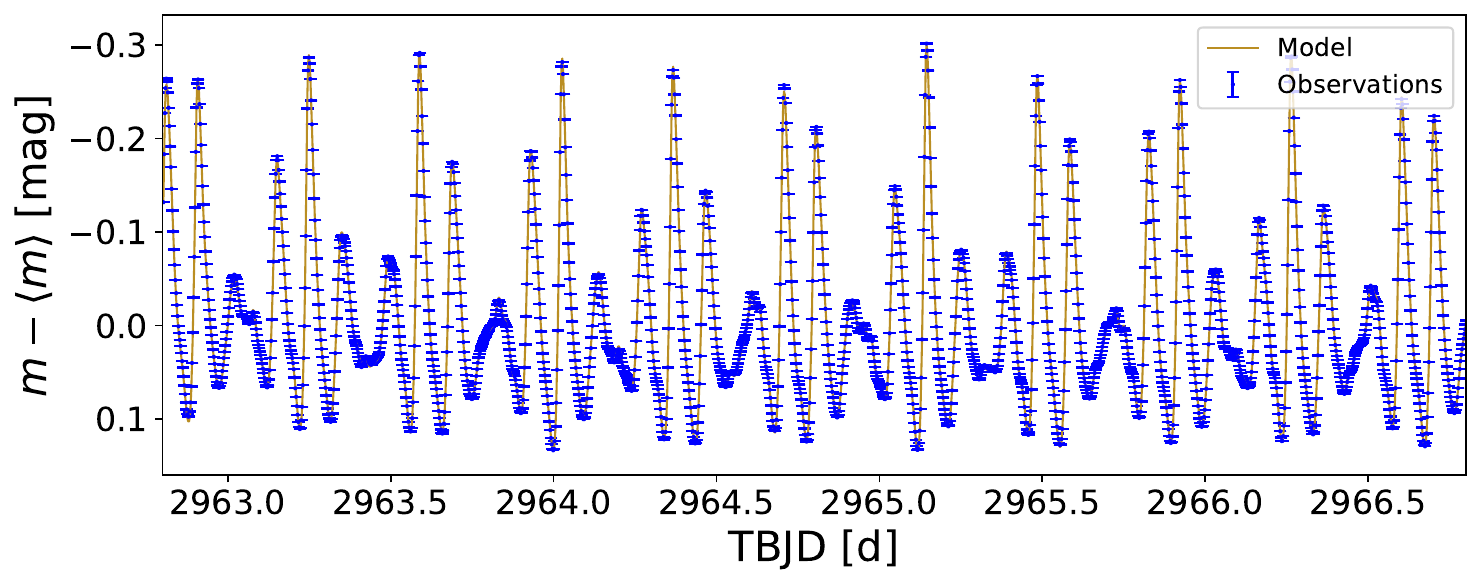}
     \caption{Segment of the \textit{TESS} light curve obtained during the first three days of observations in Sector\,61. The solid line represents the Fourier model derived from the combined Sectors\,61+62 dataset. The brightness values, $\rm{m}$, were shifted by subtracting the mean brightness, $\langle \rm{m} \rangle$. TBJD denotes the \textit{TESS} Barycentric Julian Date, defined as ${\rm TBJD} = {\rm BJD} - 2457000.0$.}
     \label{fig:lc_model}
  \end{center}
\end{figure*}

In Fig.\,\ref{fig:lc_model}, we present the \textit{TESS} light curve from the first few days of observations in Sector\,61, where the variability is clearly visible. The solid yellow line shows the model described by Eq.\,\ref{eq:m_t}, constructed using frequencies, amplitudes, and phases derived from the Fourier analysis of the combined Sectors\,61+62.    

\twocolumn
\section{Seismic models}
\label{Sect:app:sm}

\begin{center}
   \begin{figure}[ht]
     \centering
     \includegraphics[width=1\columnwidth]{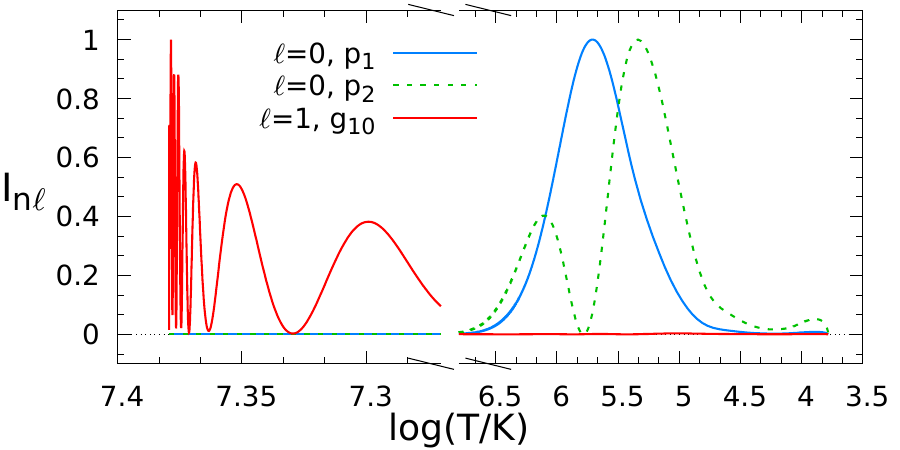}
     \caption{Mode inertia for the radial modes $(\ell=0, p_1)$ and $(\ell=0, p_2)$, as well as the dipole mode $(\ell=1, g_{10})$. Parameters of the seismic model are listed in Table\,\ref{tab:simulation_models_l1_p2}.}
     \label{fig:mode_In}
  \end{figure}
 \end{center}

\begin{center}
   \begin{figure}[ht]
       \centering
       \includegraphics[width=1\columnwidth]{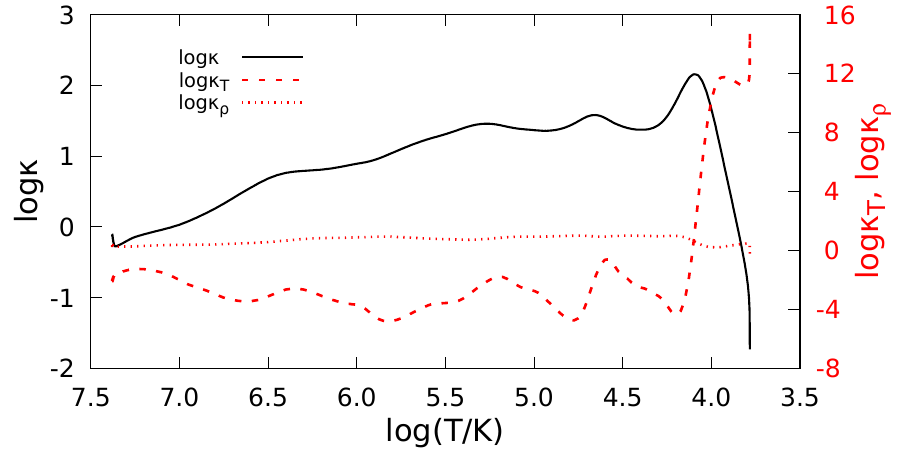}
       \includegraphics[width=1\columnwidth]{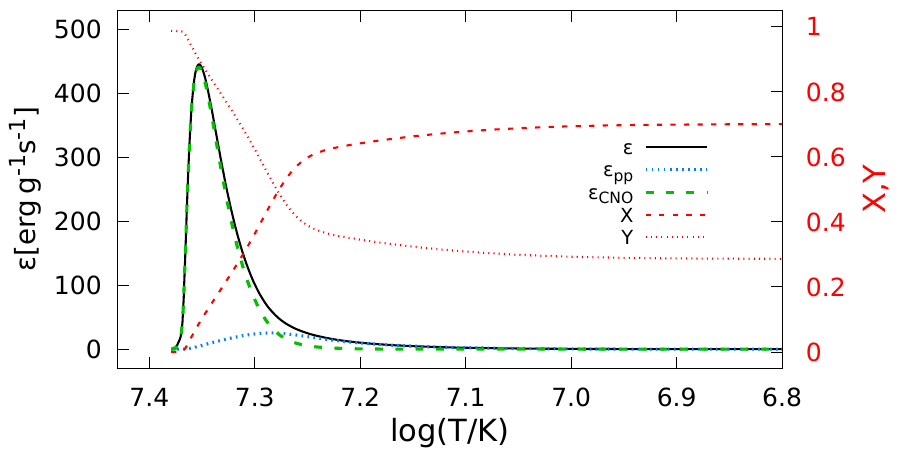}
     \caption{Upper panel: opacity coefficient, $\log\kappa$, and its derivatives with respect to temperature and density. 
     Bottom panel: left-hand axis – energy generation rates from the pp chain and CNO cycle; right-hand axis – hydrogen ($X$) and helium ($Y$) abundances. Parameters of the seismic model are listed in Table\,\ref{tab:simulation_models_l1_p2}.}
     \label{fig:kappa_L}
  \end{figure}
 \end{center}

In Fig.\,\ref{fig:mode_In}, we present the mode inertia of a representative seismic model obtained under the dipole-mode hypothesis, labelled $(\ell=1,m=0)$ in Table\,\ref{tab:simulation_models_l1_p2}. Figure\,\ref{fig:kappa_L} illustrates the internal structure of the same seismic model. The upper panel shows the Rosseland mean opacity, $\log \kappa$, together with its derivatives with respect to temperature, $\partial \log \kappa / \partial \log T$, and density, $\partial \log \kappa / \partial \log \rho$. In the bottom panel, the left-hand axis presents the energy generation rates associated with the pp chain ($\varepsilon_{\rm pp}$) and the CNO cycle ($\varepsilon_{\rm CNO}$), while the right-hand axis presents the hydrogen ($X$) and helium ($Y$) abundance profiles.

\begin{center}
   \begin{figure}[ht]
      \centering
     \includegraphics[width=0.49\columnwidth]{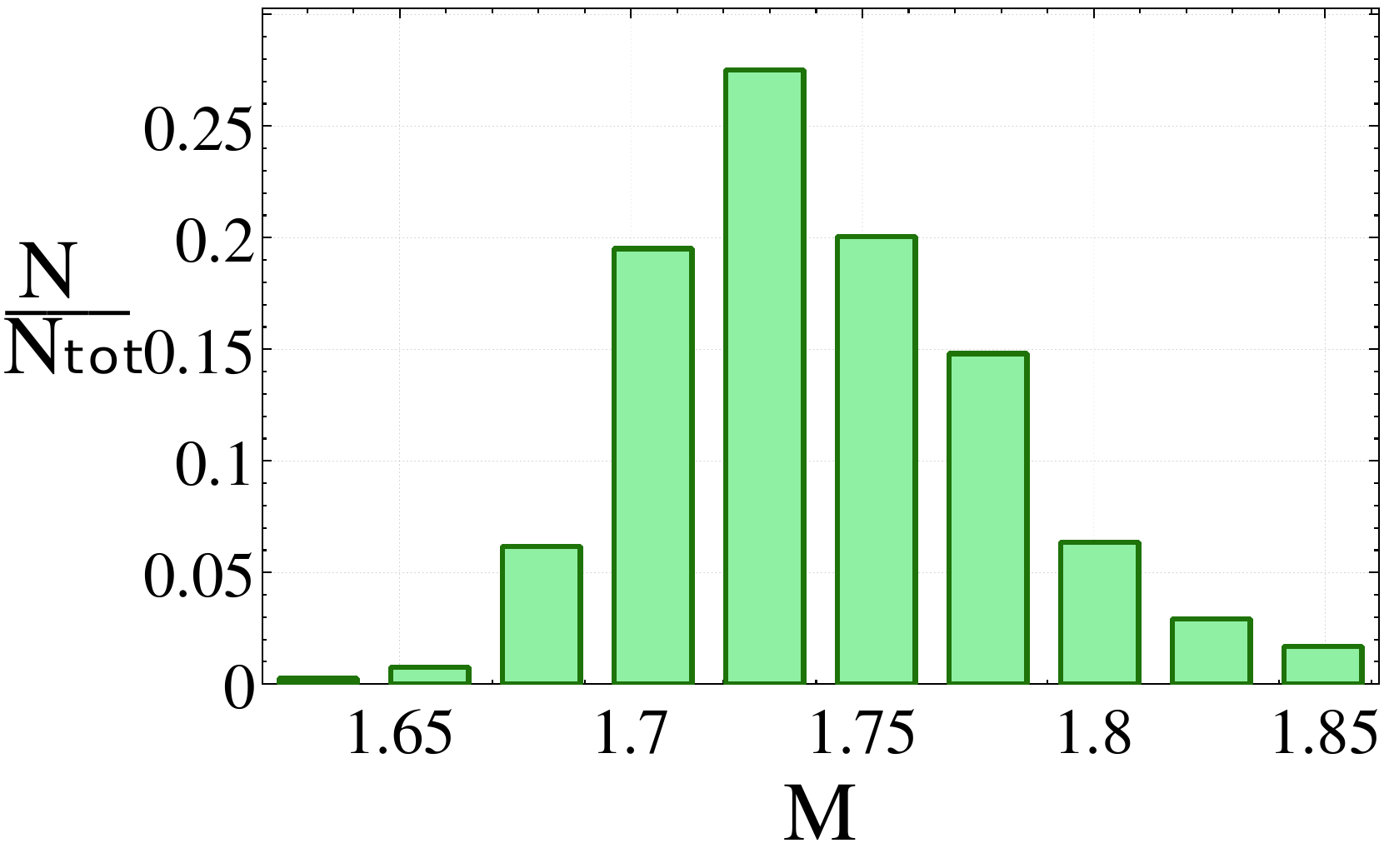}
     \includegraphics[width=0.49\columnwidth]{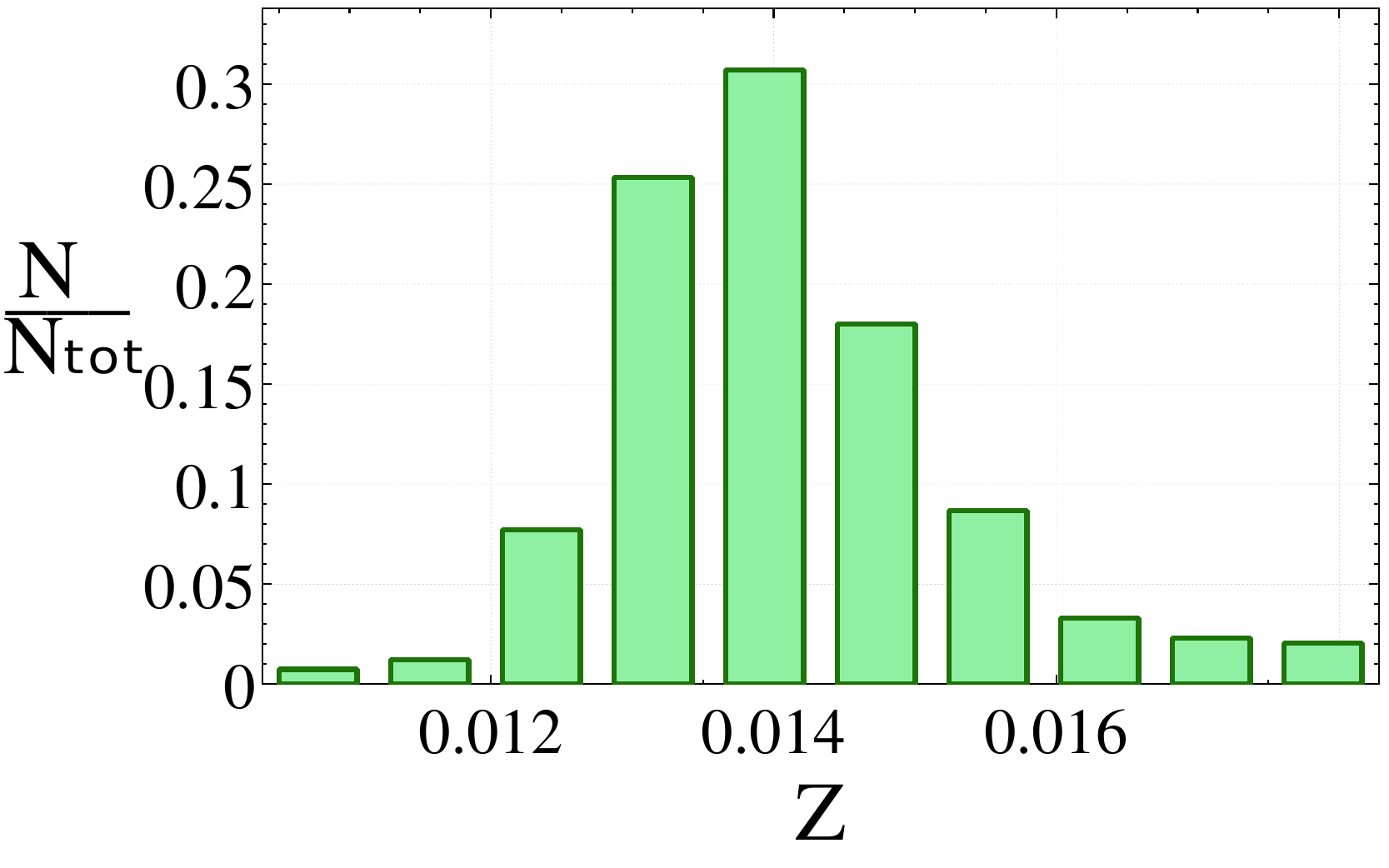}
     \includegraphics[width=0.49\columnwidth]{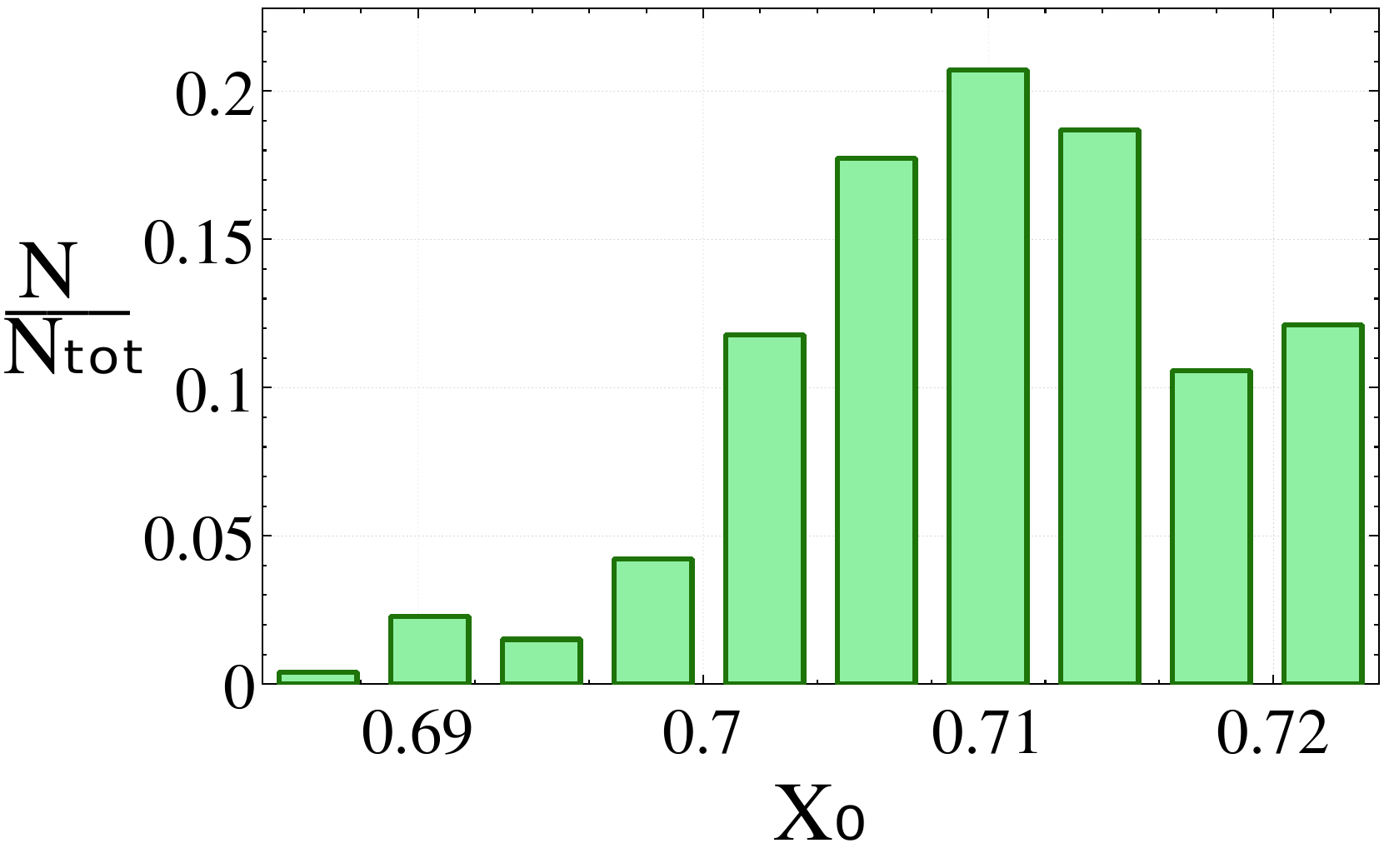}
     \includegraphics[width=0.49\columnwidth]{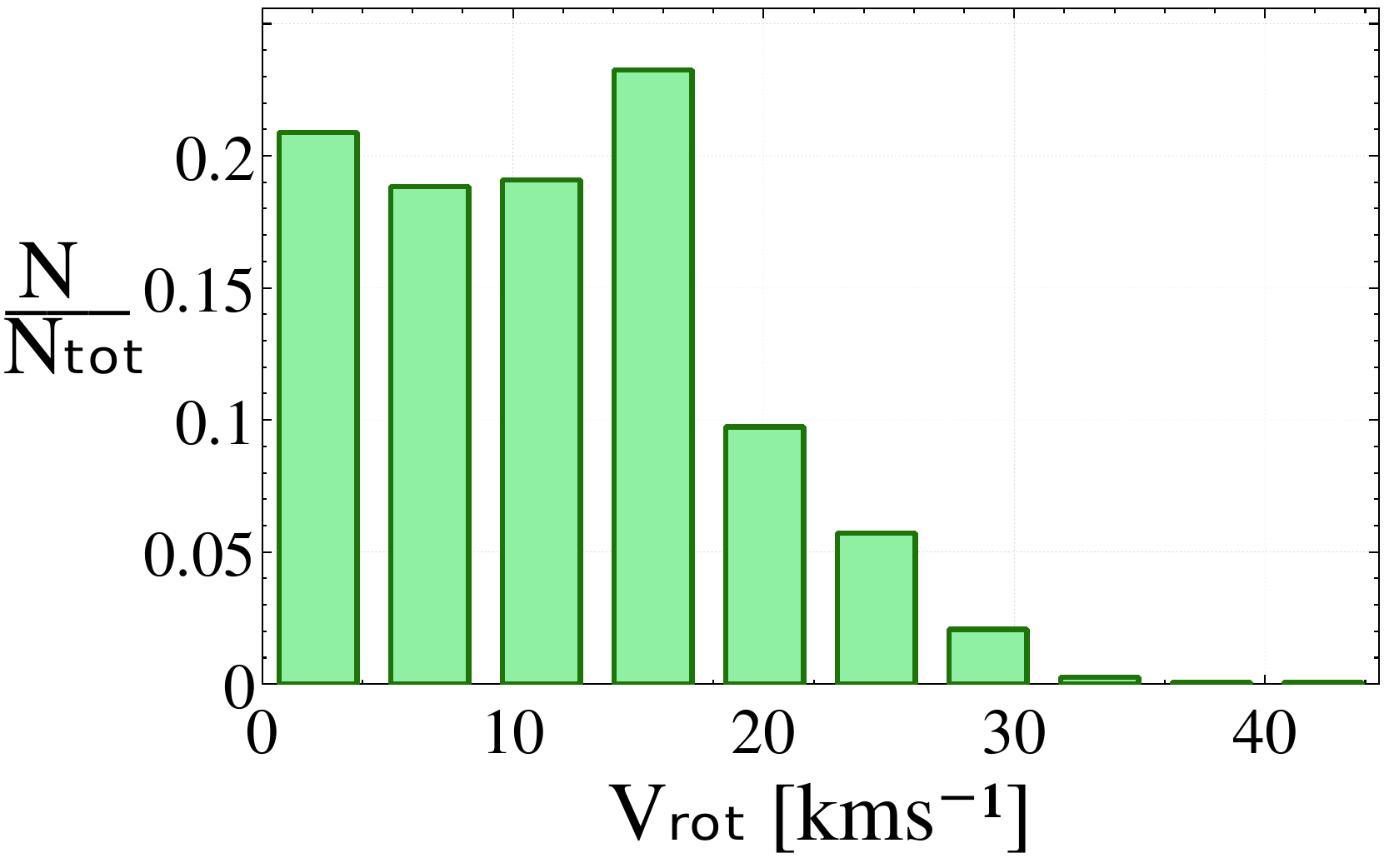}
     \includegraphics[width=0.49\columnwidth]{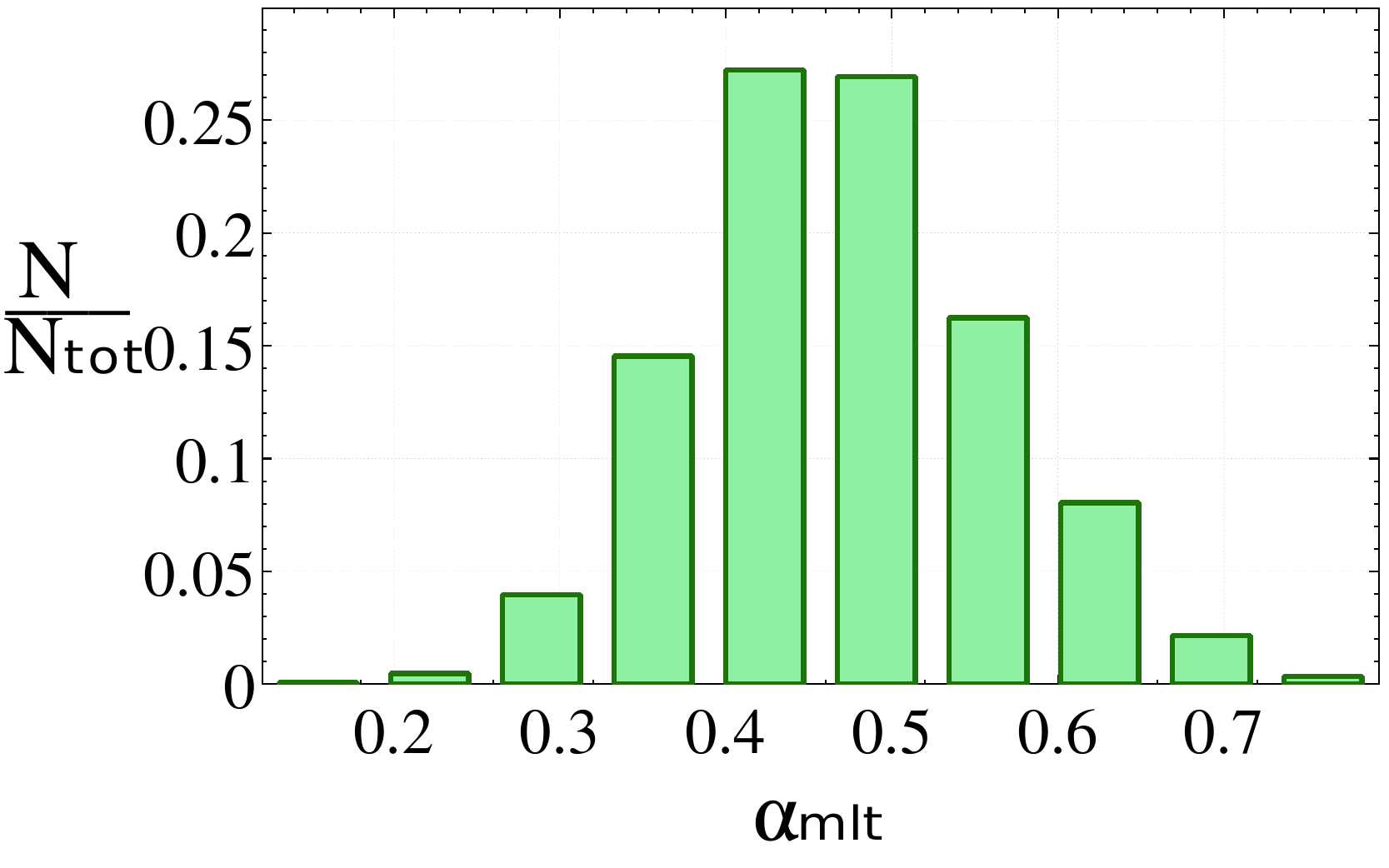}
     \includegraphics[width=0.49\columnwidth]{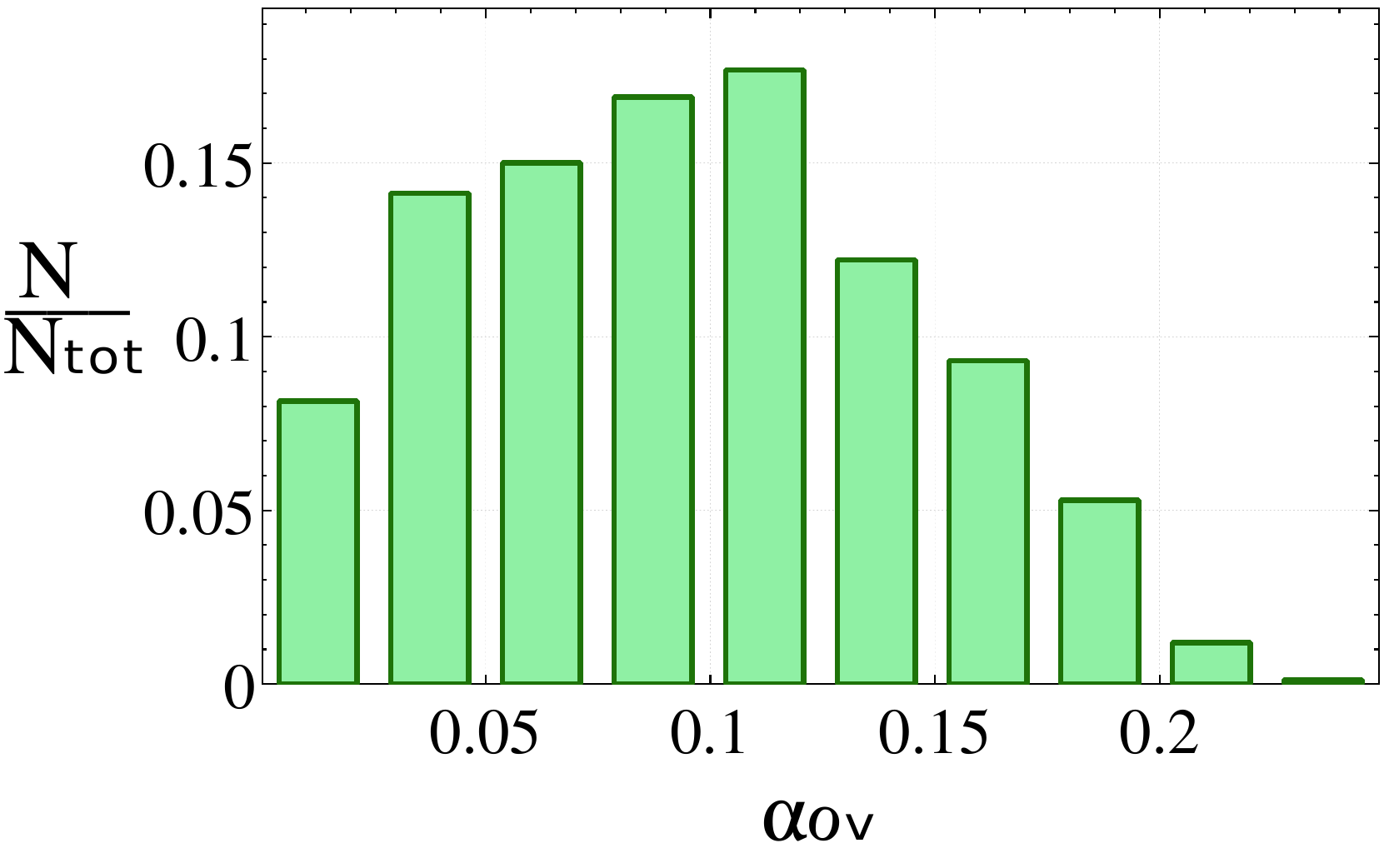}\\
     \includegraphics[width=0.49\columnwidth]{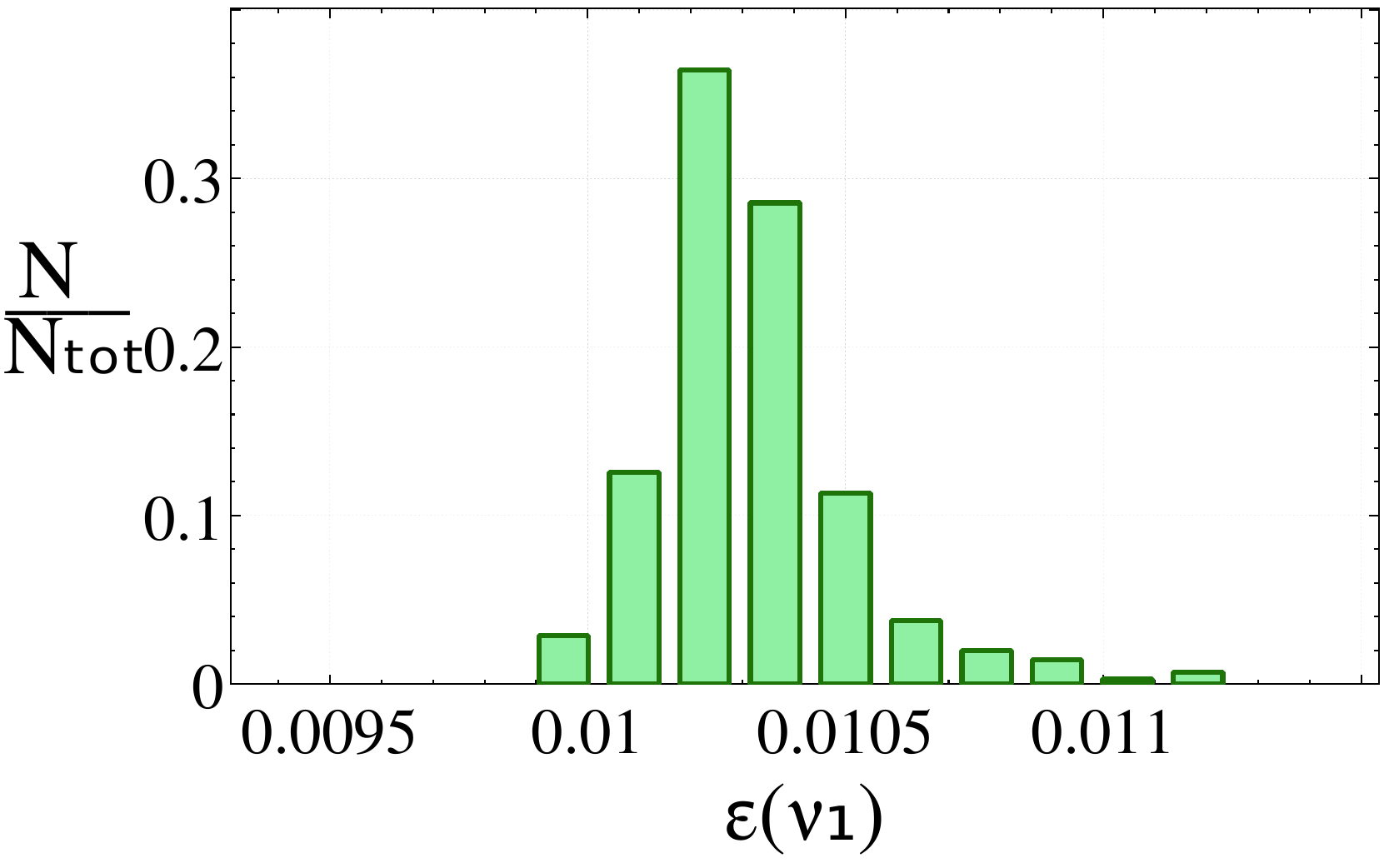}
     \includegraphics[width=0.49\columnwidth]{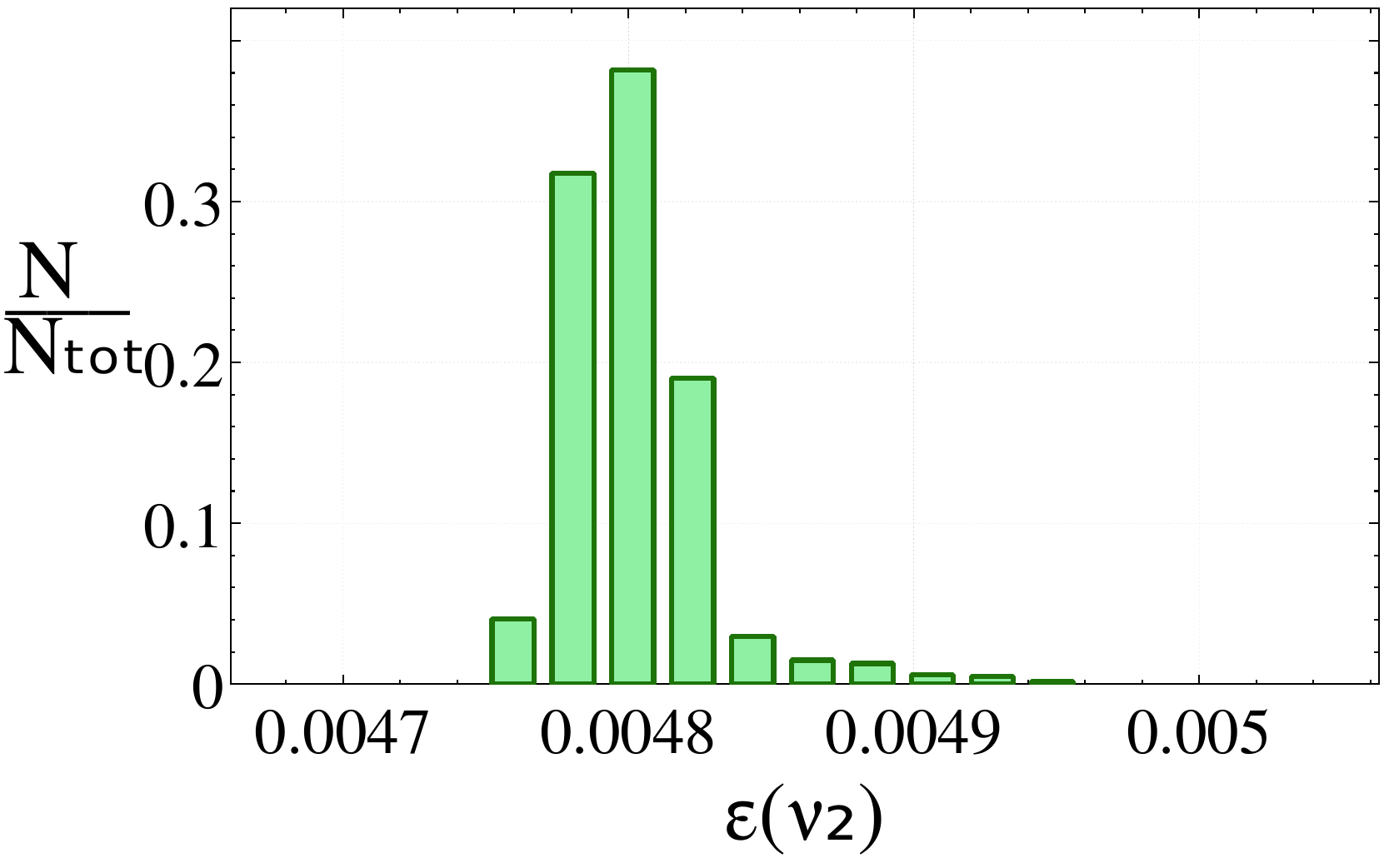}\\
     \includegraphics[width=0.49\columnwidth]{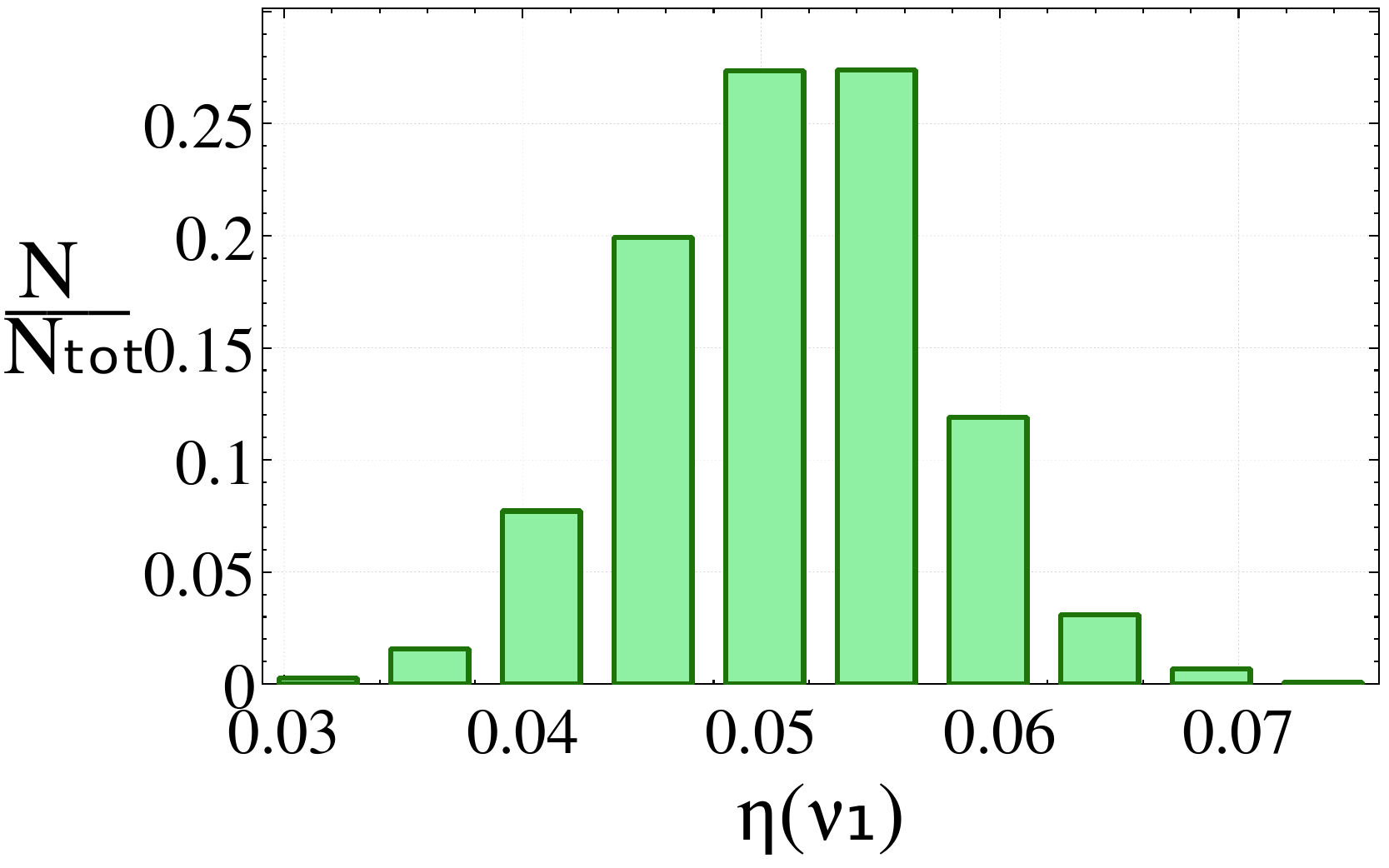}
     \includegraphics[width=0.49\columnwidth]{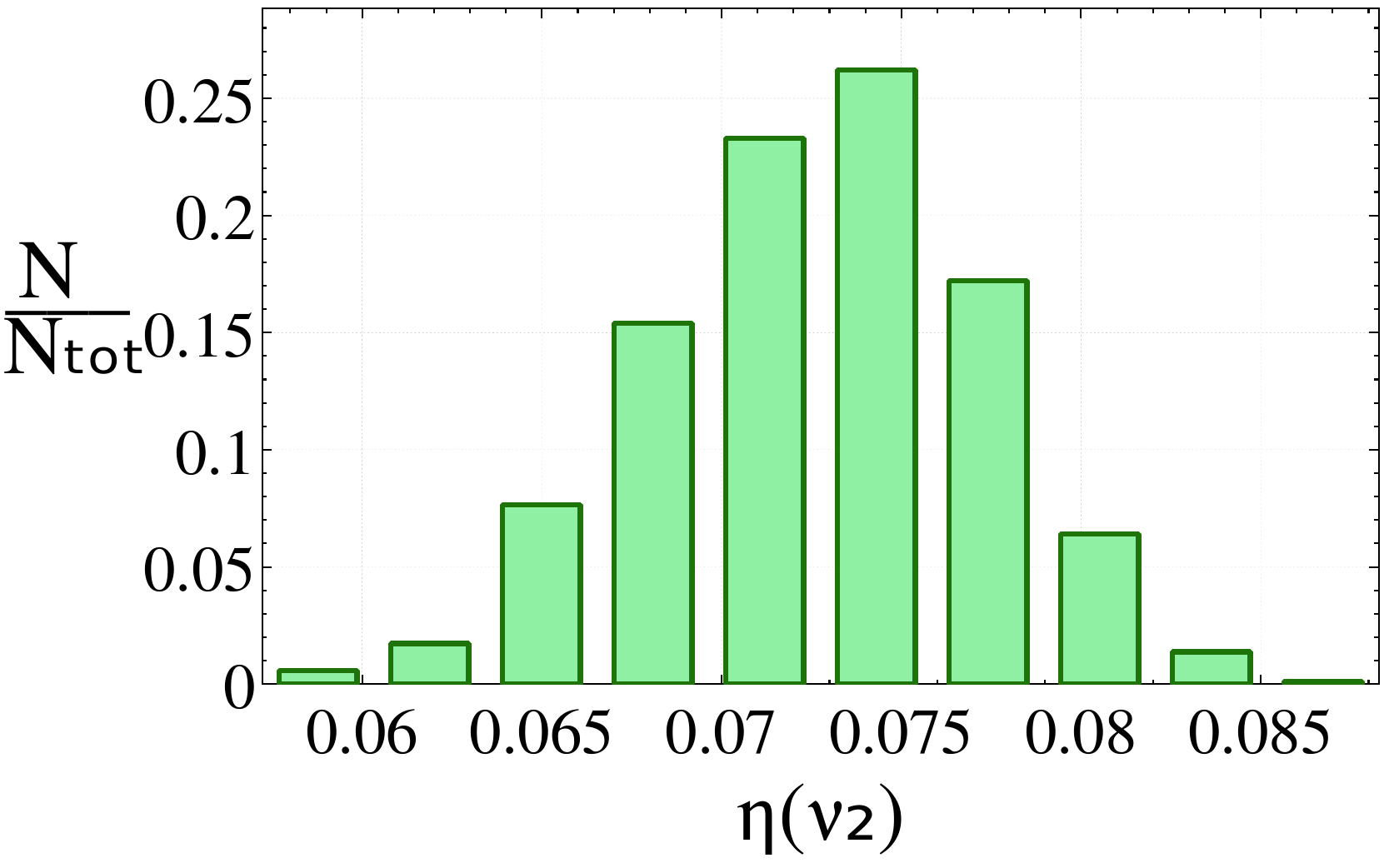}\\
     \caption{ Histograms of the adjustable stellar parameters derived from our simulations including two radial modes and OPAL opacity tables. Shown are the distributions of mass, $M$, metallicity, $Z$, initial hydrogen abundance, $X_0$, current surface rotational velocity, $V_{\rm{rot}}$, convective mixing length parameter, $\alpha_{\rm{mlt}}$, and convective core overshooting parameter, $\alpha_{\rm{ov}}$. Also included are the amplitudes of relative radius variations due to pulsation in the radial fundamental mode, $\varepsilon(\nu_1)$, and first overtone, $\varepsilon(\nu_2)$, as well as the instability parameters for the radial fundamental, $\eta(\nu_1)$, and first overtone, $\eta(\nu_2)$ modes.}
     \label{fig:simulation_his}
  \end{figure}
 \end{center}

\begin{center}
   \begin{figure}[ht]
    \begin{center}
     \includegraphics[width=0.47\columnwidth]{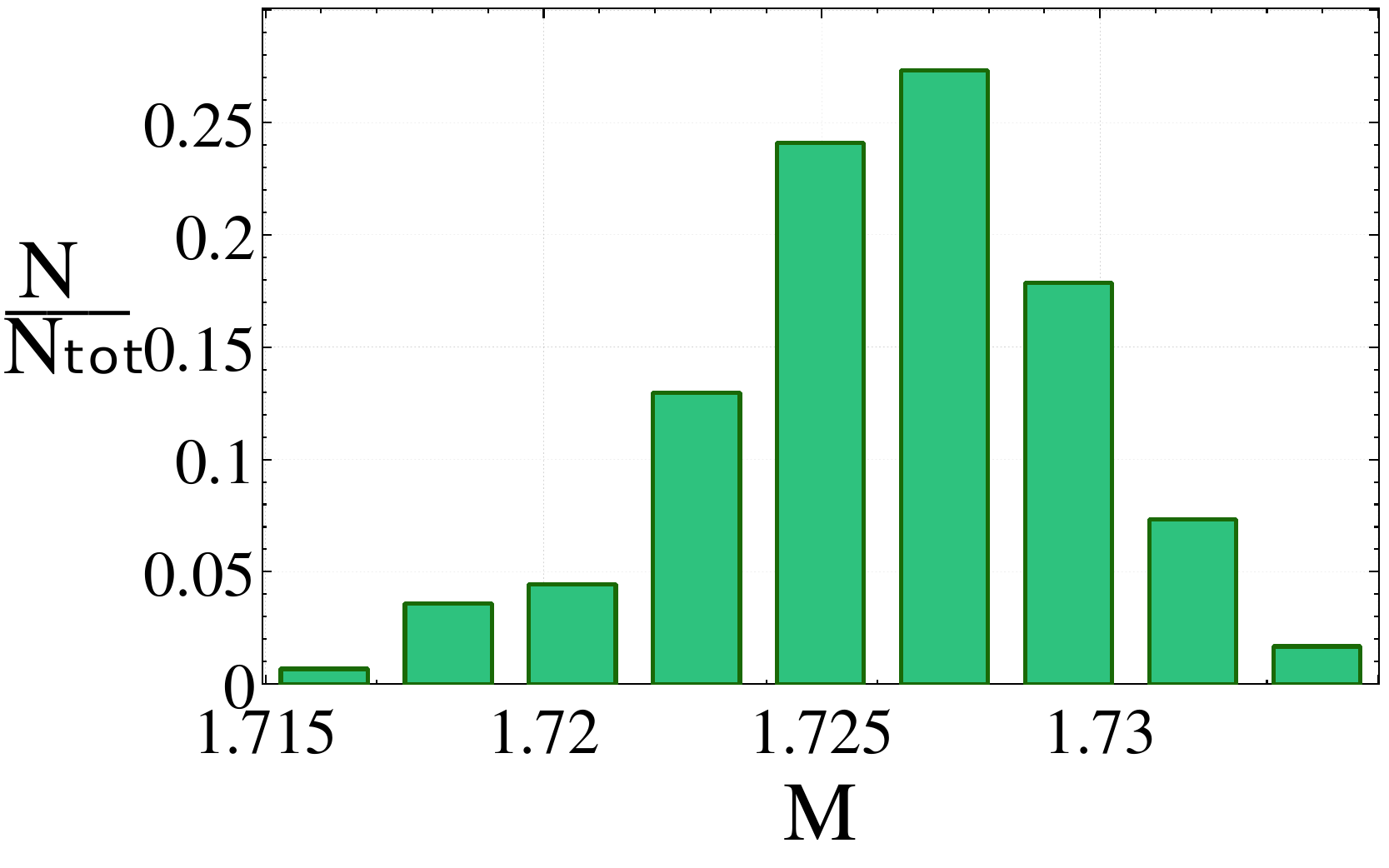}
     \includegraphics[width=0.47\columnwidth]{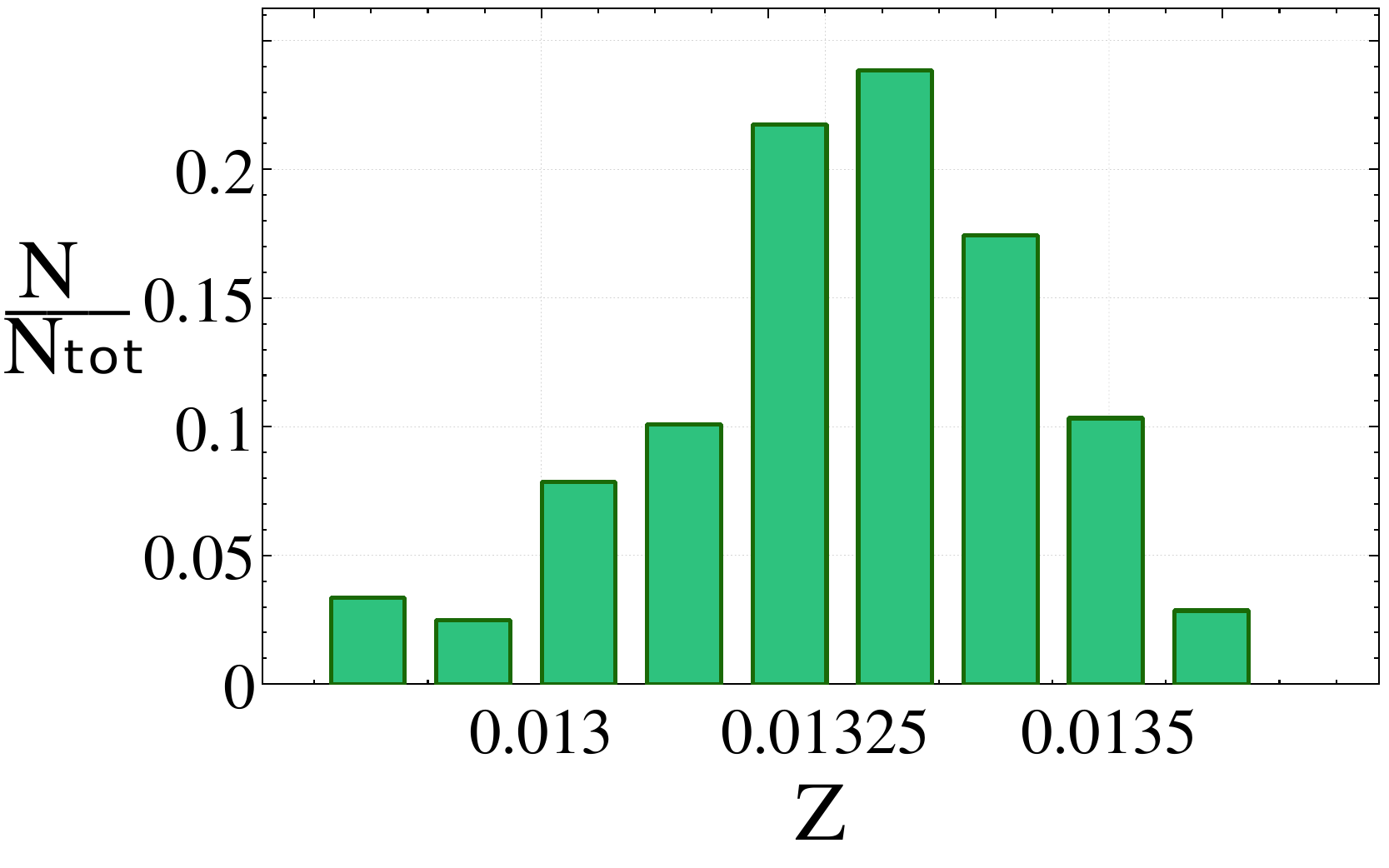}
     \includegraphics[width=0.47\columnwidth]{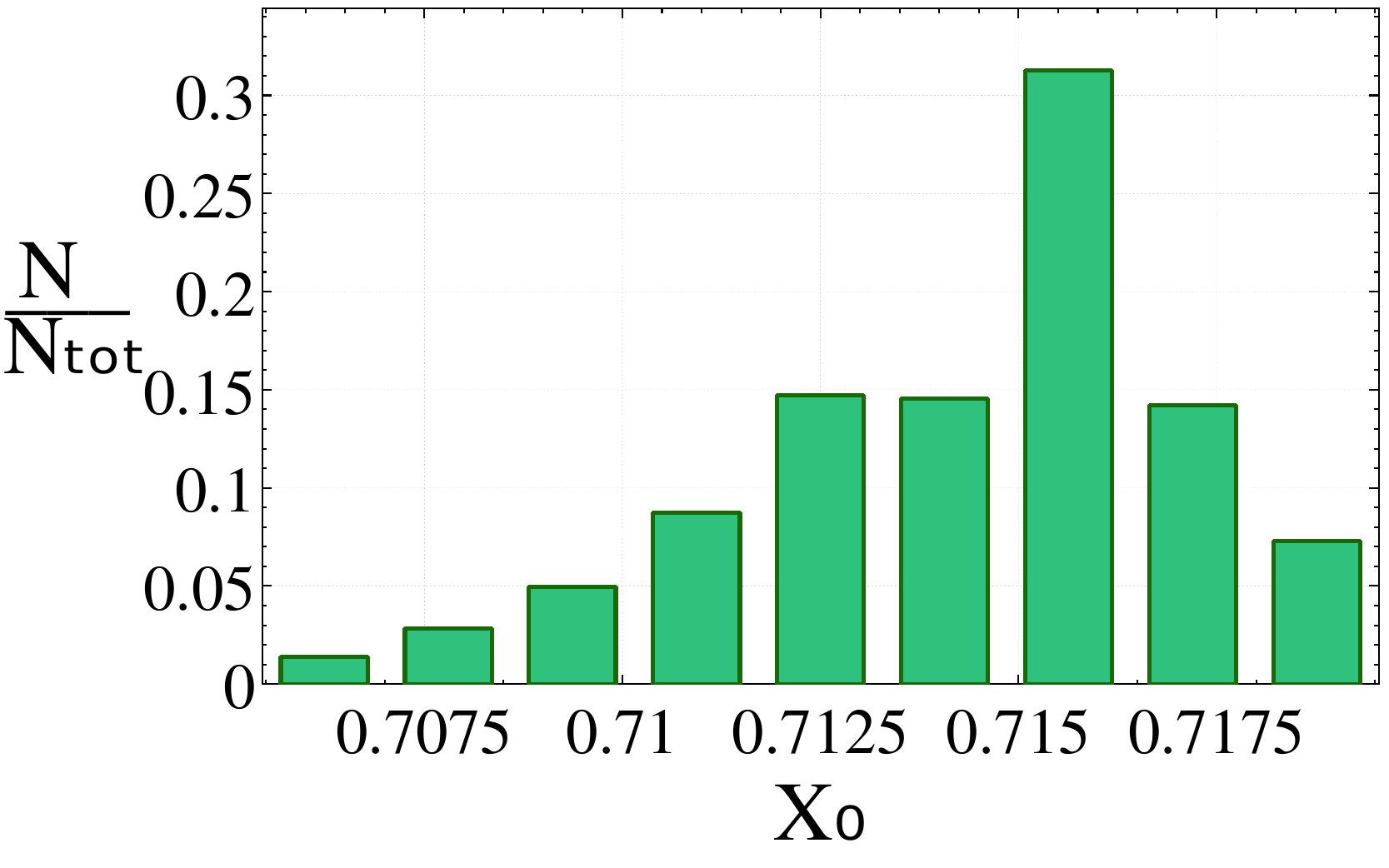}
     \includegraphics[width=0.47\columnwidth]{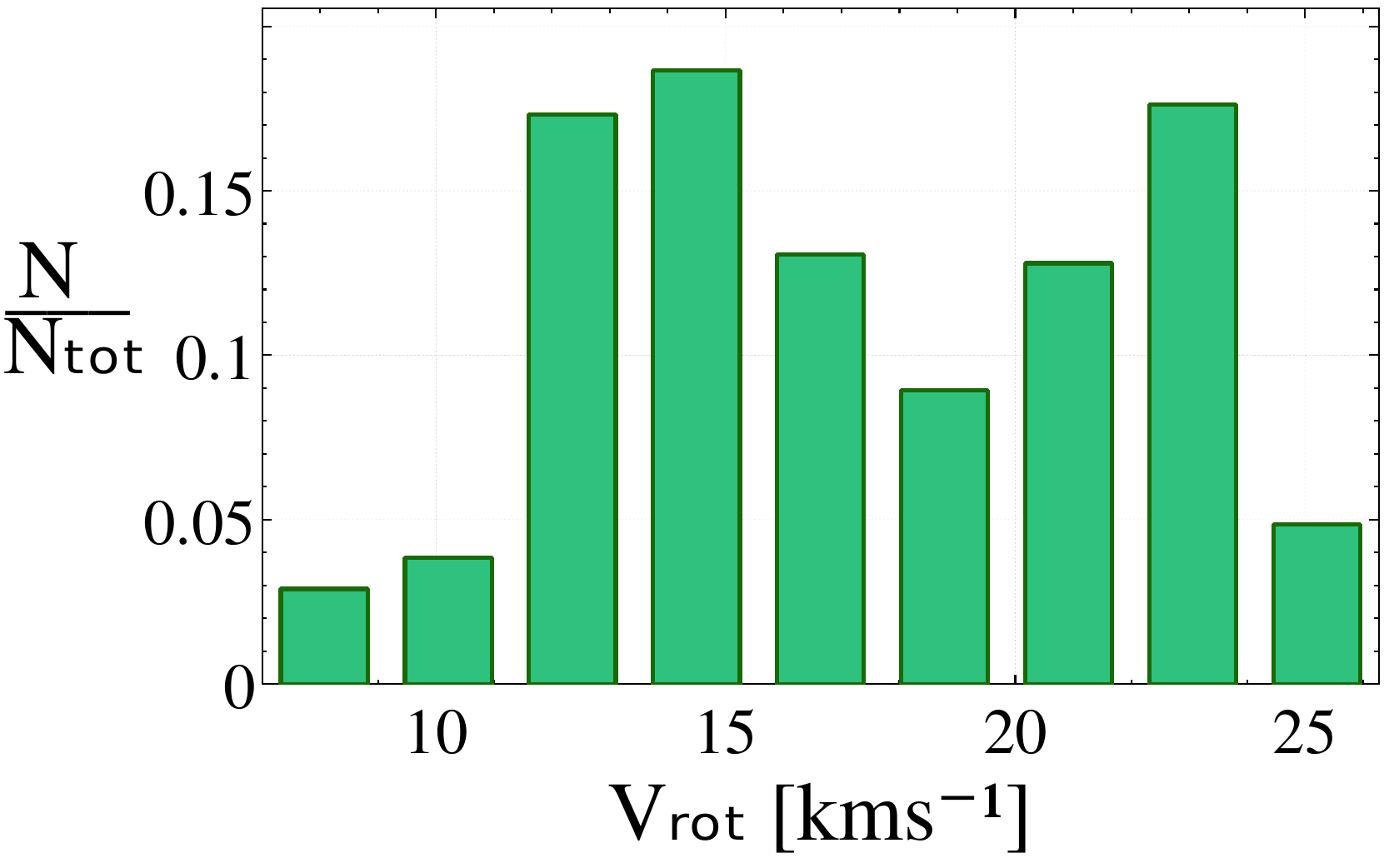}
     \includegraphics[width=0.47\columnwidth]{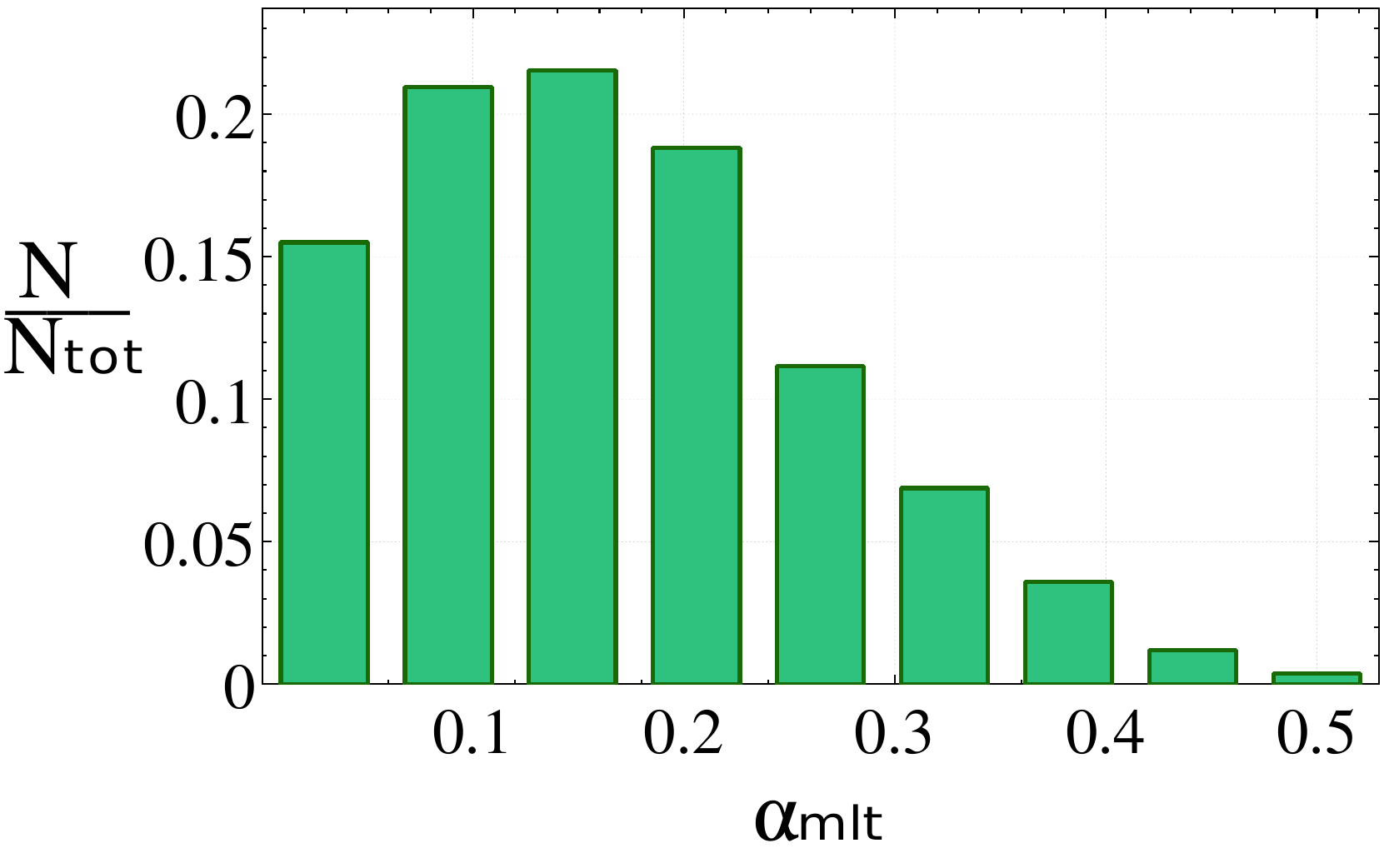}
     \includegraphics[width=0.47\columnwidth]{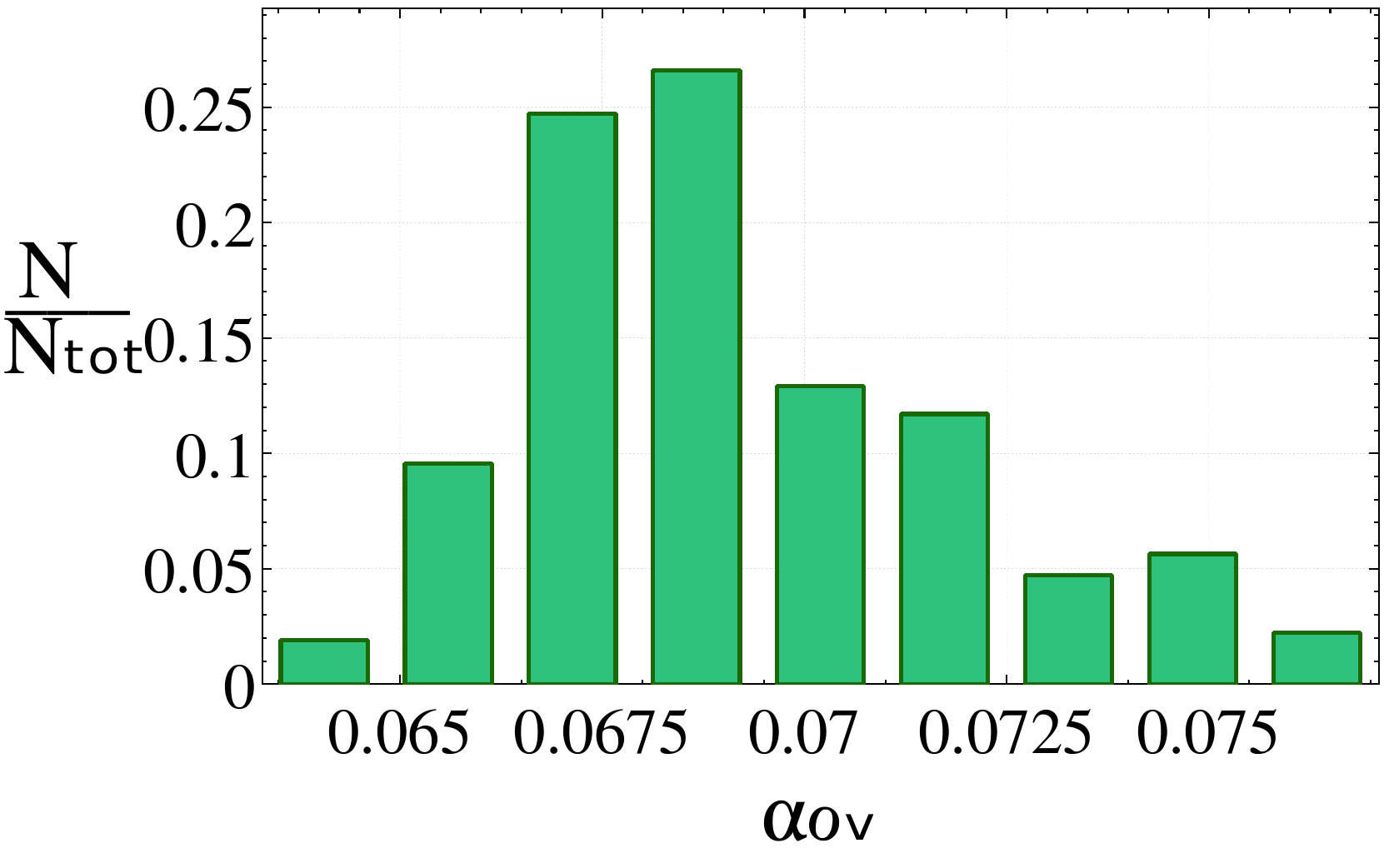}
     \includegraphics[width=0.47\columnwidth]{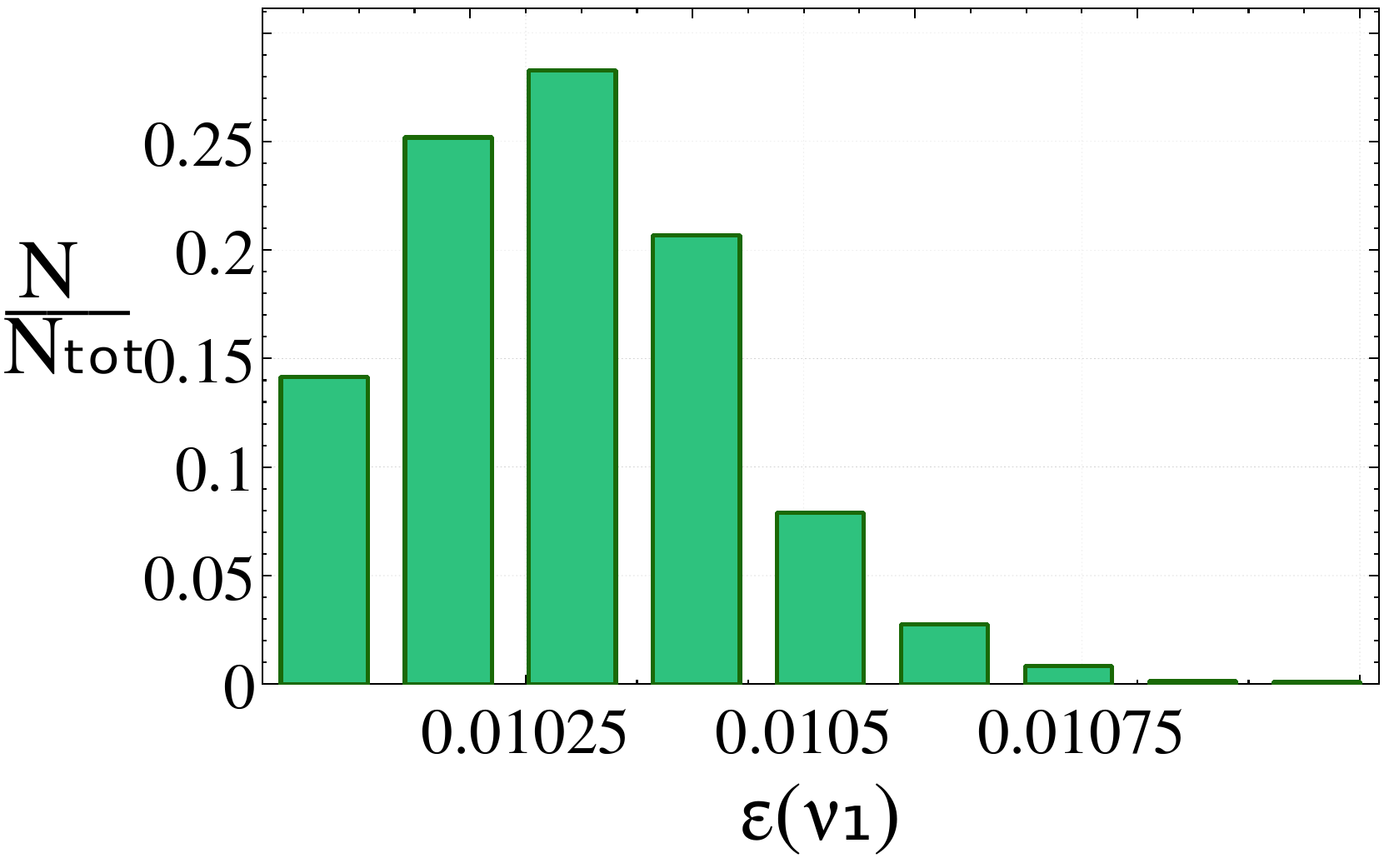}
     \includegraphics[width=0.47\columnwidth]{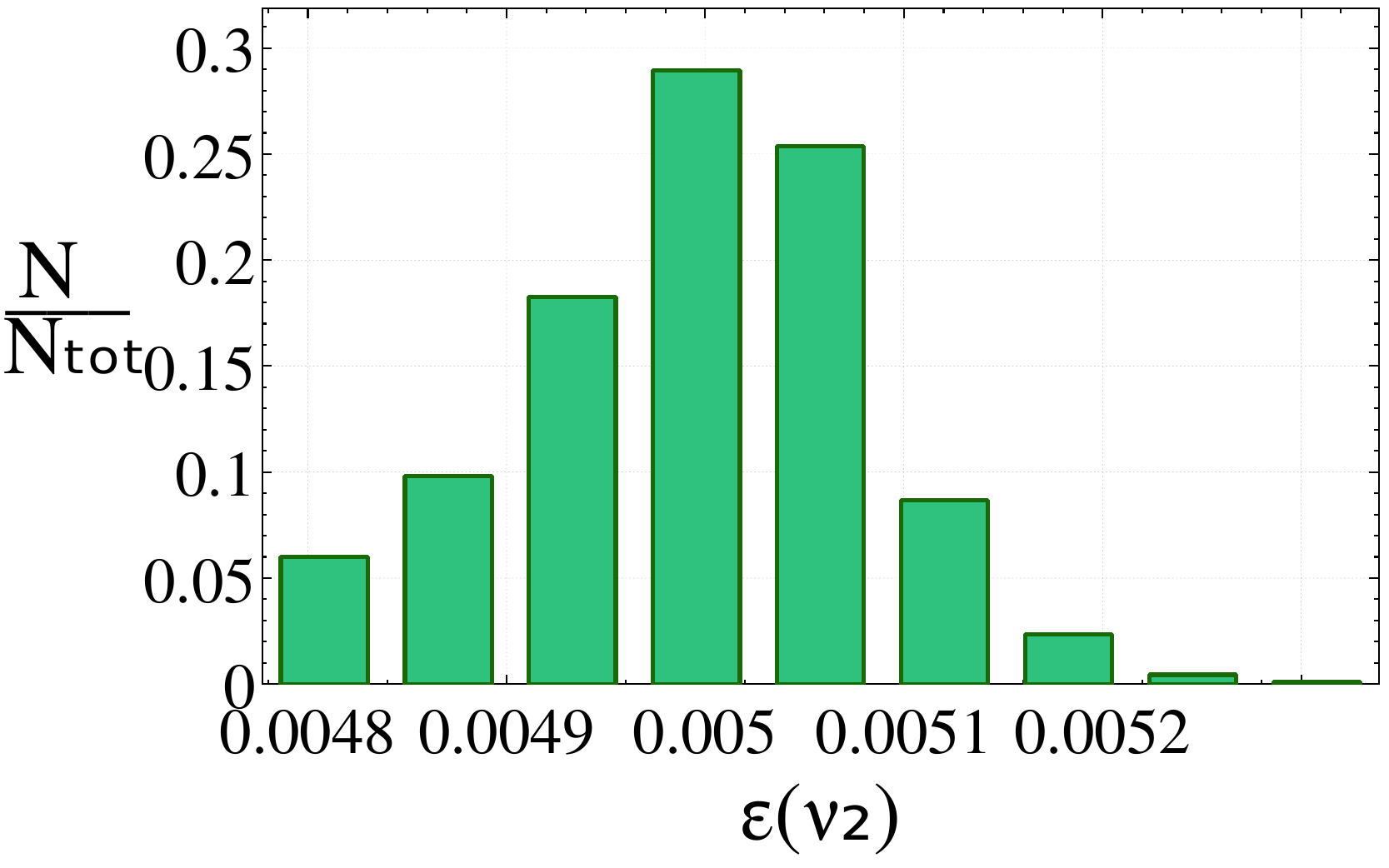}
     \includegraphics[width=0.47\columnwidth]{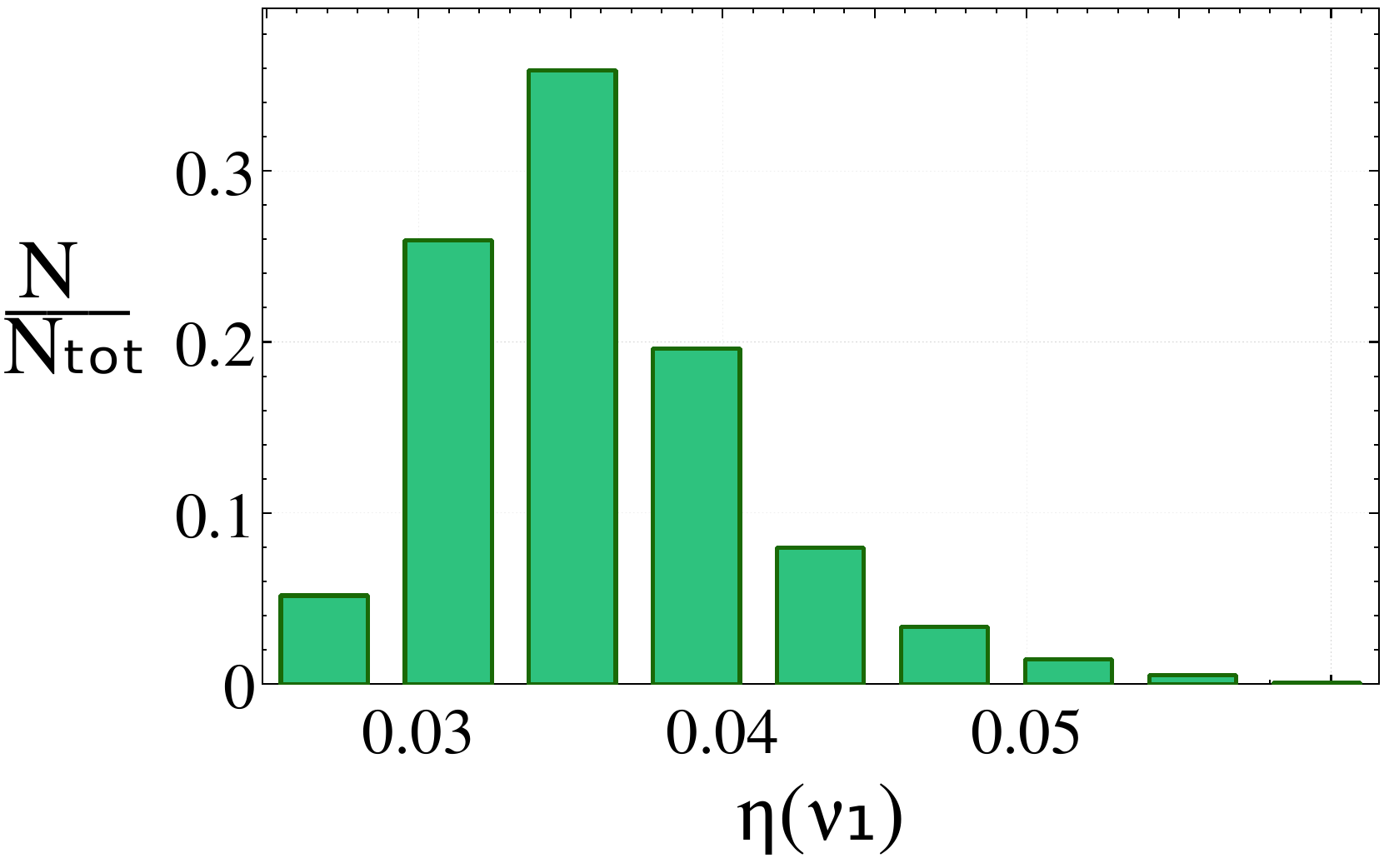}
     \includegraphics[width=0.47\columnwidth]{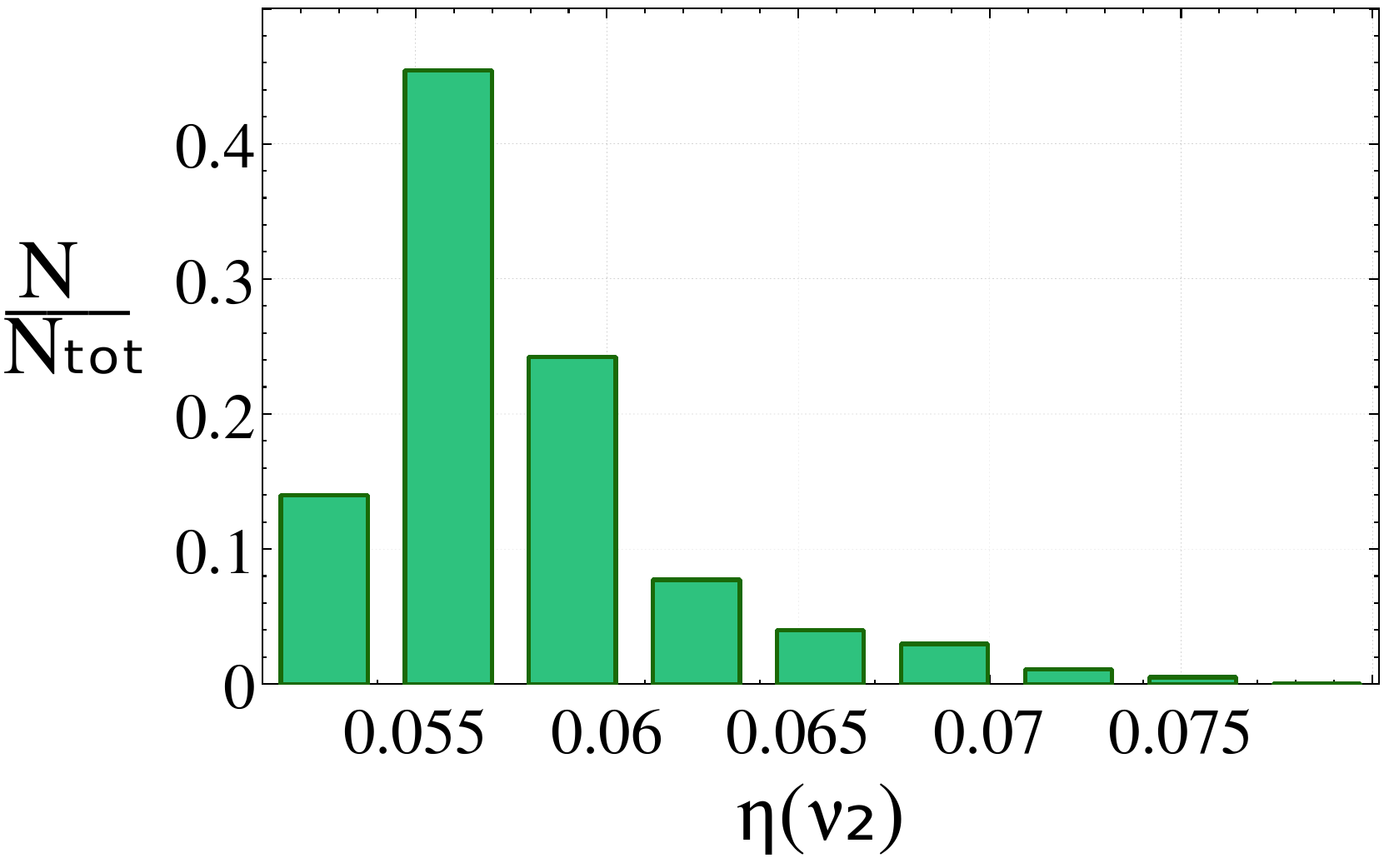}
     \includegraphics[width=0.47\columnwidth]{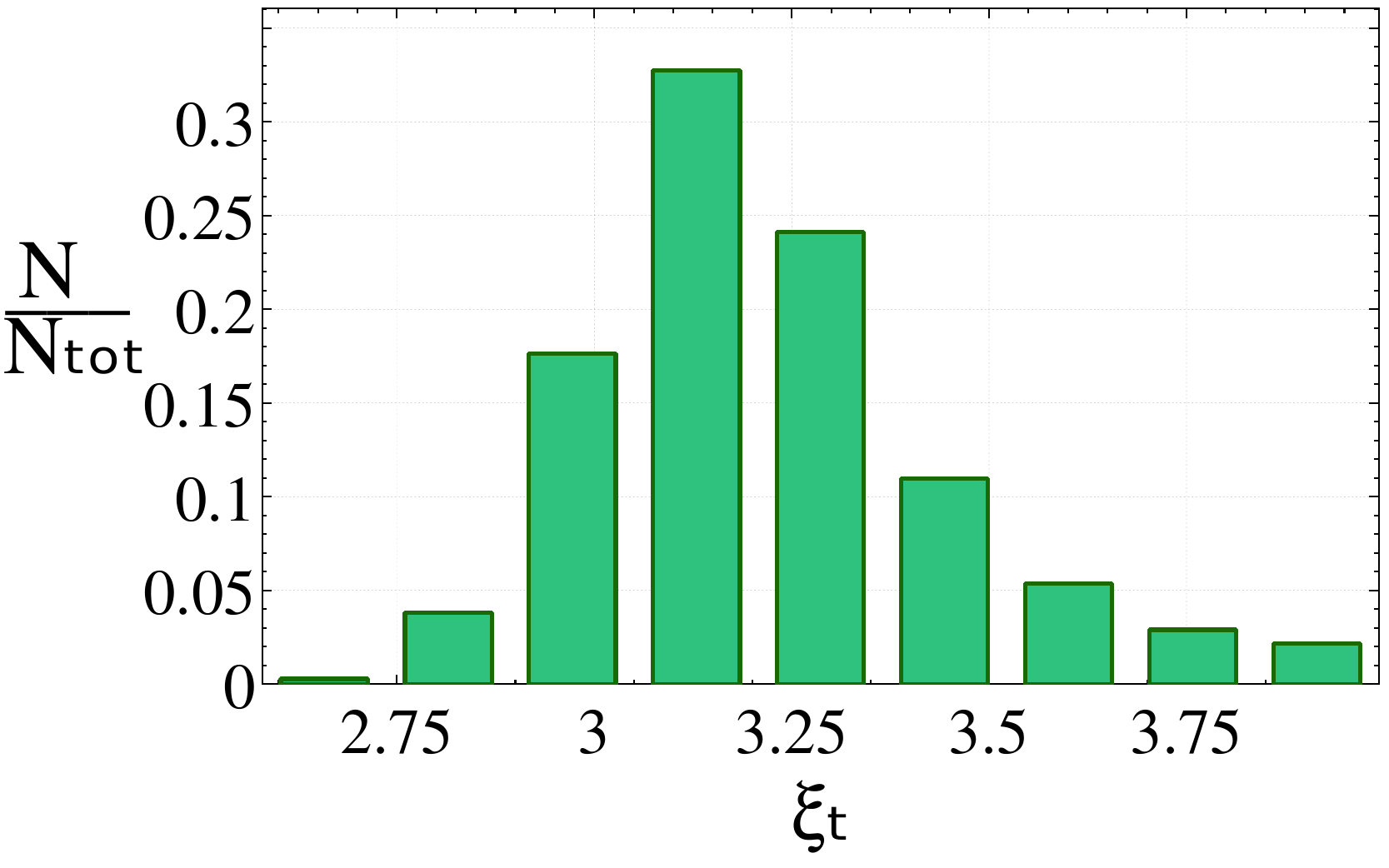}
     \includegraphics[width=0.47\columnwidth]{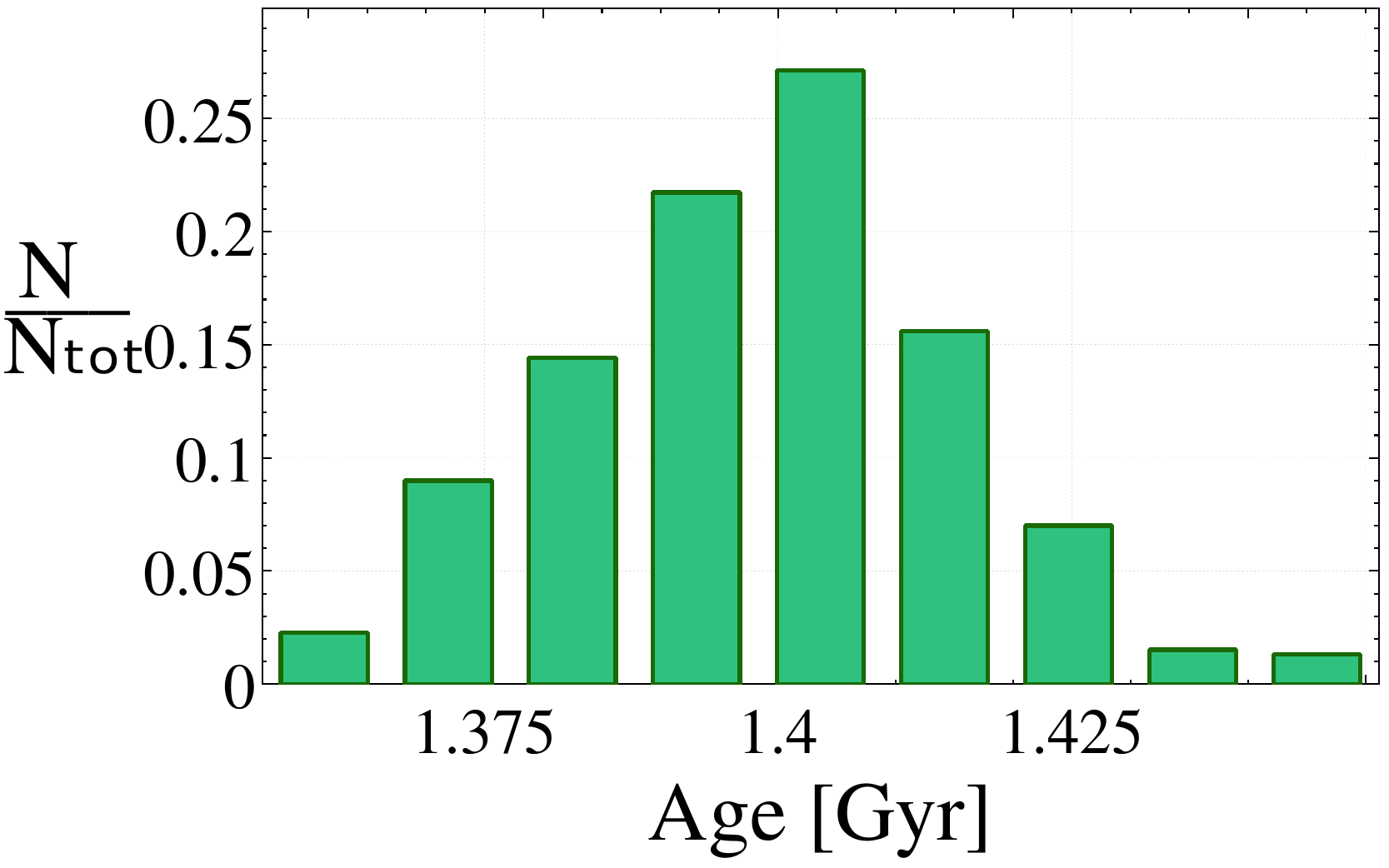}
     \caption{ Same as in Fig.~\ref{fig:simulation_his}, but for simulations including the dipole-mode identification of $\nu_3$. In addition, histograms for the microturbulent velocity, $\xi_t$, and stellar age, Age [Gyr], are also shown.}
     \label{fig:simulation_his_dipole_sel} 
      \end{center}
  \end{figure}
\end{center}

\begin{center}
   \begin{figure}[ht]
    \begin{center}
     \includegraphics[width=0.47\columnwidth]{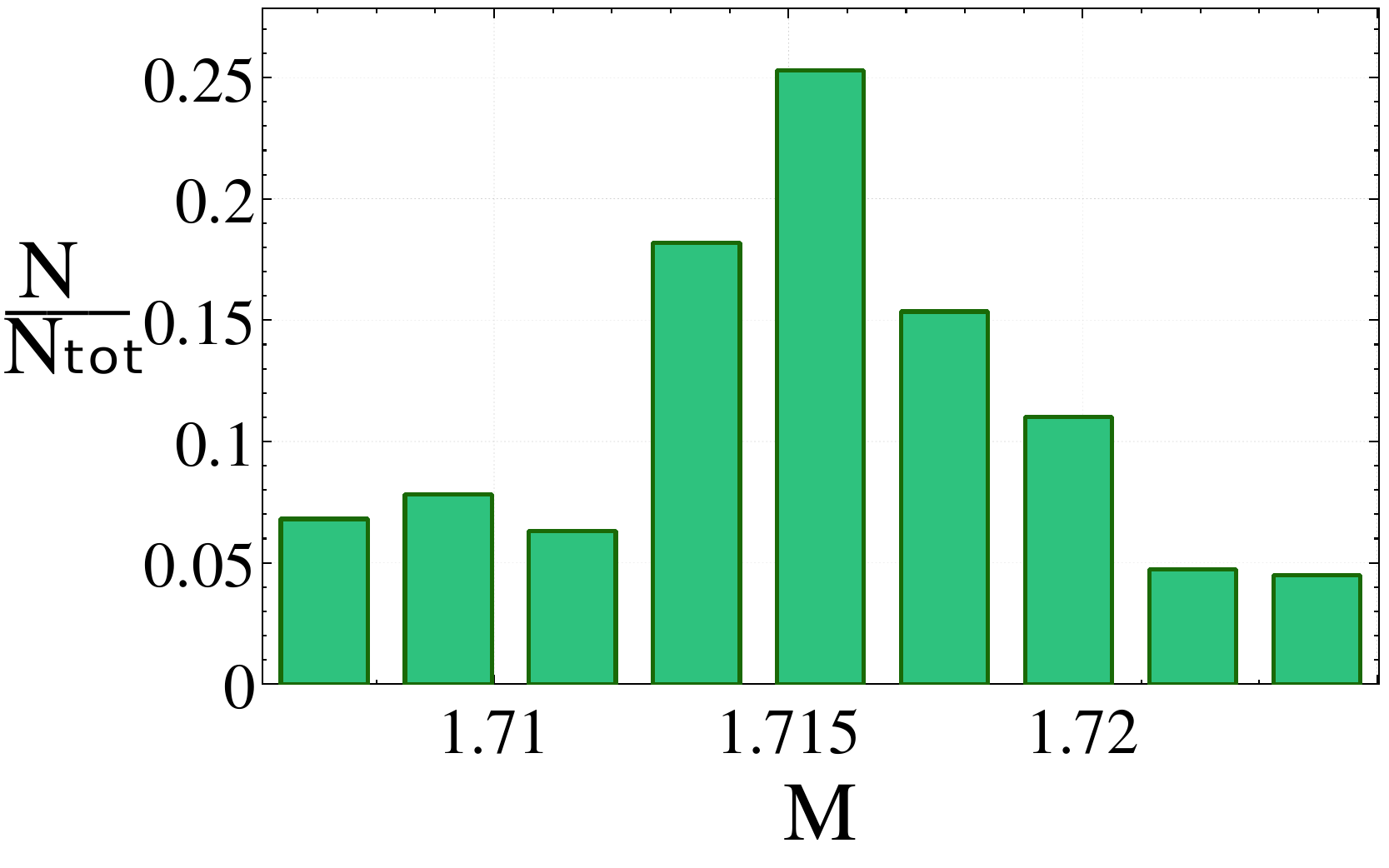}
     \includegraphics[width=0.47\columnwidth]{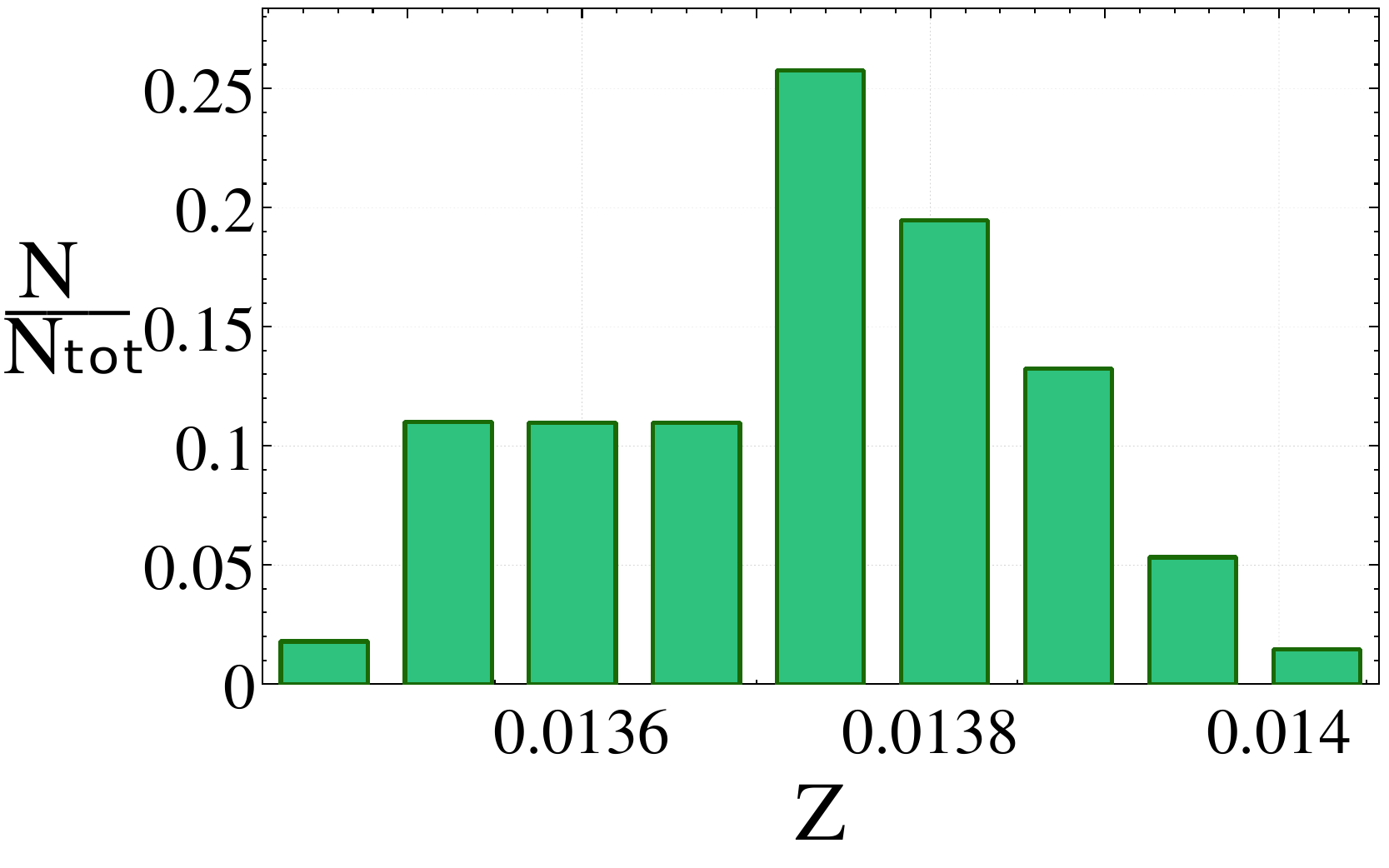}
     \includegraphics[width=0.47\columnwidth]{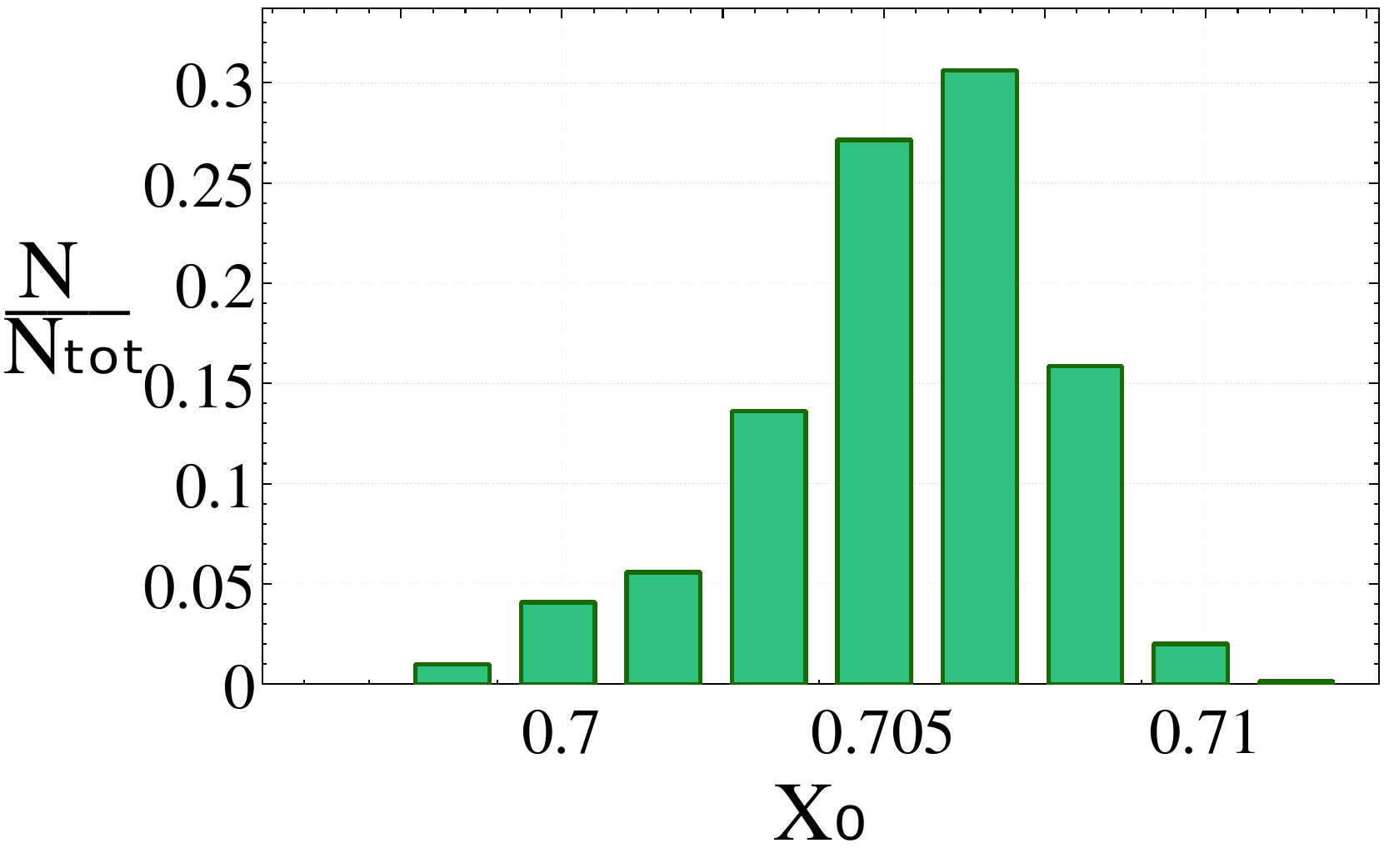}
     \includegraphics[width=0.47\columnwidth]{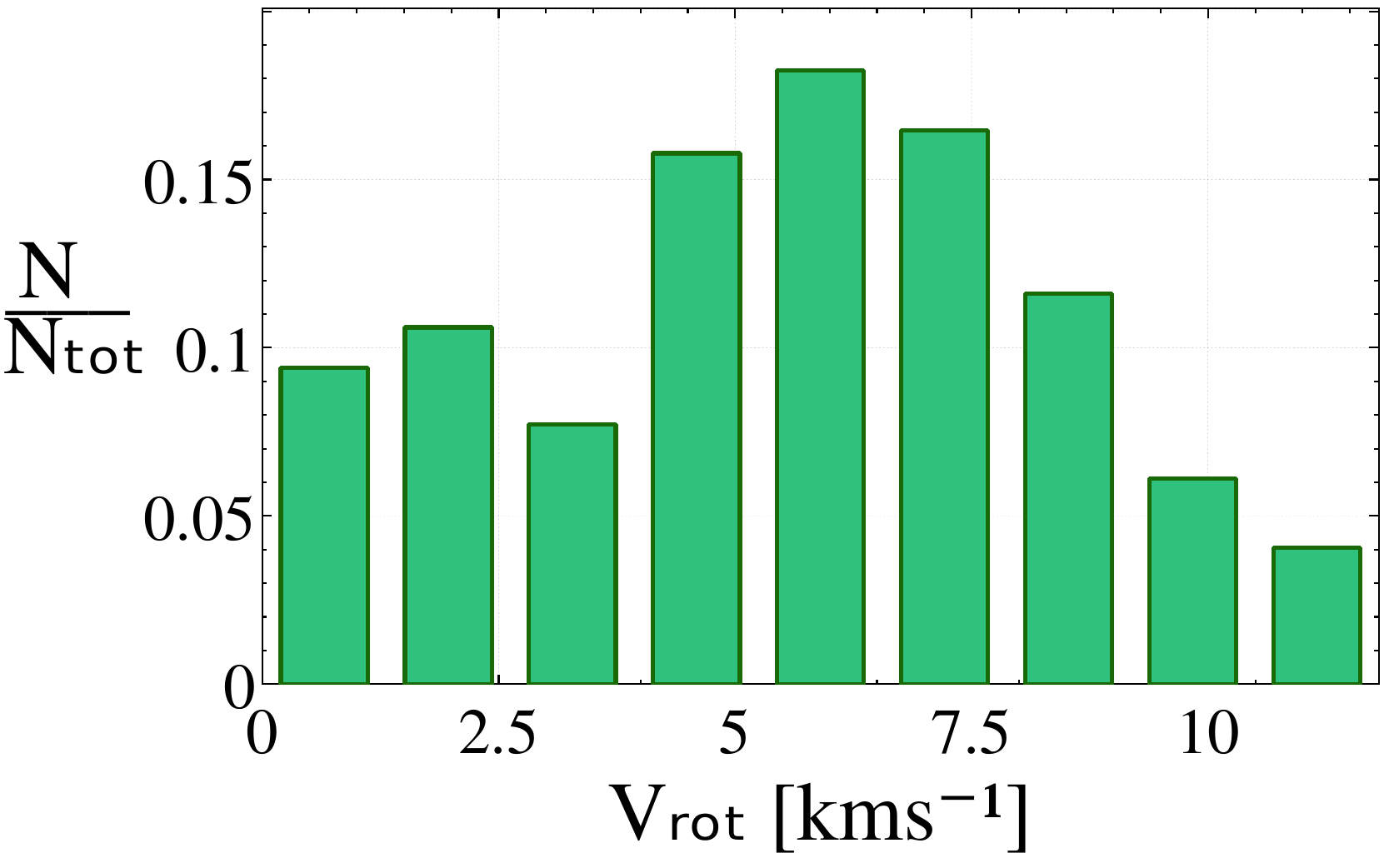}
     \includegraphics[width=0.47\columnwidth]{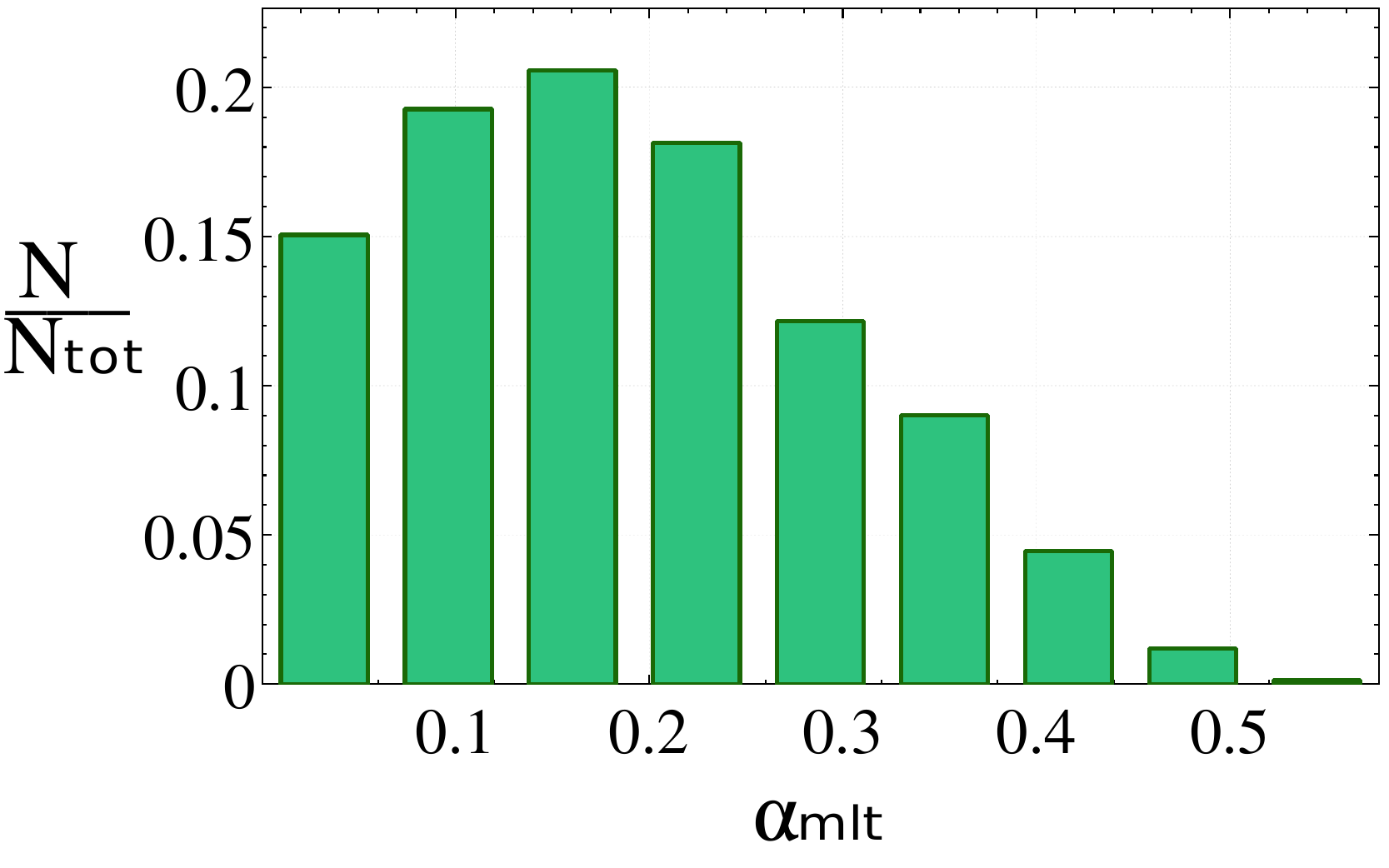}
     \includegraphics[width=0.47\columnwidth]{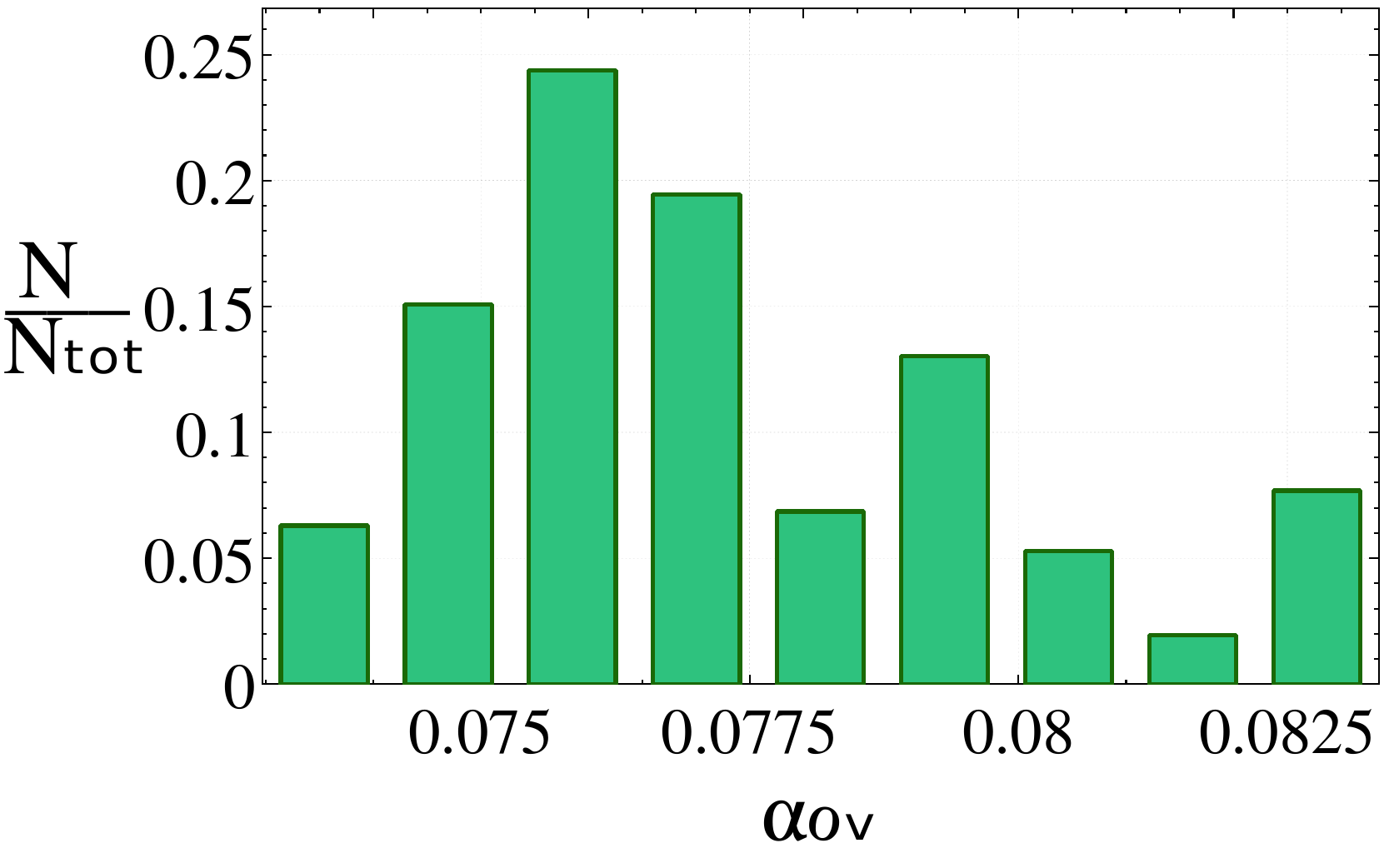}
     \includegraphics[width=0.47\columnwidth]{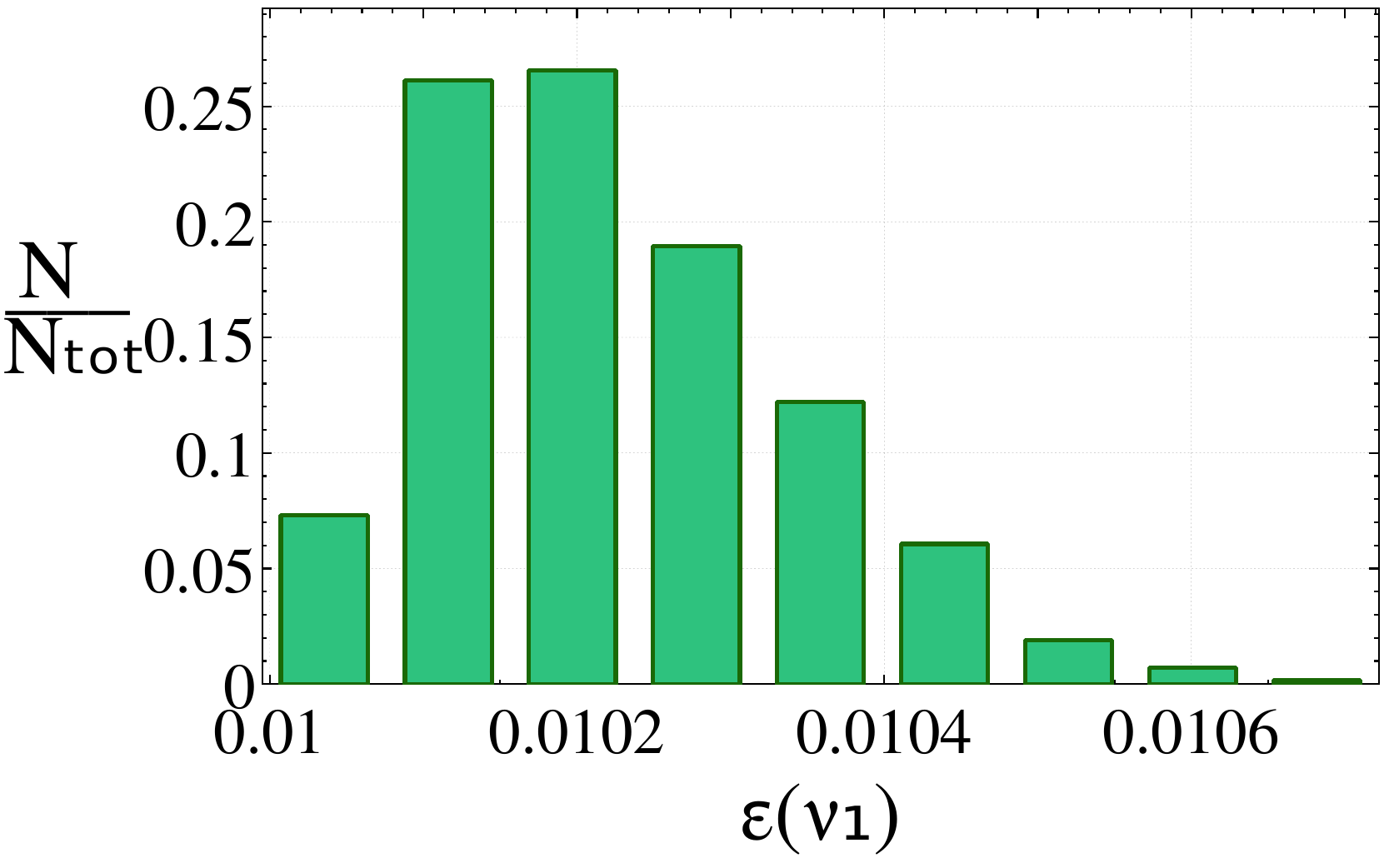}
     \includegraphics[width=0.47\columnwidth]{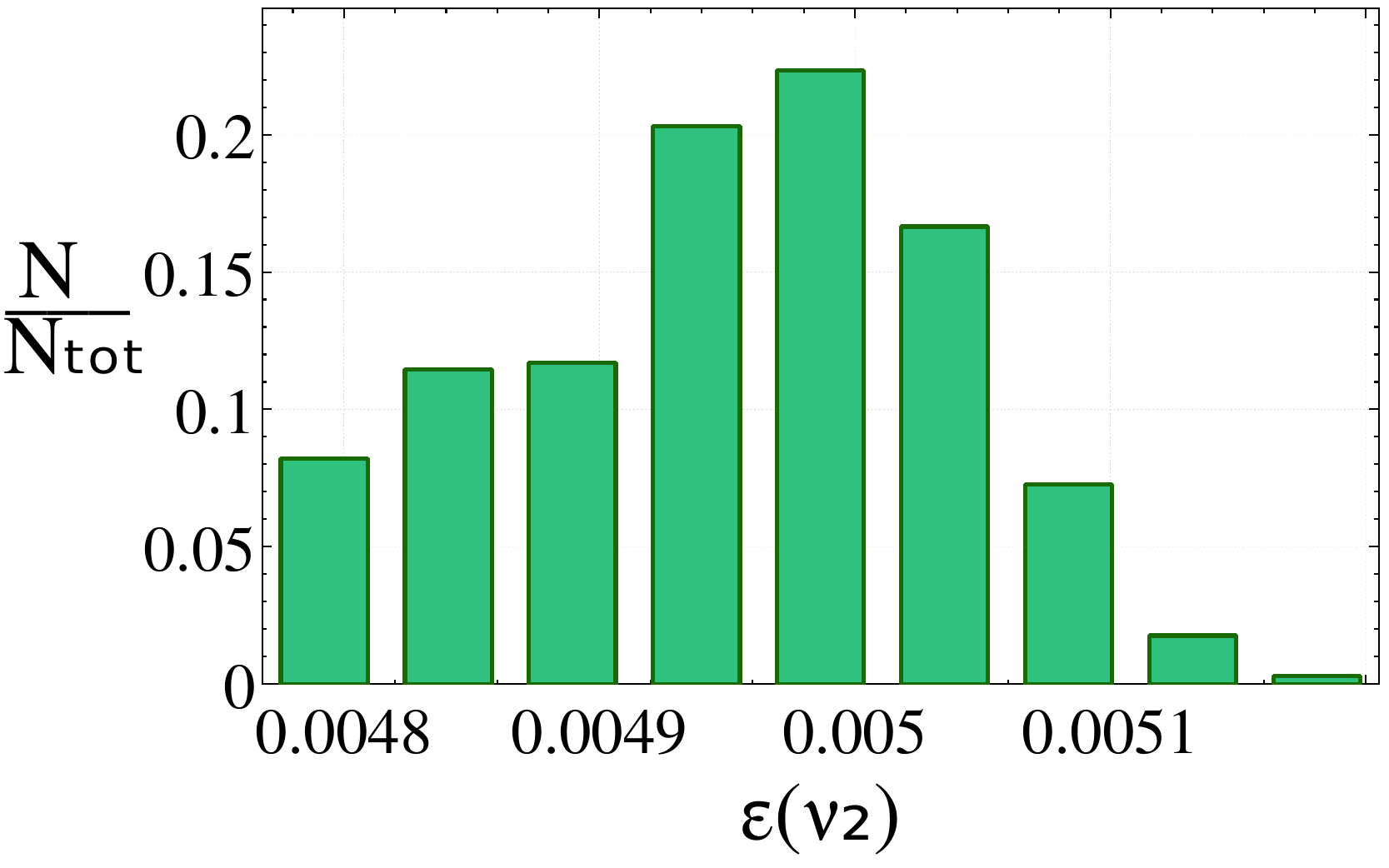}
     \includegraphics[width=0.47\columnwidth]{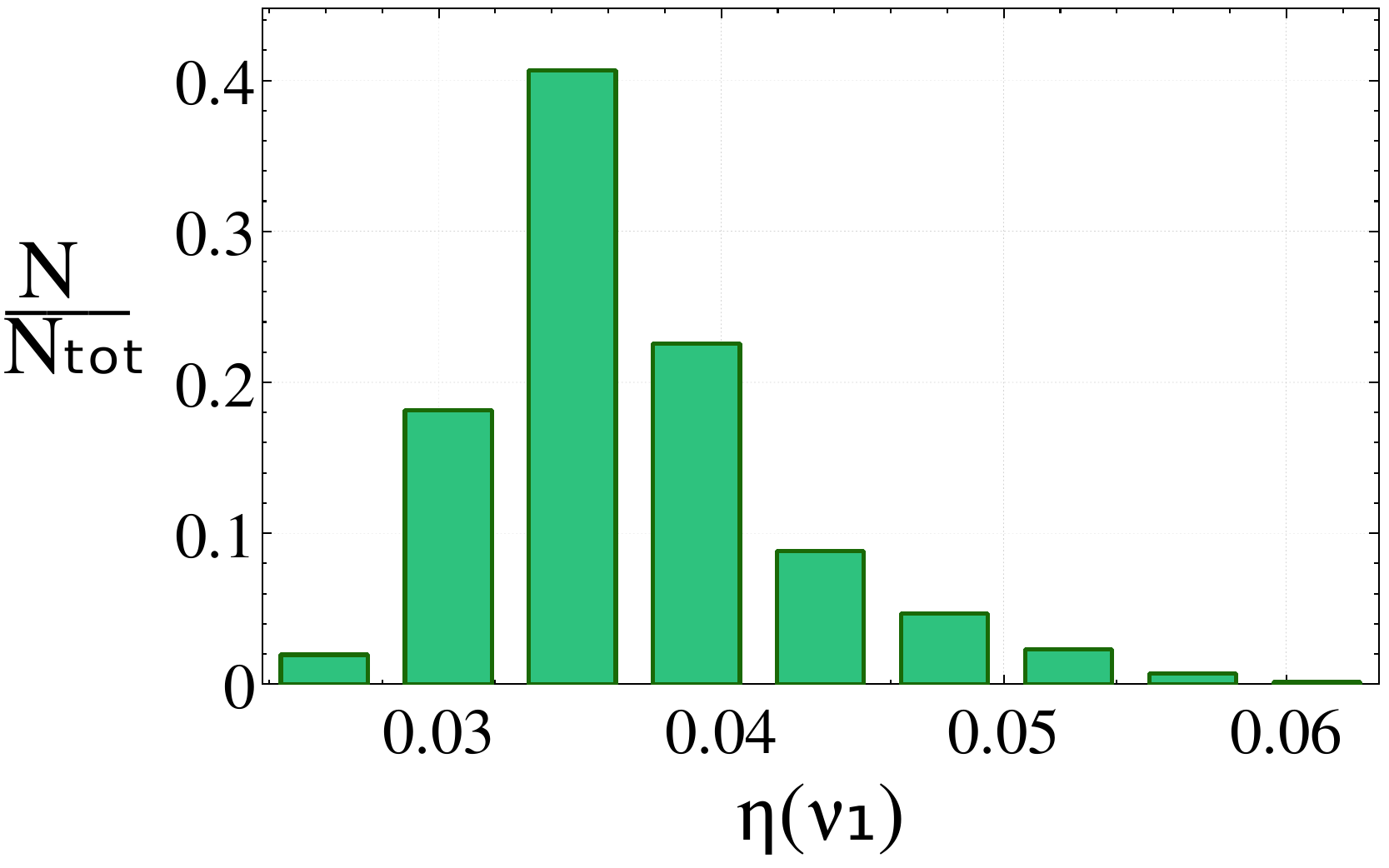}
     \includegraphics[width=0.47\columnwidth]{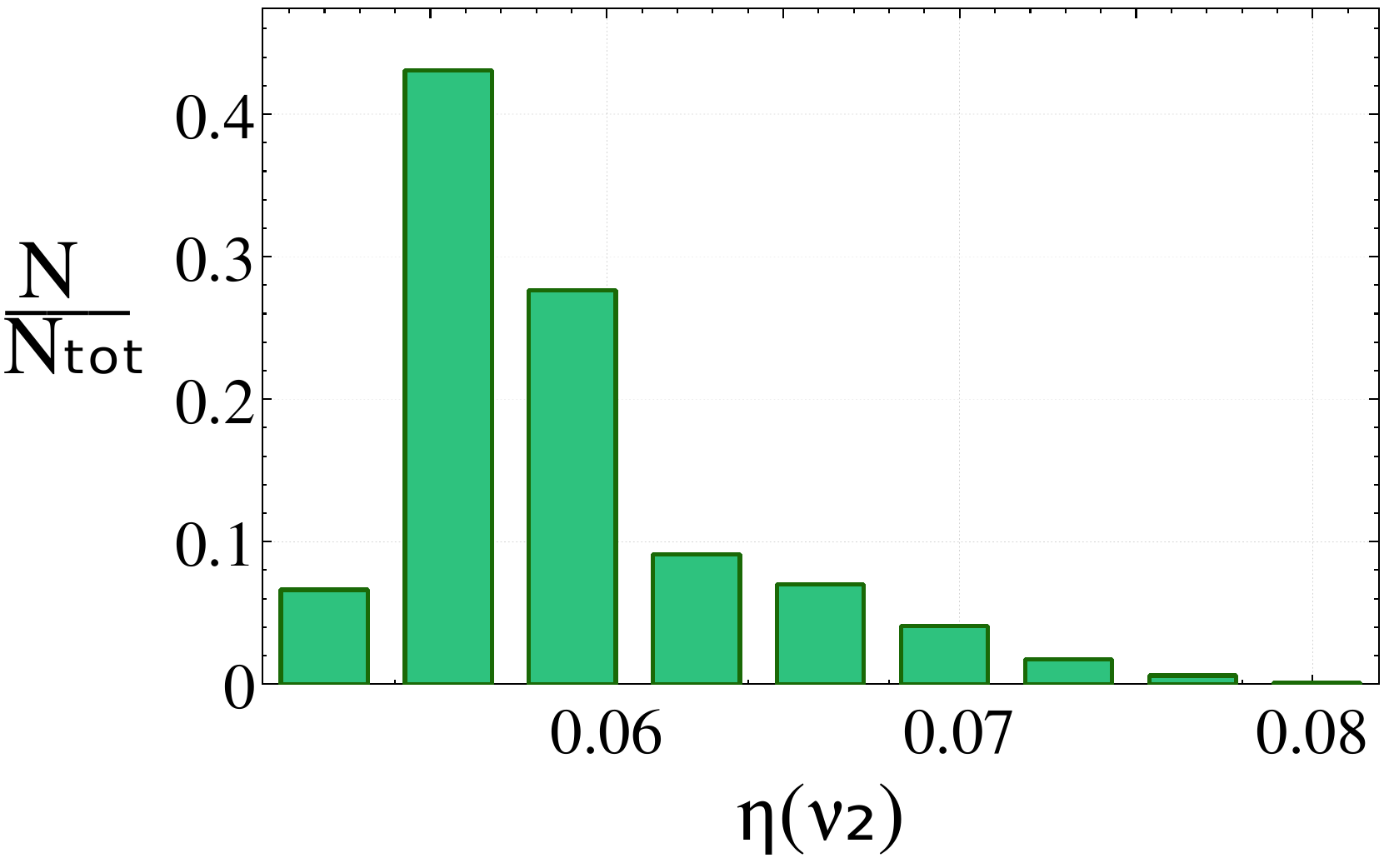}
     \includegraphics[width=0.47\columnwidth]{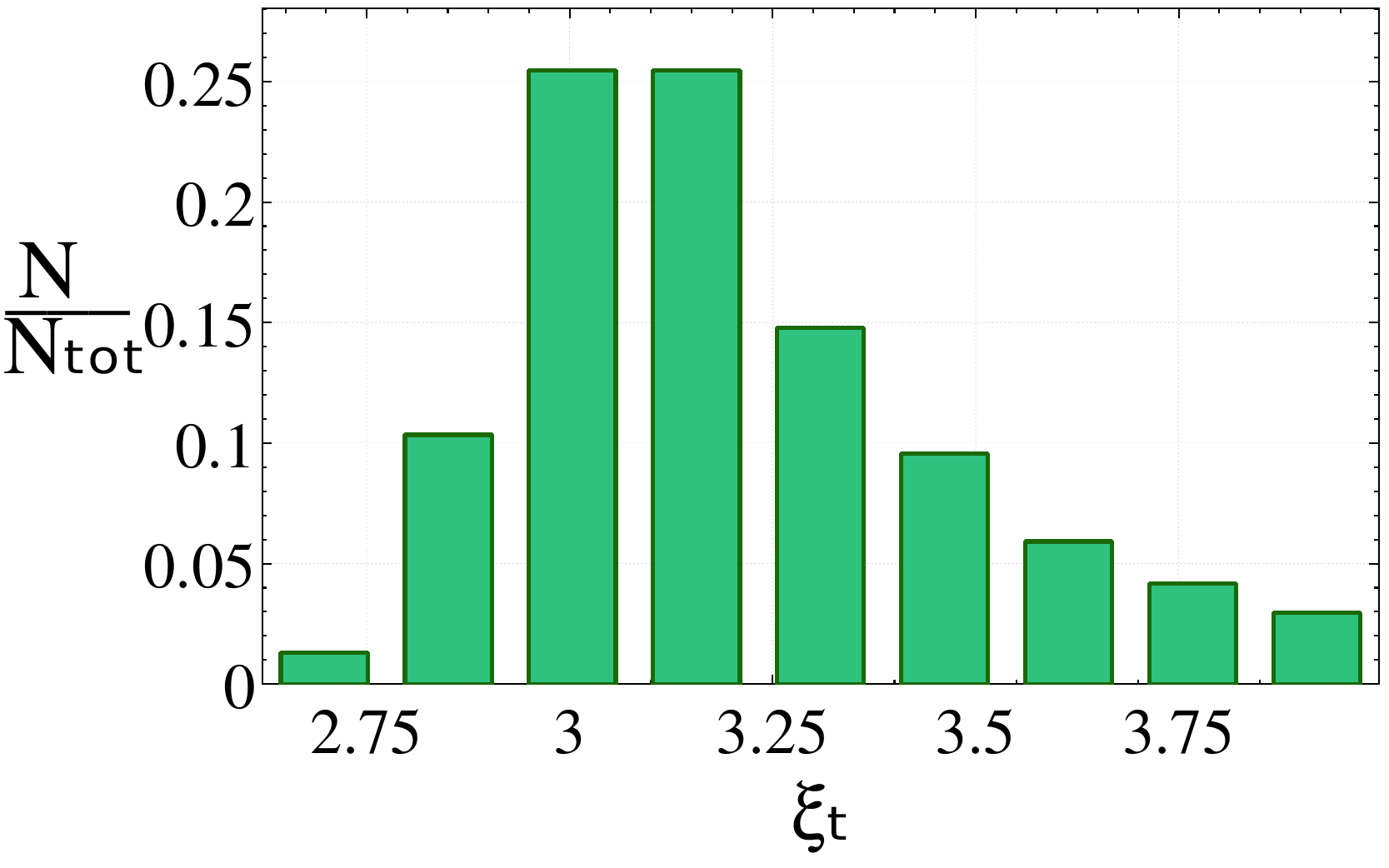}
     \includegraphics[width=0.47\columnwidth]{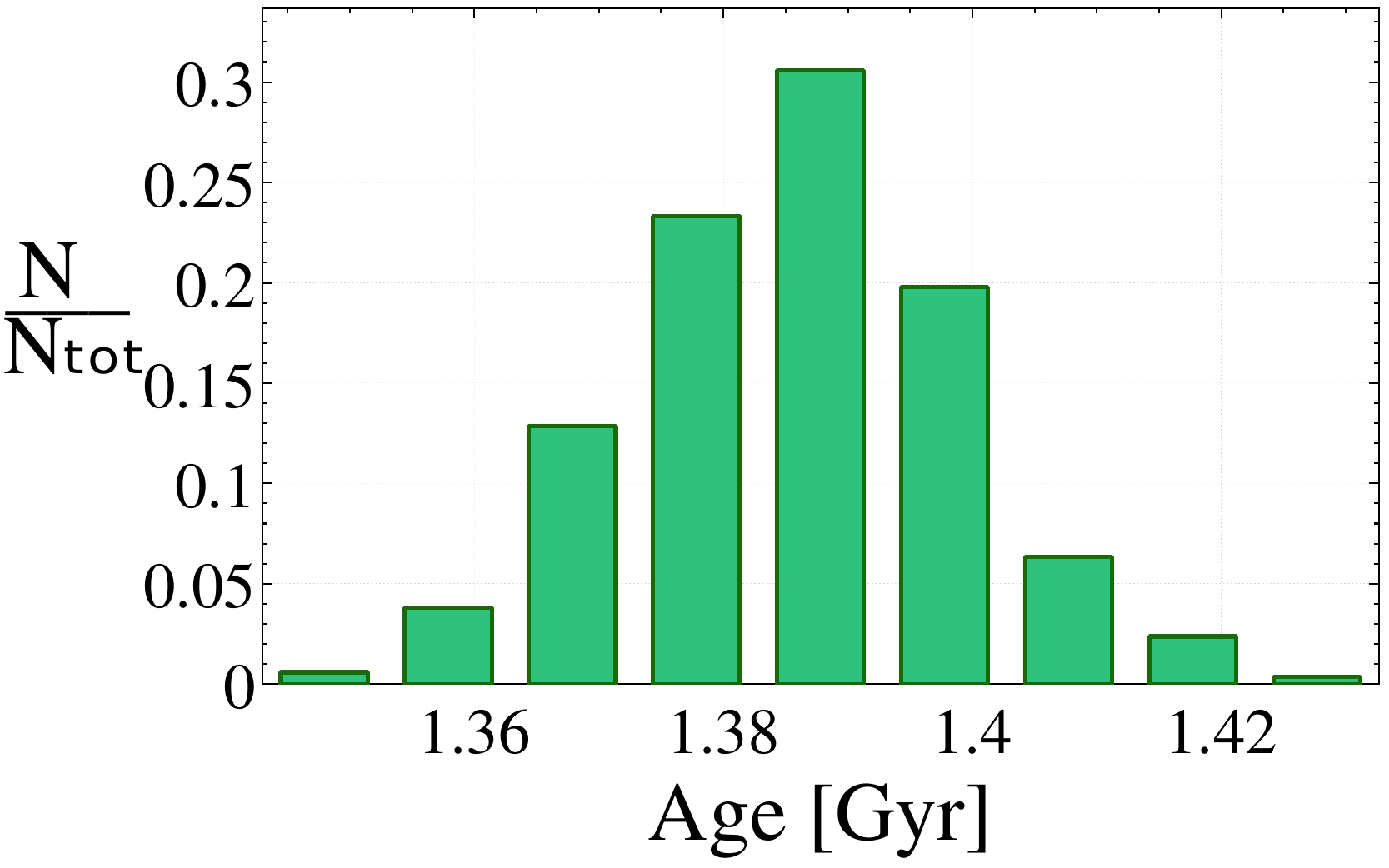}
     \caption{ Same as in Fig.~\ref{fig:simulation_his}, but for simulations including the quadrupole-mode identification of $\nu_3$. In addition, histograms for the microturbulent velocity, $\xi_t$, and stellar age, Age [Gyr], are also shown.}
     \label{fig:simulation_his_quadrupole}
      \end{center}
  \end{figure}
\end{center}

In Figs.\,\ref{fig:simulation_his}, \ref{fig:simulation_his_dipole_sel} and \ref{fig:simulation_his_quadrupole} we present the posterior distributions of selected parameters describing the seismic models of AI Vel.
The simulation shown in Fig.\,\ref{fig:simulation_his} uses as constraints the position of the star in the HR diagram, the empirical $f$-parameters, and the frequencies of the two radial modes. In Fig.\,\ref{fig:simulation_his_dipole_sel}, an additional constraint from the non-radial mode is included, assuming a dipole-mode identification. Figure\,\ref{fig:simulation_his_quadrupole} presents the corresponding results obtained under the quadrupole-mode hypothesis.

\begin{center}
   \begin{figure}[ht]
      \centering
      \includegraphics[width=1\columnwidth]{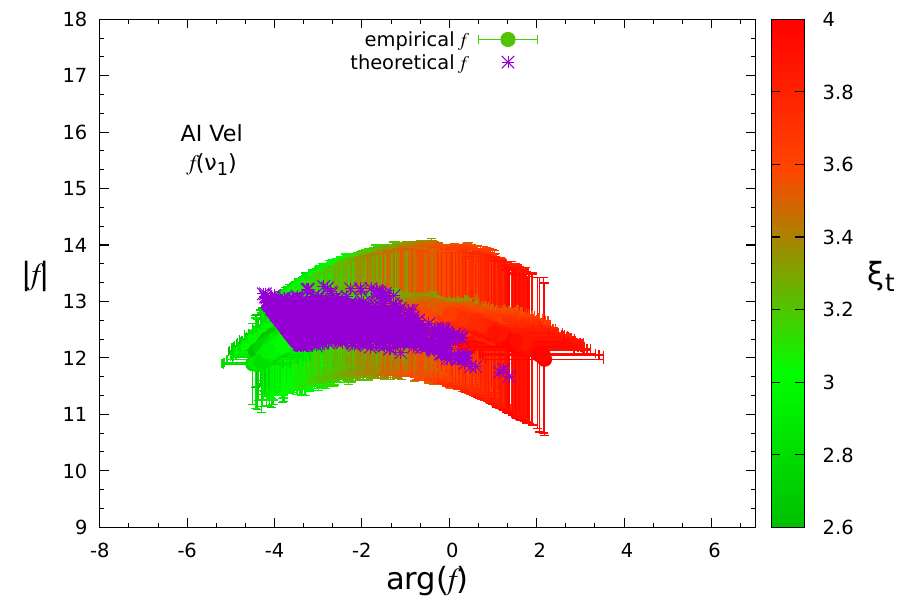}
     \includegraphics[width=1\columnwidth]{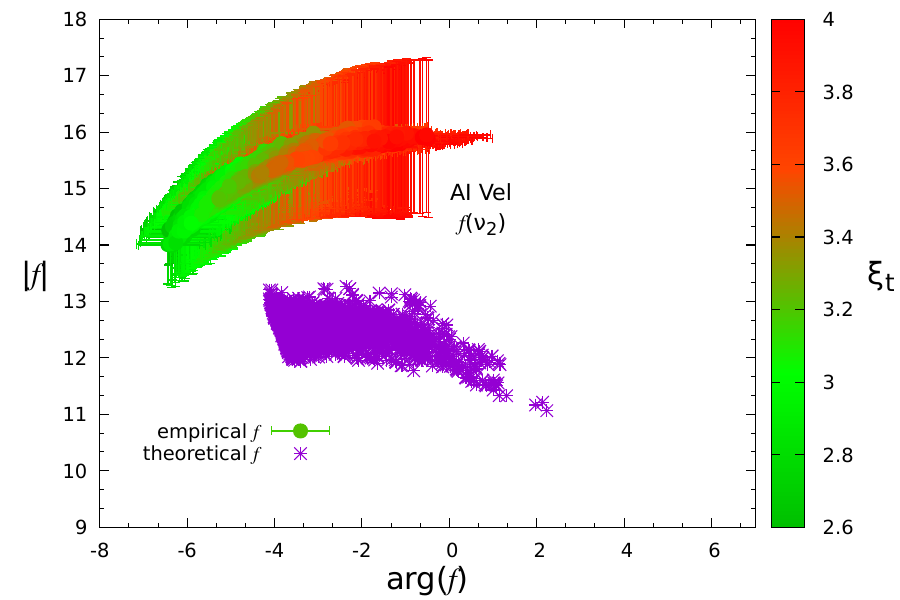}
     \caption{The same as in Fig.\,\ref{fig:empir_f_l1} but for the quadrupole mode hypothesis.}
     \label{fig:empir_f_l2}
  \end{figure}
 \end{center}

\begin{center}
   \begin{figure}[ht]
      \centering
      \includegraphics[width=1\columnwidth]{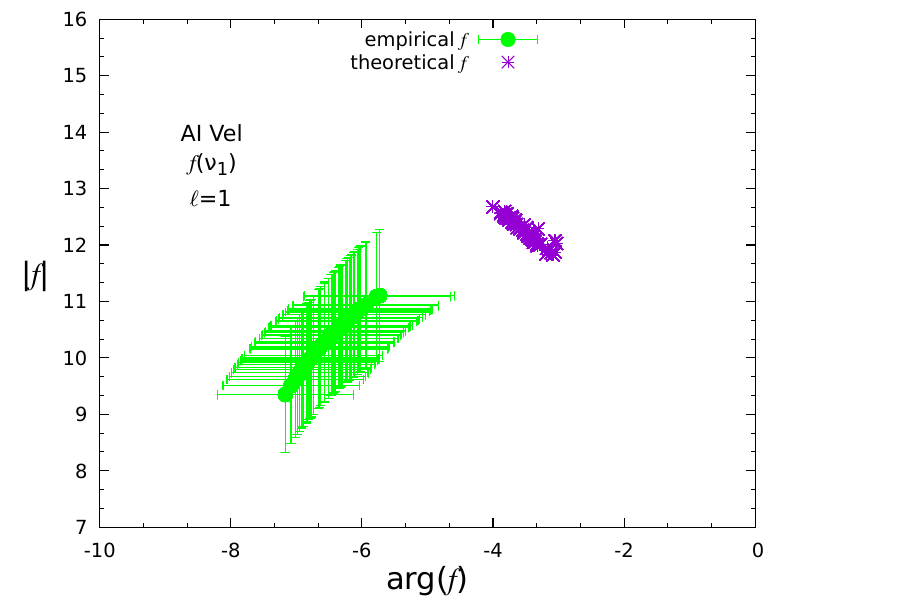}
     \includegraphics[width=1\columnwidth]{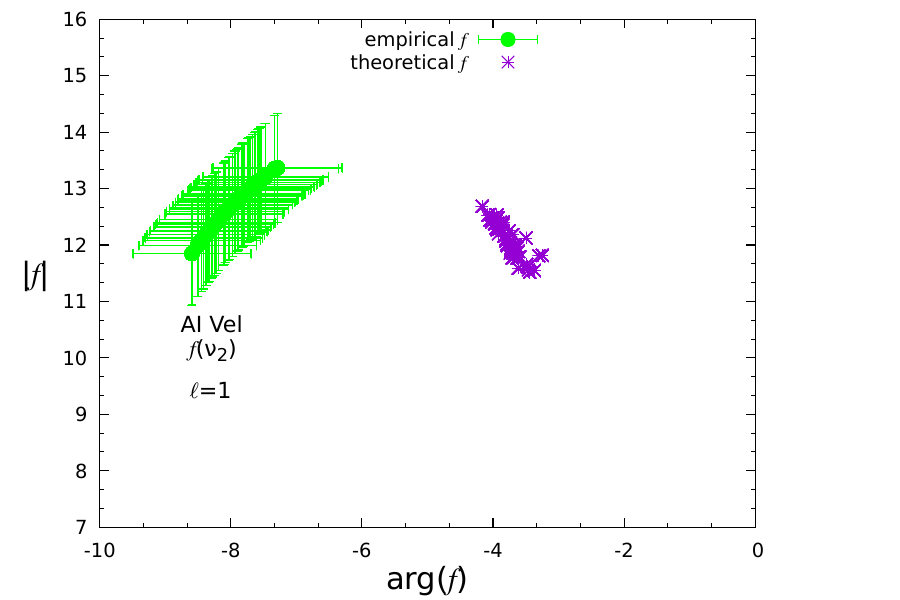}
     \caption{ The same as in Fig.\,\ref{fig:empir_f_l1} but for Kurucz atmosphere models. Microturbulent velocity was set to $\xi_t=2$\,km\,s$^{-1}$.}
     \label{fig:empir_f_Kurucz}
  \end{figure}
 \end{center}

\end{appendix}

\end{document}